\documentclass[onecolumn,pra, aps, amsmath,amssymb,superscriptaddress]{revtex4}

\usepackage{lineno}
  \usepackage{mathptmx}
\usepackage{subfigure}
\usepackage{dcolumn}
\usepackage{amsmath,amssymb}
\usepackage{bm}
\usepackage{color}
\usepackage{overpic}
\usepackage{latexsym}
\usepackage{epstopdf}
\usepackage{color}
\usepackage[english]{babel}
\usepackage{latexsym}
\usepackage{stmaryrd}
\usepackage{amsfonts}
\usepackage{psfrag,graphicx} 
\usepackage{epsf} 
\usepackage{subfigure} 
\usepackage{amsmath} 
\usepackage{amssymb} 
\usepackage{amsfonts}
\usepackage{bm}
\usepackage{natbib}
\usepackage{epstopdf}
\usepackage{appendix}

\definecolor{mygrey}{gray}{0.35}
\definecolor{myblue}{rgb}{0.2,0.2,0.8}
\definecolor{myzard}{cmyk}{0,0,0.05,0}
\definecolor{mywhite}{rgb}{1,1,1}
\definecolor{myred}{rgb}{1,0.,0.3}

\usepackage[colorlinks=true,citecolor=myblue,linkcolor=myred]{hyperref}
\def\be{\begin{equation}}
\def\ee{\end{equation}}
\def\ba{\begin{align}}
\def\enda{\end{align}}
\def\bi{\begin{itemize}}
\def\ei{\end{itemize}}

 \def\ee{\mathord{\rm e}}
 
 \def\ii{\mathord{\rm i}}

\def\half{\textstyle\frac{1}{2}}

\def\fourth{\textstyle\frac{1}{4}}

 \def\ee{\mathord{\rm e}}
 
 \def\ii{\mathord{\rm i}}

\def\half{\textstyle\frac{1}{2}}

\def\fourth{\textstyle\frac{1}{4}}

\renewcommand{\ii}{{\rm i}}
\renewcommand{\ee}{{\rm e}}

\def\beq{\begin{equation}}
\def\beq{\begin{equation}}
\def\eeq{\end{equation}}

 \newcommand{\ket}[1]{|#1\rangle}
 \newcommand{\bra}[1]{\langle #1|}

\begin{document}


\title[Short Title]{Interaction-Dependent Photon-Assisted Tunneling in Optical Lattices: \\
A Quantum Simulator of Strongly-Correlated Electrons and Dynamical Gauge Fields}

\author{Alejandro Bermudez}
\affiliation{Instituto de F\'{\i}sica Fundamental, IFF-CSIC, Calle Serrano 113b, Madrid E-28006, Spain}
\author{Diego Porras}
\affiliation{Department of Physics and Astronomy, University of Sussex, Falmer, Brighton BN19QH, United Kingdom}

\begin{abstract}
We introduce a scheme that combines  photon-assisted tunneling by a moving optical lattice with strong Hubbard interactions, and allows for  the quantum simulation of paradigmatic quantum many-body models. We show that, in a certain regime, this quantum simulator yields an effective  Hubbard Hamiltonian with tunable  bond-charge interactions, a model studied in the context of 
strongly-correlated electrons. In a different regime, we show how to exploit  a correlated destruction of tunneling to explore  Nagaoka ferromagnetism at finite Hubbard repulsion. By changing the photon-assisted tunneling parameters, we can also obtain a   
$t$-$J$ model with independently controllable   tunneling $t$, super-exchange  interaction $J$, and even a Heisenberg-Ising anisotropy. Hence, the full phase diagram of this paradigmatic  model becomes accessible to cold-atom experiments,  departing  from the region  $t\gg J$ allowed   by  standard single-band Hubbard Hamiltonians in the strong-repulsion limit. We finally show that, by generalizing the photon-assisted tunneling scheme, the quantum simulator yields models of dynamical Gauge fields, where  atoms of a given electronic state dress the tunneling of  the atoms with  a different  internal state, leading to Peierls phases that mimic a dynamical  magnetic field. 
\end{abstract}

\maketitle
\setcounter{tocdepth}{2}
\begingroup
\hypersetup{linkcolor=black}
\tableofcontents
\endgroup

\section{Introduction}

Quantum many-body physics studies systems of  interacting particles governed by the laws of quantum mechanics. This task  becomes particularly challenging in  a variety of contexts  in which  the interactions induce strong inter-particle correlations.
 For instance, this strongly-correlated behavior  appears in  condensed-matter models whenever the system cannot be divided into weakly-interacting parts, such that the whole cannot be understood as a sum of its parts and perturbative methods become futile~\cite{emergence}. This inherent complexity underlies the abundance of interesting phases of matter that  emerge at different scales,  but also  
the difficulty in understanding them  from an original microscopic model (e.g. high-$T_{\rm c}$ superconductivity~\cite{review_high_Tc}).  The same occurs at much higher temperatures and densities, where quarks and gluons interact strongly, and lead to a variety of phases  that defy our current understanding (e.g. quark matter~\cite{qcd_phase_diagram}). In the opposite regime, that of extremely low temperatures and densities, {\it ultracold atomic gases} trapped by electromagnetic fields are gradually becoming a paradigm of strongly-correlated behavior in quantum many-body physics~\cite{review_cold_atoms}. In contrast to the above condensed-matter and high-energy scenarios,  ultracold atoms have a unique property: their microscopic properties can be fully characterized and controlled in  experiments. 
This experimental control  has reached such a status  that the dream of exploiting a quantum system to understand the properties of a complex quantum many-body model (i.e.  a {\it quantum simulator}~\cite{qs_feynman}) is already  an experimental reality~\cite{qs_cold_atoms}.

{ Ultracold gases of neutral atoms} can be trapped in periodic optical potentials obtained from  the interference  of laser beams. The dynamics of the atoms in these  {\it optical lattices} resembles that of tightly-bound electrons in metals, such that this system can be considered to be a synthetic solid whose dimensionality and lattice structure can be experimentally tailored, while the nuisance of impurities, disorder, or other uncontrolled microscopic degrees of freedom present in real solids, is totally absent. Starting from this synthetic solid, it is possible to design a variety  of quantum many-body models whose microscopic parameters can be experimentally characterized and controlled. For instance, the scattering of atoms leads to a  short-range interaction that can be tuned all the way from weak to strong repulsion, such that  the superfluid-insulator quantum phase transition of the bosonic~\cite{bose_hubbard,bose_hubbard_ol,bhm_ol_exp} and fermionic~\cite{hubbard,fermi_hubbard_ol, fhm_ol_exp}  Hubbard models becomes accessible to experiments. For sufficiently strong repulsion, the half-filled Hubbard model leads to a Heisenberg antiferromagnet~\cite{superexchange_anderson}, which yields  a playground for  quantum magnetism with two-component bosonic atoms~\cite{superexchange_theory, superexchange_bhm_exp}, and the starting point to study high-$T_{\rm c}$ superconductivity with fermionic ones~\cite{rvb_cuprates,superexchange_fhm} upon controlled doping (i.e. inserting atomic vacancies with respect to the half-filled system).

In this article, we will combine this strongly-correlated behavior  with  external {\it periodic drivings} to obtain a flexible quantum simulator of quantum many-body models. In the context of optical lattices, there is a large body of relevant results  regarding periodic drivings by  modulations of the trapping optical potential.  For instance, it is  possible  to periodically modulate the phase of the laser beams forming the optical lattice, as already demonstrated in experiments of chaotic dynamics with cold atoms~\cite{phase_modulated_chaos_prop,phase_modulated_chaos_exp}. Another possibility is to modify the detuning of these  laser beams linearly in time, usually referred to as lattice acceleration, which leads to a linear gradient (i.e. constant force) in the lattice reference frame, and gives rise  to Bloch oscillations~\cite{bloch_oscilations_detuning_modulation}.  From this perspective, the previous phase modulation~\cite{phase_modulated_chaos_prop,phase_modulated_chaos_exp} may also be interpreted as a periodic forcing. The combination of these two forces permitted probing the Wannier-Stark ladder spectrum~\cite{wannier_stark_ladder_ol_exp}, and  testing the phenomenon of coherent destruction of tunneling in the absence of the gradient~\cite{cdt_ol,cdt_ol_phase_mod}. Recently~\cite{state_dependent_cdt}, a state-dependent coherent destruction of tunneling has been demonstrated in optical lattices by using a modulated magnetic field gradient instead of the phase modulation.
 In presence of an energy gradient,  one study the phenomenon of {\it photon-assisted tunneling}~\cite{cdt_ol,pat_ol_phase_mod}. We shall be particularly interested in  such photon-assisted tunneling effect, whereby the  atoms can  tunnel in the presence of an energy penalty (i.e. the linear gradient) by absorbing photons from the external driving (i.e. the periodic phase modulation). 

Photon-assisted tunneling (PAT) by phase modulation has  also turned out to be a  useful tool  for  quantum simulations. The dependence of the dressed tunneling on the modulation parameters has been used to  drive the system across the superfluid-insulator transition~\cite{pat_sf_mott,pat_sf_mott_exp,pat_sf_mott_bis}, and
  to control the  tunneling anisotropy of Bose-Hubbard models in triangular lattices leading to magnetic frustration~\cite{frustration_phase_mod,frustration_phase_mod_exp}. A subject of research that has received considerable attention recently  is the quantum simulation of orbital magnetism, whereby the  atoms  mimic the behavior of electrons in solids subjected to additional  magnetic fields~\cite{qs_cold_atoms}. Since the atoms are neutral, one must design specific schemes to simulate the effect of  artificial/synthetic magnetic fields~\cite{jz_gauge_fields,gd_gauge_fields}, and  PAT by phase modulation has also been exploited in this respect (see~\cite{goldman_review} for a recent review that also covers schemes that do not exploit  PAT). When the   phase modulation leads to an inhomogeneous periodic forcing~\cite{gauge_pat_gradient}, it is possible to dress the tunneling with  an effective complex phase. Unfortunately, this simple proposal does not allow for the quantum simulation of synthetic magnetic fields~\cite{not_gauge_pat_gradient}, and alternative schemes   have been considered. For instance, two-tone phase modulations lead to synthetic fluxes in arbitrary lattices~\cite{gauge_two_tone_drivings}, while the simpler single-tone  phase modulations yield  staggered fluxes in certain types of lattices~\cite{gauge_phase_plus_gradient,triangular_gauge,haldane_ol}.
  
   A different possibility would be to abandon the periodic  phase modulation, and investigate other types of  drivings that can lead to the aforementioned synthetic Gauge fields.
  Instead of modulating the phase of the optical lattice, one can introduce a periodic driving by considering a bi-chromatic deep optical lattice, which can lead to staggered synthetic fluxes~\cite{lim}. Alternatively, a simple periodic driving by using a pair of slightly detuned and weaker Raman beams (i.e. a shallow {\it moving optical lattice}), which has been considered in the context of PAT for trapped ions~\cite{gauge_pat},  ultracold atoms~\cite{staggered_superlattice_exp}, and generic  lattice models~\cite{gauge_pat_long} that can be applied to a variety of contexts. For ultracold atoms, this moving optical lattice yields, in a certain regime, an inhomogeneous periodic modulation of the on-site energies of the effective Hubbard model, which can be exploited as a  flexible PAT  toolbox for quantum simulations of synthetic Gauge fields~\cite{gauge_pat,staggered_superlattice_exp,gauge_pat_long, gauge_ol,nathan_pm}. Here, the atoms tunnel in the presence of an energy penalty (i.e. again, a linear gradient) by absorbing photons from the external periodic driving (i.e. this time, the moving optical lattice), and acquire a Peierls phase that plays the role of a synthetic Gauge field, and  depends upon the wavevectors of the Raman beams. 

In this work, we explore a modification of this scheme by considering that the energy penalty can also be caused by the on-site Hubbard interactions, yielding a {\it Hubbard blockade} that inhibits the tunneling of atoms  involving double occupation of a lattice site. The combination of this Hubbard blockade with the periodic driving  by a moving optical lattice will induce an
 {\it interaction-dependent photon-assisted tunneling}. Let us note that the interplay of Hubbard interactions, linear gradients, and phase modulation of the optical lattice, has been shown to be responsible for interaction-shifted resonances in the PAT of  Bose-Hubbard dimers~\cite{interaction_pat_theory} and chains~\cite{interaction_resonances_cdt_theory}. Similar effects have been observed experimentally by considering a periodic modulation of the intensity of the optical-lattice laser beams~\cite{pat_bhm,superexchange_pat}, rather than the aforementioned phase modulation. This interaction-dependent PAT can lead to new schemes to control effective magnetic Hamiltonians~\cite{pat_bhm,superexchange_pat}, or to methods that enhance the effects of three-body interactions~\cite{int_gradient_3_body}. We should also mention other proposals that are  relevant for the particular subject of our work. These concern the engineering of density-dependent tunnelings by either combining laser-assisted schemes with state-dependent  lattices~\cite{density_dependent_gauge_fields}  in the spirit of the original proposal~\cite{jz_gauge_fields},  or a periodic modulation of the Hubbard interactions~\cite{interaction_CDT,interaction_modulations,cold_atoms_int_modulation_fermions,peierls_cold_atoms_int_modulation,cold_atoms_double_int_modulation_fermions}.

In this work, we will show that  the  interaction-dependent PAT by a moving optical lattice offers a very flexible quantum simulator for paradigmatic models of strongly-correlated electrons, and can even allow for the quantum simulation of synthetic Gauge fields that are dynamical, in contrast to the static ones mentioned above. As explained below, such synthetic Gauge fields evolve under a free Hamiltonian that is not Gauge invariant, and thus depart from the standard Lattice  theory approach to Gauge theories.
 
 This article is organized as follows. In Sec.~\ref{scheme}, we introduce the scheme to implement the interaction-dependent PAT with ultracold  atoms in optical lattices, and derive a set of effective Hamiltonians that depend on the specific driving,  lattice dimensionality, and  fermionic/bosonic quantum statistics. The scope of the many-body phenomena that can be studied through these effective Hamiltonians is discussed in Sec.~\ref{qs_applications}. Finally, we present our conclusions and outlook in Sec.~\ref{conclusions}.

\section{Interaction-dependent photon-assisted tunneling}
\label{scheme}

In this section, we present a detailed proposal to combine PAT by periodic drivings with strong Hubbard interactions in experiments of ultracold alkali atoms in optical lattices. We  show that by controlling {\it (i)} the atomic interactions by Feshbach resonances, and {\it (ii)} an additional moving optical lattice, one can exploit an interaction-dependent PAT  to delve into interesting quantum many-body models that arise in the condensed-matter and high-energy scenarios. 

The starting point is, as customary~\cite{review_cold_atoms, lewenstein_review}, a trapped atomic gas  described in second quantization
\beq
\label{eq:originalH}
H=\sum_{\sigma}\int{\rm d}^3r\Psi_\sigma^\dagger({\bf r})\left(\frac{-\boldsymbol{\nabla}^2}{2m_\sigma}+\epsilon_\sigma\ket{\sigma}\bra{\sigma}+V_{\rm ot}(\bf r)\right)\Psi_\sigma^{\phantom{\dagger}}({\bf r})+\frac{1}{2}\sum_{\sigma,\sigma'}\int{\rm d}^3r\int{\rm d}^3r'\Psi_\sigma^\dagger({\bf r})\Psi_{\sigma'}^\dagger({\bf r}')V_{\rm int}^{\sigma\sigma'}({\bf r}-{\bf r}')\Psi_{\sigma'}^{\phantom{\dagger}}({\bf r}')\Psi_\sigma^{\phantom{\dagger}}({\bf r}),
\eeq
where we set $\hbar=1$ henceforth. Here, $\Psi_\sigma^\dagger({\bf r}),\Psi_\sigma^{\phantom{\dagger}}({\bf r})$  create-annihilate atoms with mass $m_\sigma$ at the position {\bf r}, and in the electronic state $\ket{\sigma}$ corresponding to a particular energy  level $\epsilon_\sigma$ of the atomic groundstate manifold. To remain as general as possible, we consider that the components labelled by $\sigma$ may correspond to the states of a bosonic gas, a fermionic one, or a mixture of both, which will determine the particular algebraic relations of the  creation-annihilation operators. 

We have introduced  an optical  trapping potential  $V_{\rm ot}(\boldsymbol{r})=\sum_{\alpha}V_{0,\alpha}\sin^2(kr_{\alpha})+\half m \omega_{\rm t,\alpha}^2r^2_{\alpha}$ that consists of: 
{\it (i)} A state-independent periodic potential, where   $V_{0,\alpha}$ are the ac-Stark shifts of independent  pairs of retro-reflected laser beams of wavelength $\lambda=2\pi/k$, which propagate 
 along the axis $\alpha\in\{x,y,z\}$, and are far detuned with respect to the excited atomic states. To obtain state-independent potentials, we assume that the detunings of the laser beams with respect to the excited states are much larger than the energy splittings in $\epsilon_\sigma$,   and that the retro-reflected  beams along each axis have parallel linear polarizations~\cite{ol_review}. To obtain independent potentials along each axis, the  pairs of interfering beams must have orthogonal polarizations, or detuned frequencies, with respect to other pairs of  beams propagating along a different axis. Therefore, it is possible to  tune the lattice depths $V_{0,\alpha}$  independently by controlling the beam intensities, which allows to tailor the effective dimensionality of the system.
  {\it (ii)} A   harmonic  trapping caused by a combination of the laser Gaussian profile and the retro-reflection scheme, where  $m\omega_{\rm t,\alpha}\lambda^2$ is the characteristic trapping energy assumed to be sufficiently weak   $\omega_{\rm t,\alpha}\ll \omega^0_{\sigma,\alpha}=2E_{\rm R,\sigma}\sqrt{V_{0,\alpha}/E_{\rm R,\sigma}}$, where  $E_{\rm R,\sigma}=k^2/2m_\sigma$ is the so-called recoil energy.   
  
The final ingredient of the cold-atom Hamiltonian~\eqref{eq:originalH} is  the  $s$-wave scattering, which dominates at sufficiently low temperatures. This is described by  a contact pseudo-potential $V_{\rm int}^{\sigma\sigma'}({\bf r}-{\bf r}')=4\pi a_{\sigma\sigma'}\delta({\bf r}- {\bf r}')/2\mu_{\sigma\sigma'}$ characterized by the reduced masses $\mu_{\sigma\sigma'}=m_\sigma m_{\sigma'}/(m_\sigma+m_{\sigma'})$, and the scattering lengths $a_{\sigma\sigma'}$ for the collisions of two atoms in the internal state $\ket{\sigma,\sigma'}$. Such scattering lengths can be modified experimentally through an external magnetic field via the  so-called Feshbach resonances~\cite{feshbach}.  In the following sections, we  show how to  exploit  an interaction-dependent PAT as new tool to engineer  quantum many-body Hamiltonians by tuning these scattering lengths appropriately in the presence of a weak moving optical lattice.

\subsection{Scheme for a periodically-modulated  ultracold Fermi gas}
\label{sec:fermions}

Let us consider a single-species  gas  of fermionic atoms with two hyperfine states $\ket{{\uparrow}}=\ket{F,M}, \ket{{\downarrow}}=\ket{F',M'}$, such that there is a unique mass $m_\uparrow=m_\downarrow=:m$ and recoil energy $E_{\rm R,\uparrow}=E_{\rm R,\downarrow}=:E_{\rm R}$. We introduce the Wannier basis, $\Psi_\sigma({\bf r})=\sum_{{\bf i}}w({\bf r}-{\bf R}_{\bf i})f_{{\bf i},\sigma}$, where $w({\bf r}-{\bf R}_{\bf i})$ are the Wannier functions, and $f_{{\bf i},\sigma}$ are the  fermionic operators that annihilate an atom of pseudospin $\sigma=\{\uparrow,\downarrow\}$ at the minima of an  optical lattice potential   ${\bf R}_{\bf i}$ labelled by the vector of integers $\bf i$. We shall consider cubic optical lattices, although we note that the scheme detailed below can be directly applied to any other lattice geometry.  In this basis,  the general Hamiltonian~\eqref{eq:originalH} can be expressed in terms of  the standard Fermi-Hubbard model~\cite{fermi_hubbard_ol}, namely
\beq
\label{eq:hubbard}
H_{\rm FH}=H_{\rm loc}+H_{\rm kin}+V_{\rm int}=\sum_{{\bf i},\sigma}\epsilon_{{\bf i},\sigma}f_{{\bf i},\sigma}^{\dagger}f_{{\bf i},\sigma}^{\phantom{\dagger}}-\sum_{{\bf i},\alpha}\sum_{\sigma}\left(t_{\alpha}f_{{\bf i},\sigma}^{\dagger}f_{{\bf i}+{\bf e}_{\alpha},\sigma}^{\phantom{\dagger}}+{\rm H.c.}\right)+\frac{1}{2}\sum_{{\bf i},\sigma}U_{\sigma\overline{\sigma}}f_{{\bf i},\sigma}^{\dagger}f_{{\bf i},\overline{\sigma}}^{\dagger}f_{{\bf i},\overline{\sigma}}^{\phantom{\dagger}}f_{{\bf i},\sigma}^{\phantom{\dagger}},
\eeq
where we have introduced  the  unit vectors ${\bf e}_{\alpha}$, and the notation $\overline{\sigma}=\{\downarrow,\uparrow\}$ for $\sigma=\{\uparrow,\downarrow\}$. Here, $\epsilon_{{\bf i},\sigma}=\epsilon_\sigma+\sum_\alpha\half m \omega_{\rm t,\alpha}^2({ R}_{\bf i,\alpha})^2$ includes the hyperfine energies and the weak parabolic trapping potential,  $t_{\alpha}$ is the tunneling strength of atoms between neighboring potential wells along the $\alpha$-axis, and $U_{\sigma\overline{\sigma}}=U_{\overline{\sigma}\sigma}$ stands for the on-site interaction due to s-wave scattering, which only allows for interactions between fermions of a different state. As customary, we have neglected long-range tunnelings and interactions, which requires  sufficiently deep optical lattices  $\{V_{0,x},V_{0,y},V_{0,z}\}\gg E_{\rm R}$. This can be  justified considering that 
\beq
\label{eq:gaussian}
t_{\alpha}=\frac{4}{\sqrt{\pi}}E_{\rm R}\left(\frac{V_{0,\alpha}}{E_{\rm R}}\right)^{3/4}\ee^{-2\sqrt{\frac{V_{0,\alpha}}{E_{\rm R}}}}, \hspace{2ex} U_{\sigma{\sigma'}}=\sqrt{\frac{8}{\pi}}ka_{\sigma\sigma'}E_{\rm R}\left(\frac{V_{0,x}V_{0,y}V_{0,z}}{E_{\rm R}^3}\right)^{1/4},
\eeq
while longer range terms are exponentially suppressed with the distance by ${\rm exp}\{-m\omega^{0}_{\sigma,\alpha}(R_{\bf i,\alpha}-R_{\bf j,\alpha})^2/4\}$ according to a Gaussian approximation. Let us also note that, for $\omega_{\rm t,\alpha}\ll \omega^0_{\sigma,\alpha}$, the harmonic trapping does not modify the tunneling, but simply leads to a local term in the Wannier basis that has been incorporated in the local on-site energies $\epsilon_{{\bf i},\sigma}$ of the Fermi-Hubbard model~\eqref{eq:hubbard}.

\begin{figure*}
\centering
\includegraphics[width=1\columnwidth]{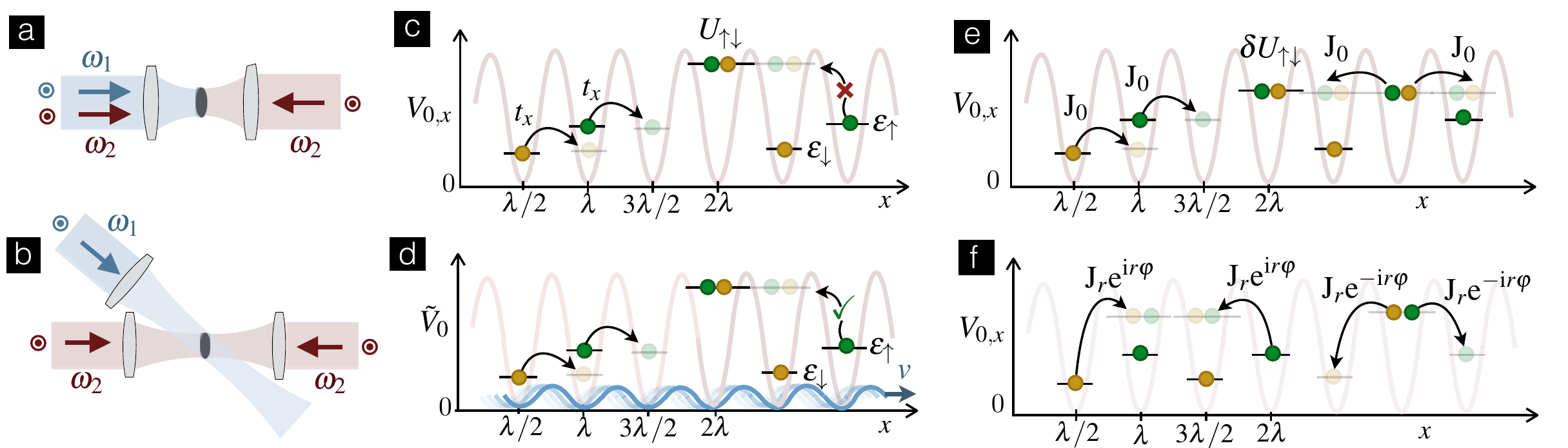}
\caption{ {\bf Scheme of the spin-independent PAT for fermions: }{\bf (a)} Laser scheme corresponding to the static optical lattice formed by the retro-reflected beams (red arrows) of frequency $\omega_2$, and the moving optical lattice formed by a slightly detuned  Raman beam of frequency $\omega_1$ (blue arrow) and the counter-propagating laser beam of frequency  of frequency $\omega_2$ (red arrow). All the laser beams have a linear polarization perpendicular to the plane of the figure, as illustrated by the filled circles. {\bf (b)} Same as before, but considering that the detuned Raman beam (blue arrow) forms an angle with respect to the static optical lattice beams (red arrows). {\bf (c)} Fermionic atoms in two hyperfine states $\ket{{\uparrow}}$ (green circles), $\ket{{\downarrow}}$  (orange circles) are trapped at the nodes of a static optical lattice potential (red lines). In the regime of strong $s$-wave scattering,  atoms  tunnel with strength $t_{x}$ between unoccupied sites centered at the energies $\epsilon_{\uparrow},\epsilon_{\downarrow}$. Conversely, tunneling of one atom to an already-occupied site is inhibited by the large energy penalty $t_{x}\ll U_{\uparrow\downarrow}$ (i.e. Hubbard blockade). {\bf (d)} Moving optical lattice potential  (blue lines for snapshots of the wave traveling at speed $v=\Delta\omega/\Delta k$). The tunneling involving doubly-occupied sites can be reactivated when the atoms absorb $r$ photons from the moving lattice, providing the required energy to overcome the interaction penalty. {\bf (e,f)} The PAT contains diagonal {\bf (e)} and off-diagonal {\bf (f)} correlated  events. The diagonal terms correspond to a dressed tunneling within the subspaces of single- or doubly-occupied sites, and is controlled by the Bessel function ${\rm J}_0:={\rm J}_0(\eta)$. The off-diagonal  tunneling connects the subspaces of single- and doubly-occupied sites, and is controlled by the Bessel function ${\rm J}_r:={\rm J}_r(\eta)$ and dressed by the tunneling phase $\ee^{\pm \ii r\varphi}$. Also note that the residual dressed  interaction $\delta U_{\uparrow\downarrow}$ is changed with respect to the  original bare one $ U_{\uparrow\downarrow}$.}
\label{fig_pat_scheme}
\end{figure*}

We consider the limit of very strong repulsion $U_{\uparrow\downarrow}\gg t_{\alpha}$, such that the bare tunneling events  connecting single-occupied sites to  doubly-occupied ones  are energetically inhibited, as depicted in Fig.~\ref{fig_pat_scheme}{\bf (c)}. We shall refer to this tunneling suppression as a Hubbard blockade by reminiscence of the Coulomb blockade that inhibits the sequential tunneling of electrons through quantum dots. The idea is to overcome this Hubbard blockade via the phenomenon of PAT (i.e. the fermions obtain the  required energy for tunneling  by absorbing photons from an external periodic driving). As shall be shown below,  the tunneling of fermions between two lattice sites will depend on the density of fermions of the opposite pseudospin populating those sites, which shall be exploited to build a quantum simulator. We now discuss two possible periodic drivings that lead to such PAT, and organize our presentation  by introducing the less demanding schemes first, adding  more complexity gradually.

\subsubsection{Two-component fermions in spin-independent moving optical lattices}

{\it (i) One-dimensional scheme:} To introduce the main ideas in the simpler setting, let us start by considering a one-dimensional (1D) Fermi-Hubbard model obtained from Eq.~\eqref{eq:hubbard} for $\{V_{0,y},V_{0,z}\}\gg V_{0,x}$, such that only tunneling along the $x$-axis is relevant
\beq
\label{eq:hubbard1d}
H_{\rm FH}=H_{\rm loc}+H_{\rm kin}+V_{\rm int}=\sum_{{i},\sigma}\epsilon_{{i},\sigma}f_{{i},\sigma}^{\dagger}f_{ i,\sigma}^{\phantom{\dagger}}-\sum_{{ i},\sigma}\left(t_{x}f_{{ i},\sigma}^{\dagger}f_{{ i}+1,\sigma}^{\phantom{\dagger}}+{\rm H.c.}\right)+\frac{1}{2}\sum_{{ i},\sigma}U_{\sigma\overline{\sigma}}f_{{ i},\sigma}^{\dagger}f_{{ i},\overline{\sigma}}^{\dagger}f_{{ i},\overline{\sigma}}^{\phantom{\dagger}}f_{{ i},\sigma}^{\phantom{\dagger}}.
\eeq
 As an external periodic driving, we consider a  moving optical lattice stemming from  a pair of non-copropagating laser beams  along the $x$-axis. These beams  are slightly detuned with respect to each other (i.e. traveling wave   as opposed to the standing wave   of  the static optical lattice,  Fig.~\ref{fig_pat_scheme}{\bf (d)}), but again far detuned with respect to the excited states (i.e. Raman beams). Moreover, they have the same linear polarization as the laser beams of the static optical lattice to ensure an spin-independent potential~\cite{ol_review}. Since this  moving lattice could induce a spurious tunneling due to the recoil kick imparted by the lasers, we 
assume that its intensity  is much weaker  $\tilde{V}_{0}\ll V_{0,x}$ (i.e. $\tilde{t}_x=\tilde{V}_{0}{\rm exp}\{-\frac{\pi^2}{4}(V_{0,\alpha}/E_{\rm R})^{1/2}\}\ll t_x$ in the Gaussian approximation). In this regime, the effect of the moving lattice is a periodic spin-independent modulation of the trapping frequencies of each potential well 
\beq
\label{eq:modulation}
H_{\rm mod}(t)=\sum_{i,\sigma}\frac{\tilde{V}_{0}}{2}\cos\left(\Delta k X_i-\Delta\omega t+\varphi\right)n^{\phantom{\dagger}}_{i,\sigma}, \hspace{2ex} n_{i,\sigma}^{\phantom{\dagger}}=f_{i,\sigma}^{{\dagger}}f_{i,\sigma}^{\phantom{\dagger}},
\eeq
where $\Delta k=({\bf k}_1-{\bf k}_2)\cdot {\bf e}_x$ is the wavevector difference, $X_i=\frac{\lambda}{2}i$ stands for the minima of the original optical-lattice potential,  $\Delta\omega=\omega_1-\omega_2$ is the  detuning of the  laser beams, and $\varphi$ is the relative phase with respect to the static optical lattice. By setting $\Delta\omega\approx U_{\uparrow\downarrow}/r$ for a positive integer $r\in\mathbb{Z}$, the above Hubbard blockade   for $U_{\uparrow\downarrow}\gg t_{\alpha}$ can be overcome through the absorption of $r$ photons from the periodic driving (see Fig.~\ref{fig_pat_scheme}{\bf (d)}). To be more precise, as the driving comes from a two-photon ac-Stark shift, the process involves absorbing $r$ photons from one laser beam and  subsequently emitting   them onto the other laser beam.

To provide explicit expressions for this { interaction-dependent PAT}, we move to the interaction picture with respect to $U_0(t)=\mathcal{T}\left(\exp\{\ii \int_0^t{\rm d}\tau(V_{\rm int}+H_{\rm mod}(\tau))\}\right)$, such that the fermionic  annihilation operators become
\beq
\label{eq:int_picture}
U_0(t)f_{i,\sigma}U^\dagger_0(t)=\ee^{-\ii t U_{\uparrow\downarrow}n_{i,\overline{\sigma}}}\ee^{\ii\frac{\tilde{V}_{0}}{2\Delta\omega}\sin(\Delta k X_i-\Delta\omega t+\varphi)}f_{i,\sigma}=\ee^{-\ii t U_{\uparrow\downarrow}n_{i,\overline{\sigma}}}\sum_{n\in\mathbb{Z}}{\rm J}_n\left(\frac{\eta}{2}\right)\ee^{\ii n(\Delta k X_i-\Delta\omega t+\varphi)}f_{i,\sigma},
\eeq
where we have gauged away an irrelevant phase by transforming the fermion operators~\cite{comment_gauge}. The second part of the equality is obtained after introducing   the important parameter
\beq
\eta=\tilde{V}_{0}/\Delta\omega,
\eeq
 and using the Jacobi-Anger expansion for  first-order Bessel functions ${\rm J}_n(z)$, namely $\ee^{\ii z\sin\theta}=\sum_{n\in\mathbb{Z}}{\rm J}_n(z)\ee^{\ii n \theta}$~\cite{book_bessel}. For simplicity, we  set  $\Delta k X_i=(\half \Delta k \lambda)i=\pi i$, which can be achieved with laser beams of  the standing and moving  lattices of the same wavelength, and both propagating along the $x$-axis.  In this configuration, it thus suffices to add a single laser beam detuned with respect to the optical-lattice laser beams  (see Fig.~\ref{fig_pat_scheme}{\bf (a)}). However, this could be generalized to $\Delta k X_i=(\half \Delta k \lambda)i=\pi i/r$, which may be relevant if the detuned Raman beam does not propagate along the $x$-axis, but makes some angle with respect to that axis (e.g. $r$=2 for an angle $\alpha=\pi/6$, see Fig.~\ref{fig_pat_scheme}{\bf (b)}).

By substituting the expression~\eqref{eq:int_picture} in the kinetic Hamiltonian $H_{\rm kin}(t)=U_0(t)H_{\rm kin}U^{\dagger}_0(t)$, one finds
 \beq
 H_{\rm kin}(t)=-\sum_{i,\sigma}\left(t_{x,\overline{\sigma}}(t)f_{i,\sigma}^{\dagger}f_{i+1,\sigma}^{\phantom{\dagger}}+{\rm H.c.}\right),\hspace{2ex}t_{x,\overline{\sigma}}(t)=t_{x}\ee^{-\ii t U_{\uparrow\downarrow}\Delta n_{i+1,\overline{\sigma}}}\mathrm{f}(t),
\eeq
where we have introduced the population difference operator 
\beq
\Delta n_{i+1,\overline{\sigma}}= n_{i+1,\overline{\sigma}}-n_{i,\overline{\sigma}}, 
\eeq
and a dynamical dressing function
\beq
\label{eq:mod_function}
\mathrm{f}(t)=\sum_{n,m}{\rm J}_n\left(\frac{\eta}{2}\right){\rm J}_m\left(\frac{\eta}{2}\right)\ee^{-\ii (n\pi i-m\pi(i+1))}\ee^{-\ii (n-m)\varphi}\ee^{\ii (n-m)\Delta\omega t}.
\eeq
 As announced earlier, the tunneling that connects single-occupied sites to  doubly-occupied ones, yielding  $\langle \Delta n_{i+1,\overline{\sigma}}\rangle =\pm 1$, is negligible in the absence of the driving $\tilde{V}_{0}=0$. In this limit,  the dressing function is $\mathrm{f}(t)=1$, such that  the dressed tunneling  can be neglected $\langle  t_{x,\overline{\sigma}}(t)\rangle=t_{x}\ee^{\mp\ii t U_{\uparrow\downarrow}}\approx0$ in a rotating-wave approximation for $t_{x}\ll U_{\uparrow\downarrow}$ (see Fig.~\ref{fig_pat_scheme}{\bf (c)}). By switching on the periodic driving $\tilde{V}_{0}\neq 0$, this tunneling becomes assisted by the harmonics of the dressing function that are close to resonance with the Hubbard interaction, namely for the integers fulfilling  $\pm U_{\uparrow\downarrow}=(n-m)\Delta\omega$  (Fig.~\ref{fig_pat_scheme}{\bf (d)}). In particular, by   assuming that 
\beq
\label{pat_regime}
t_{x}, \delta U_{\uparrow\downarrow}=(U_{\uparrow\downarrow}-r\Delta\omega)\ll U_{\uparrow\downarrow}\approx r\Delta\omega,
\eeq
we can neglect the majority of tunneling events using a similar rotating-wave argument, except for those that satisfy $n=r\Delta n_{i+1,\overline{\sigma}}+m$. Accordingly,  the dressing function becomes simplified  
\beq
\label{eq_RWA}
\mathrm{f}(t)=\sum_{m}{\rm J}_m\left(\frac{\eta}{2}\right){\rm J}_{m+r\Delta n_{i+1,\overline{\sigma}}}\left(\frac{\eta}{2}\right)\ee^{\ii m\pi}\ee^{-\ii \pi r\Delta n_{i+1,\overline{\sigma}}i}\ee^{-\ii r\varphi \Delta n_{i+1,\overline{\sigma}}}.
\eeq
We  further assume that the laser detuning is chosen in such a way  that $r$ is an even integer, and  set   $\ee^{-\ii \pi r\Delta n_{i+1,\overline{\sigma}}i}=1$ for any population difference. Making use of the Neumann-Graf addition formula for Bessel functions~\cite{book_bessel}, namely $\sum_{n\in\mathbb{Z}}{\rm J}_n(z){\rm J}_{n+\nu}(z)\ee^{\ii n \theta}={\rm J}_{\nu}\left(2|z\sin(\theta/2)|\right)\ee^{\ii(\pi-\theta)\nu/2}$, we can express the PAT in terms of a single Bessel function
\beq
t_{x,\overline{\sigma}}(t)=t_{x}\ee^{-\ii t \delta U_{\uparrow\downarrow}\Delta n_{i+1,\overline{\sigma}}}{\rm J}_{r\Delta n_{i+1,\overline{\sigma}}}\left(\eta\right)\ee^{-\ii r\varphi \Delta n_{i+1,\overline{\sigma}}},
\eeq
which should be understood in terms of its Taylor series expansion.

The total time-evolution operator $U(t)=U_0^{\dagger}(t)\ee^{-\ii t\sum_{i}\delta U_{\uparrow\downarrow}n_{i\uparrow}n_{i\downarrow}}\ee^{-\ii H_{\rm eff}t},$ can thus be expressed in terms of a time-independent Hubbard Hamiltonian of the form~\eqref{eq:hubbard}. However, the dressed tunneling strengths   now depend on  the density of fermions of the opposite pseudospin, and the residual Hubbard interaction depends on the resonance condition in Eq.~\eqref{pat_regime}, such that
\beq
\label{eq:eff_H}
 H_{\rm eff}=\sum_{i,\sigma}\epsilon_{i,\sigma}f_{i,\sigma}^{\dagger}f_{i,\sigma}^{\phantom{\dagger}}-\sum_{i,\sigma}\left(t_{x}{\rm J}_{r\Delta n_{i+1,\overline{\sigma}}}\left(\eta\right)\ee^{-\ii r\varphi \Delta n_{i+1,\overline{\sigma}}}f_{i,\sigma}^{\dagger}f_{i+1,\sigma}^{\phantom{\dagger}}+{\rm H.c.}\right)+\frac{1}{2}\sum_{i,\sigma}\delta U_{\sigma\overline{\sigma}}f_{i,\sigma}^{\dagger}f_{i,\overline{\sigma}}^{\dagger}f_{i,\overline{\sigma}}^{\phantom{\dagger}}f_{i,\sigma}^{\phantom{\dagger}}.
\eeq
As announced in the introduction, the Hubbard-blockaded tunneling becomes activated through a PAT phenomenon, and leads to a density-dependent tunneling  that can be written as follows
\beq
\label{eq:tunn_op}
\begin{split}
{\rm J}_{r\Delta n_{i+1,\overline{\sigma}}}\left(\eta\right)&={\rm J}_0\left(\eta\right)h_{i,\overline{\sigma}}h_{i+1,\overline{\sigma}}+{\rm J}_0\left(\eta\right)n_{i,\overline{\sigma}}n_{i+1,\overline{\sigma}}+{\rm J}_{r}\left(\eta\right)n_{i,\overline{\sigma}}h_{i+1,\overline{\sigma}}+{\rm J}_{r}\left(\eta\right)h_{i,\overline{\sigma}}n_{i+1,\overline{\sigma}}.
\end{split}
\eeq
where we have defined the hole number operators $h_{i,\overline{\sigma}}=1-n_{i,\overline{\sigma}}$. The first term in Eq.~\eqref{eq:tunn_op} describes the tunneling within the subspace of single-occupied sites  $\mathcal{H}_{\rm s}$, whereas the second one corresponds to   tunneling within the subspace of doubly-occupied sites $\mathcal{H}_{\rm d}$ (see Fig.~\ref{fig_pat_scheme}{\bf (e)}). These  subspaces can be described as two Hubbard sub-bands centered around $\epsilon_{\rm s}=0$ and $\epsilon_{\rm d}=\delta U_{\uparrow\downarrow}$. Finally, the third and fourth terms stand for tunneling events  connecting the single-occupied to the doubly-occupied subspaces (see Fig.~\ref{fig_pat_scheme}{\bf (f)}). These four terms can be thus understood   as  diagonal and off-diagonal  tunnelings.  We note that a similar classification of the tunneling events of the original Hubbard model~\eqref{eq:hubbard1d} can be performed by using $H_{\rm kin}\to \sum_{i,\sigma}(n_{i,\overline{\sigma}}+h_{i,\overline{\sigma}})H_{\rm kin}^{\sigma,i}(n_{i,\overline{\sigma}}+h_{i,\overline{\sigma}})$.
Let us emphasize, however, that the ratio of these diagonal/off-diagonal processes cannot be controlled, which contrasts  the PAT Hamiltonians~\eqref{eq:tunn_op}, where  one can  adjust the intensity of the moving optical lattice $\tilde{V}_{0}$ such that the ratio of the Bessel functions  attains the desired value. This  will be crucial to obtain a {\it tunable $t$-$J$ model} with  fully controllable parameters in Sec.~\ref{tJ_model}. In Sec.~\ref{bond_charge_model}, we will use this formulation to connect the effective model to the so-called {\it bond-charge interactions}, which leads to a quantum simulator of exotic Hubbard models. Moreover,  the tunneling of one pseudospin acquires a complex phase that depends on the density of the other pseudospin, which will be crucial for the quantum simulation of {\it dynamical Gauge fields} in Sec.~\ref{gauge_fields}, when complemented with additional terms that allow us to control each pseudospin independently.

\begin{figure*}
\centering
\includegraphics[width=.9\columnwidth]{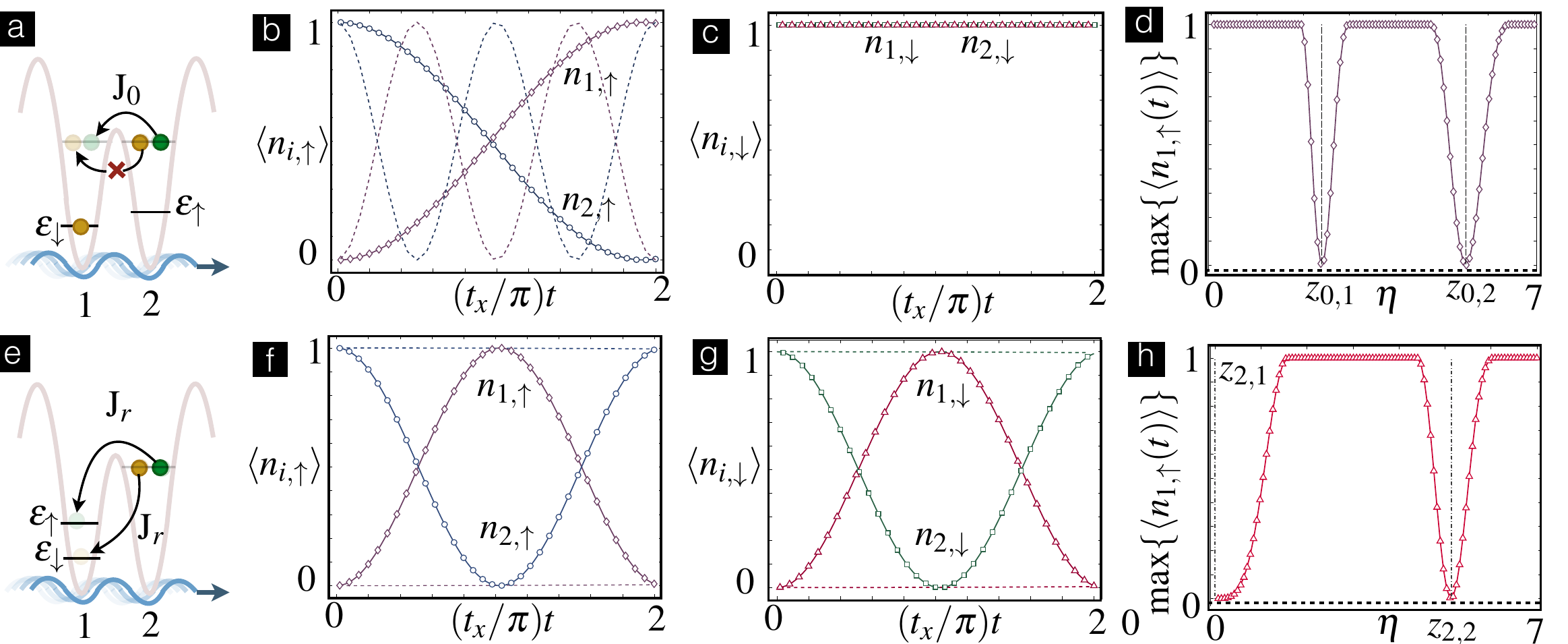}
\caption{ {\bf  Interaction-dependent resonant PAT for two-component fermions:} Population dynamics of a periodically-driven Fermi-Hubbard dimer with parameters $t_x=0.1$, $U_{\uparrow\downarrow}=20$, and $\varphi=0$ for different density distributions:  {\bf (a-d)} Initial state with one spin-up atom on the right well in a background of spin-down atoms $\ket{\downarrow_1,\uparrow\downarrow_2}=f_{1\downarrow}^\dagger f_{2\uparrow}^\dagger f_{2\downarrow}^\dagger\ket{0}$; {\bf (e-f)} initial state with a pair of atoms on the right well $\ket{0_1,\uparrow\downarrow_2}= f_{2\uparrow}^\dagger f_{2\downarrow}^\dagger\ket{0}$.  Dashed lines correspond to the tunneling for the un-driven $\tilde{V}_0=0$ dimer. Solid lines stand for the resonantly-driven dimer  $\Delta\omega=U_{\uparrow\downarrow}/2$ (i.e. two-photon assisted tunneling $r=2$) with $\tilde{V}_0=3\Delta\omega$ dictated by the exact Hamiltonian~\eqref{eq:hubbard1d}-\eqref{eq:modulation}. Symbols  stand for the dynamics under the effective Hamiltonian~\eqref{eq:eff_H}-\eqref{eq:tunn_op} for the same parameters. This criterion of symbols, dashed and solid lines, is kept for other figures.   {\bf (a, b, c)} Photon-assisted tunneling for the spin-up atom, and Pauli blockade of the spin-down atoms which cannot tunnel due to the exclusion principle. {\bf (d)} Maximal arrival density of the spin-up atom ${\rm max}\{\langle n_{1,\uparrow}(t)\rangle:0<t<2\pi/ t_x\}$, which displays minima  exactly at  the zeros of the Bessel function ${\rm J_0}(z_{0,n})=0$ (i.e. coherent destruction of tunneling). {\bf (e, f, g)} Photon-assisted tunneling for both the spin-up and spin-down atoms. {\bf (h)} Maximal arrival density of the spin-up atom ${\rm max}\{\langle n_{1,\uparrow}(t)\rangle:0<t<2\pi/ t_x\}$, which displays minima  exactly at  the zeros of the Bessel function ${\rm J_2}(z_{2,n})=0$. In comparison to {\bf (d)}, we observe that the coherent destruction of tunneling depends on the background of spin-down atoms, leading to the correlated coherent destruction of tunneling exploited in Sec.~\ref{qs_applications}.   }
\label{fig_pat_fermions_spin_indep}
\end{figure*}

In order to test the validity of our derivations, we   compare numerically  the dynamics obtained from  the effective Hamiltonian~\eqref{eq:eff_H}-\eqref{eq:tunn_op}, and the periodically driven one~\eqref{eq:hubbard1d}-\eqref{eq:modulation} in the simplest setting: a Fermi-Hubbard dimer (see Figs.~\ref{fig_pat_fermions_spin_indep}{\bf (a,e)}). In Fig.~\ref{fig_pat_fermions_spin_indep}, we explore the real-time dynamics for different configurations of atoms in the initial state. {\it (i) Pauli-blockaded regime}: Figs.~\ref{fig_pat_fermions_spin_indep}{\bf (b,c)} represent the dynamics of the initial atomic configuration  $\ket{\downarrow_1,\uparrow\downarrow_2}=f_{1\downarrow}^\dagger f_{2\uparrow}^\dagger f_{2\downarrow}^\dagger\ket{0}$, which does not display Hubbard blockade as the tunneling preserves the number of doubly-occupied sites. Nonetheless, the bare tunneling for the spin-up atoms (see dashed lines of Fig.~\ref{fig_pat_fermions_spin_indep}{\bf (b)}) is renormalized due to the periodic driving, as shown by the different population dynamics displayed by the solid lines (exact) and the symbols (effective). The excellent agreement between the solid lines and the symbols proves the validity of our derivations, and the accuracy of the interaction-dependent PAT Hamiltonian~\eqref{eq:eff_H}-\eqref{eq:tunn_op}. In particular, it shows that provided the constrains in Eq.~\eqref{pat_regime} are carefully fulfilled by the system paramaters, terms beyond the rotating-wave approximation leading to Eq.~\eqref{eq_RWA}, within the single-band approximation, do not lead to additional errors departing from the desired target Hamiltonian evolution. Regarding the dynamics of the down-spin atoms, we note that these cannot tunnel due to the Pauli exclusion principle, as depicted in  Fig.~\ref{fig_pat_fermions_spin_indep}{\bf (c)}. {\it (ii) Hubbard-blockaded regime}: Figs.~\ref{fig_pat_fermions_spin_indep}{\bf (f,g)} represent the dynamics of the initial atomic configurations $\ket{0_1,\uparrow\downarrow_2}= f_{2\uparrow}^\dagger f_{2\downarrow}^\dagger\ket{0}$, which suffers a Hubbard blockade as the tunneling must change the number of doubly-occupied sites. Hence, in the absence of the driving, the atomic tunneling is totally forbidden (see dashed lines of Fig.~\ref{fig_pat_fermions_spin_indep}{\bf (f,g)}). By switching on the driving, we observe that the tunneling of both spin-up and spin-down atoms is reactivated, as shown  by the solid lines (exact) and the symbols (effective), which again show an excellent agreement  supporting our analytical results.  

Let us now address the phenomenon of {\it correlated  destruction of tunneling} exploited in Sec.~\ref{qs_applications} for the quantum simulation of strongly-correlated models. According to Eq.~\eqref{eq:tunn_op}, the tunneling is dressed by a different Bessel function depending on the particle-hole densities, and it can  get coherently suppressed when the driving parameter $\eta$ coincides with a zero of the corresponding Bessel function. In Fig.\ref{fig_pat_fermions_spin_indep}{\bf (d)}, we observe this effect for the first pair of zeros, $\eta=z_{0,n}$ with $n=1,2$,  of the Bessel function ${\rm J}_0(z_{0,n})=0$, which are displayed by the  dashed dotted lines. We see how the maximal average population that reaches the left site of the Hubbard dimer vanishes when the driving ratio coincides with any of the zeros.  In Fig.~\ref{fig_pat_fermions_spin_indep}{\bf (h)}, we see that for a different particle-hole distribution, the coherent destruction of tunneling takes place at the zeros of a different Bessel function, namely  $\eta=z_{r,n}$ for $n=1,2$ zeros of the Bessel function ${\rm J}_r(z_{r,n})=0$ for the chosen $r=2$. Since the zeros of the two Bessel function do not coincide, we can independently suppress the tunneling correlated to a particular particle-hole distribution (i.e.  correlated  destruction of tunneling ), which will be relevant in in Sec.~\ref{qs_applications}.

\begin{figure*}
\centering
\includegraphics[width=.75\columnwidth]{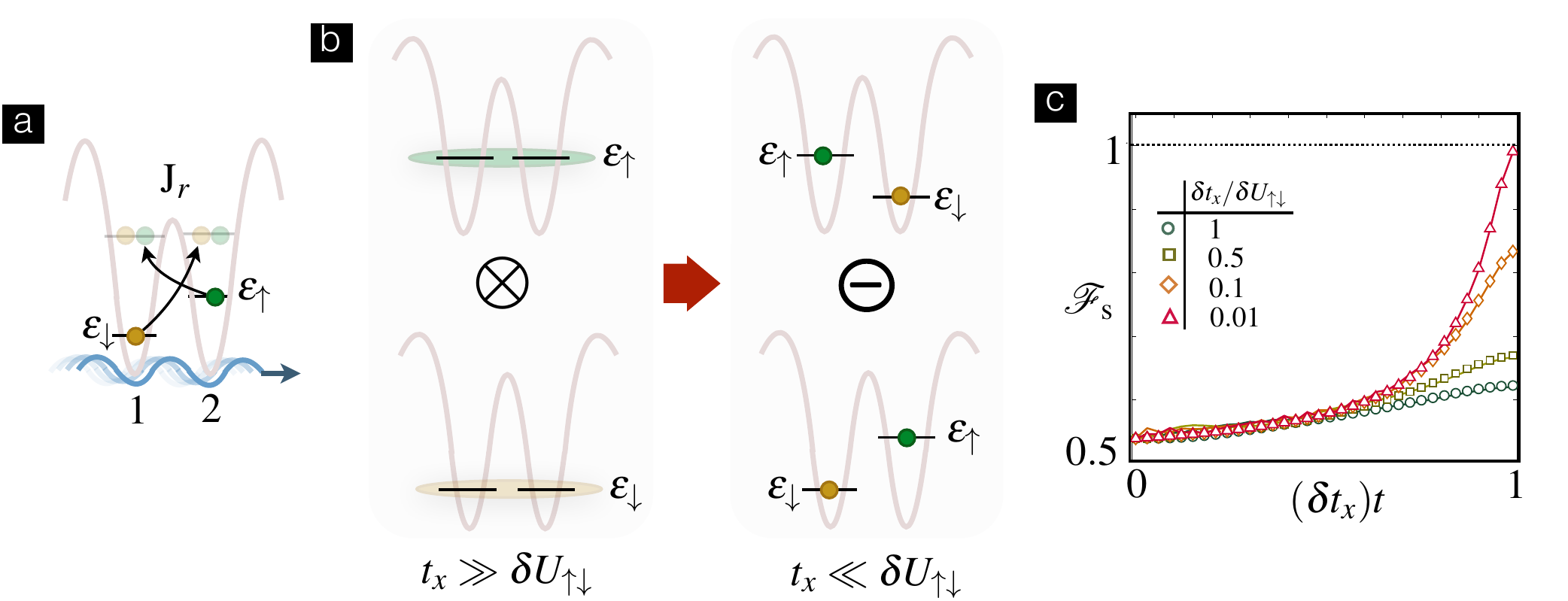}
\caption{ {\bf Adiabatic evolution of an off-resonant PAT dimer at half filling:}  {\bf (a)} Scheme of the interaction-dependent PAT for a half-filled two-site Fermi-Hubbard model. {\bf (b)}  Scheme for the adiabatic evolution in the half-filled Fermi-Hubbard dimer. At $t=0$, the system is initialized in the Fermi sea with  spin-up/down atoms delocalized along the dimer, which is the groundstate for $t_x\gg\delta U_{\uparrow\downarrow}$. By adiabatically switching off the tunneling, the final state for $t_x\ll\delta U_{\uparrow\downarrow}$ should be  the spin singlet, which corresponds to the groundstate of an antiferromagnetic Heisenberg dimer.  {\bf (c)} Singlet fidelity $\mathcal{F}_{\rm s}$ for a periodically-driven Fermi-Hubbard dimer with parameters $t_x=0.1$, $U_{\uparrow\downarrow}=20$, $\tilde{V}_0=1.84\Delta\omega$, $\varphi=0$, and $\delta U_{\uparrow\downarrow}=0.1t_x$, with a slow ramp of the tunneling strength $t_x\to t_x(1-(\delta t_x )t)$. The symbols correspond to the results given by the effective description~\eqref{eq:eff_H}-\eqref{eq:tunn_op} for different quench rates $\delta t_x$, while the solid lines correspond to the periodically-driven description~\eqref{eq:hubbard1d}-\eqref{eq:modulation}.    }
\label{fig_pat_fermions_adiab}
\end{figure*}

We have so far  presented numerical tests supporting the validity  of the resonant PAT,  $\Delta\omega=U_{\uparrow\downarrow}/r$, such that the residual interactions of the dressed Fermi-Hubbard model~\eqref{eq:eff_H} vanish $\delta U_{\uparrow\downarrow}=0$. However, our analytical results show that finite Hubbard interactions $\delta U_{\uparrow\downarrow}=(U_{\uparrow\downarrow}-r\Delta\omega)$ can be achieved by changing the velocity of the moving optical lattice $\Delta\omega\neq U_{\uparrow\downarrow}/r$, which will be crucial for several quantum simulations in Sec.~\ref{qs_applications}. Let us test this result  by numerically integrating an adiabatic evolution according to the effective~\eqref{eq:eff_H}-\eqref{eq:tunn_op} and  periodically-driven~\eqref{eq:hubbard1d}-\eqref{eq:modulation} Hamiltonians for a half-filled dimer (Fig.~\ref{fig_pat_fermions_adiab}{\bf (a)}). We study the evolution of the system (see  Fig.~\ref{fig_pat_fermions_adiab}{\bf (b)}), for a slow ramp of the tunneling strength $t_x\to t_x(1-(\delta t_x) t)$ with a rate $\delta t_x$. Initially, the dimer is prepared in the groundstate, which resembles   a Fermi sea with the spin-up/down atoms delocalized along the dimer, corresponding to the groundstate of the Fermi-Hubbard dimer for $t_x\gg\delta U_{\uparrow\downarrow}$. After the quench $t_{\rm f}\approx  1/\delta t_x$, the dimer should be in a spin singlet state corresponding to the groundstate of an antiferromagnetic Heisenberg model that arises for $t_x\ll\delta U_{\uparrow\downarrow}$
\beq
\label{sup_ad}
\ket{\Psi_0}\approx\ket{\rm FS}=\frac{1}{2}\left(f_{1\uparrow}^\dagger+f_{2\uparrow}^\dagger\right)\left(f_{1\downarrow}^\dagger+f_{2\downarrow}^\dagger\right)\ket{0} \longrightarrow \ket{\Psi_{\rm f}}\approx\ket{\rm HS}=\frac{1}{\sqrt{2}}\left(f_{1\uparrow}^\dagger f_{2\downarrow}^\dagger-f_{1\downarrow}^\dagger f_{2\uparrow}^\dagger\right)\ket{0},
\eeq
In Fig.~\ref{fig_pat_fermions_adiab}{\bf (c)}, we represent the numerical results for the Heisenberg-singlet fidelity $\mathcal{F}_{\rm s}(t)=|\langle{\rm HS}\ket{\Psi(t)}|^2$ as a function of the ramp time, and for different ramp rates. We observe that the fidelity approaches $\mathcal{F}_{\rm s}(1/\delta t_x)\approx1$ for the very slow ramps, where the adiabatic evolution is expected to be more accurate. Once again, the good agreement between the effective~\eqref{eq:eff_H}-\eqref{eq:tunn_op} and  periodically-driven~\eqref{eq:hubbard1d}-\eqref{eq:modulation} Hamiltonians, support our claim that one can study the effects of finite Hubbard interactions, and their interplay with the dressed PAT tunneling.

At this point, it is worth commenting on the effect of higher excited bands that would be present in the optical-lattice setup, but are not contained in the single-band approximation implicit to Eq.~\eqref{eq:hubbard}, and the rest of our treatment. The periodic modulation  may also assist inter-band transition by multi-photon resonances where $n\Delta\omega=\Delta E$, where $n\in\mathbb{Z}$ and $\Delta E$ is the energy gap between the lowest and some higher band. To avoid such processes, one must ensure that these resonances are avoided for the lowest-lying bands where the number of absorbed photons $n$ can be the lowest. Eventually, this parameter choice may lead to a resonance with a much higher band for $n\ll 1$, but provided that $\Delta E\ll \tilde{V}_0$, the population transferred will be exponentially slower than the inter-band tunneling $t_{\rm inter-band}\sim t_{x}(\Delta E/ \tilde{V}_0)^n\ll t_x$. One must thus make sure that these inter-band population transfer is much slower than the time required for the experiment, which becomes essential for the cases that deal with the slower super exchange~\eqref{sup_ad}. In addition to these possible errors in the simulation, one also has to consider the effect of working in a different periodically-modulated picture, which can lead to micro-motion contributions at the driving frequency that can alter the experimental measurements~\cite{nathan_pm}.

\begin{figure*}
\centering
\includegraphics[width=.8\columnwidth]{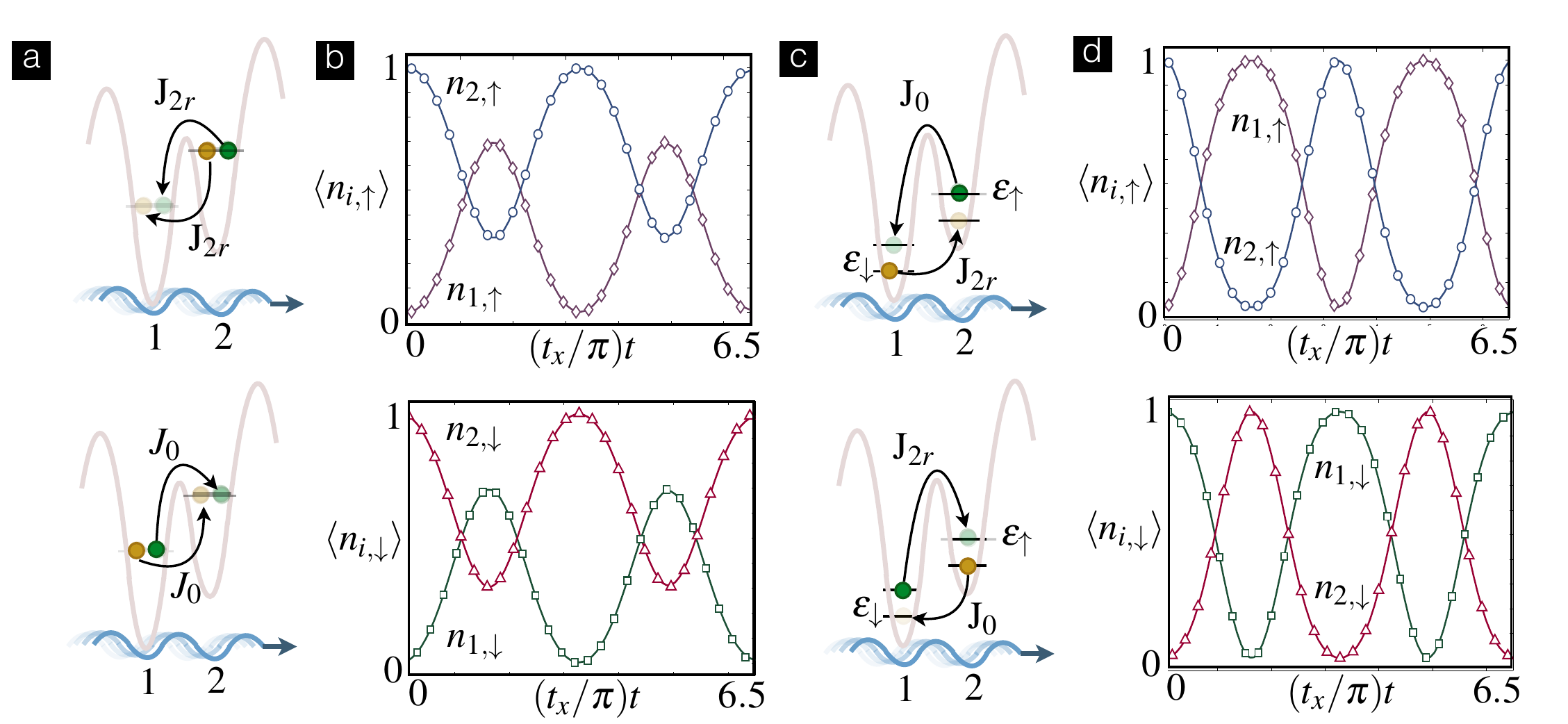}
\caption{ {\bf  Interaction-dependent resonant PAT for two-component fermions in a gradient:} Population dynamics of a periodically-driven Fermi-Hubbard dimer with parameters $t_x=0.1$, $U_{\uparrow\downarrow}=20$, $\tilde{V}_0=3.5\Delta\omega$ , $\varphi=0$, and subjected to an additional gradient $\Delta=U_{\uparrow\downarrow}$.   {\bf (a)} Scheme for PAT for states with double occupancies $\ket{0_1,\uparrow\downarrow_2}=f_{2\uparrow}^\dagger f_{2\downarrow}^\dagger\ket{0}$ (upper diagram), and $\ket{\uparrow\downarrow_1, 0_2}=f_{1\uparrow}^\dagger f_{1\downarrow}^\dagger\ket{0}$ (lower diagram), showing two different tunneling dressings that lead to the two-tone beating displayed in  {\bf (b)} for the initial state $\ket{\Psi_0}=\ket{0_1,\uparrow\downarrow_2}$. As usual, the solid lines correspond to  the numerical solution of the exact Hamiltonian~\eqref{eq:hubbard1d}-\eqref{eq:modulation} in the presence of an additional gradient, while the symbols stand for the numerical solution of the effective Hamiltonian~\eqref{eq:eff_H} with the modified tunnelings~\eqref{eq:tunn_op_grad} due to the  presence of the gradient. {\bf (c)} Scheme for PAT for states with single occupancies $\ket{\downarrow_1,\uparrow_2}=f_{1\downarrow}^\dagger f_{2\uparrow}^\dagger\ket{0}$ (upper diagram), and $\ket{\uparrow_1,\downarrow_2}=f_{1\uparrow}^\dagger f_{2\downarrow}^\dagger\ket{0}$ (lower diagram), showing two different tunneling dressings that yield  the two-tone beating displayed in  {\bf (d)}  for the initial state $\ket{\Psi_0}=\ket{\downarrow_1,\uparrow_2}$.   }
\label{fig_pat_fermions_gradient}
\end{figure*}

Before moving to the PAT in higher dimensions, let us  mention that we could gain additional flexibility in the scheme by introducing an additional linear gradient, which may come from a lattice acceleration, an external electric field, or a magnetic-field gradient. If we tune the gradient such that it coincides with the on-site interaction, we can generalize Eq.~\eqref{eq:eff_H} by substituting
\beq
\label{eq:tunn_op_grad}
\begin{split}
{\rm J}_{r\Delta n_{i+1,\overline{\sigma}}}\left(\eta\right)\to{\rm J}_{r(1+\Delta n_{i+1,\overline{\sigma}})}\left(\eta\right)&={\rm J}_r\left(\eta\right)h_{i,\overline{\sigma}}h_{i+1,\overline{\sigma}}+{\rm J}_r\left(\eta\right)n_{i,\overline{\sigma}}n_{i+1,\overline{\sigma}}+{\rm J}_{0}\left(\eta\right)n_{i,\overline{\sigma}}h_{i+1,\overline{\sigma}}+{\rm J}_{2r}\left(\eta\right)h_{i,\overline{\sigma}}n_{i+1,\overline{\sigma}}.
\end{split}
\eeq
According to this expression, the off-diagonal tunnelings connecting doubly- to single-occupied sites (i.e. fourth term in Eq.~\eqref{eq:tunn_op_grad} represented in the upper panel of Fig.~\ref{fig_pat_fermions_gradient}{\bf (a)}), and single- to doubly-occupied sites (i.e. third term in Eq.~\eqref{eq:tunn_op_grad} represented in the upper panel of Fig.~\ref{fig_pat_fermions_gradient}{\bf (c)}),  depend on different Bessel functions.  This leads to a two-tone beating in the tunneling dynamics, as shown in Figs.~\ref{fig_pat_fermions_gradient}{\bf (b),(d)}, which also serve as tests of the validity of our analytical derivations.

\vspace{1ex}
{\it (ii) Higher-dimensional scheme:} 
The scheme presented above can be directly generalized   beyond 1D. The static optical lattice should be modified such that it allows for tunneling along two ($V_{0,z}\gg \{V_{0,x},V_{0,y}\}\gg E_{\rm R}$) or three ($ \{V_{0,x},V_{0,y},V_{0,z}\}\gg E_{\rm R}$) directions. As can be observed from Eqs.~\eqref{eq:int_picture}-\eqref{eq:mod_function}, to assist the tunneling along a given direction, it is crucial that the periodic modulation~\eqref{eq:modulation} has a phase that varies along that particular direction.  Therefore, we would need to include additional moving optical lattices that propagate along the remaining axes, dressing the corresponding tunneling along two  $\alpha=\{x,y\}$, or three $\alpha=\{x,y,z\}$ directions. One may consider  adding one independent detuned laser beam per axis, paralleling the construction of the one-dimensional case. Otherwise, one could simply tilt the laser beam of the one-dimensional case, such that it has a non-vanishing projection propagating along each axis. The former scheme would lead to independent moving lattices along each axis whose intensity and frequency can be tuned separately, whereas the latter would lead to a non-separable moving lattice that dresses all the different tunnellings with the same intensity and frequency, albeit one could play with the propagation angle. 

For simplicity, we consider the first  situation, such that the periodic driving is 
\beq
\label{eq:modulation_3d}
H_{\rm mod}(t)=\sum_{{\bf i},\sigma}\sum_{\alpha}\frac{\tilde{V}_{0,\alpha}}{2}\cos\left(\Delta k_\alpha R_{{\bf i},\alpha}-\Delta\omega_\alpha t+\varphi_\alpha\right)n^{\phantom{\dagger}}_{{\bf i},\sigma}, 
\eeq
where  we have introduced the labeling indexes ${\bf i}=(i_x,i_y)$ for 2D, and   ${\bf i}=(i_x,i_y, i_z)$ for 3D. As before, we have assumed that for 2D ($ \tilde{V}_{0,x}\ll V_{0,x}$, and $\tilde{V}_{0,y}\ll V_{0,y}$), and for 3D ($ \tilde{V}_{0,x}\ll V_{0,x}$, $\tilde{V}_{0,y}\ll V_{0,y}$, and $\tilde{V}_{0,z}\ll V_{0,z}$), such that the moving lattices do not modify the bare tunneling and only lead to a periodic modulation of the on-site  energies. Each of these moving lattices assists the tunneling along a given direction, and does not interfere with the tunnelings along the remaining axes. Accordingly, the  interaction-dependent PAT is a direct generalization of~\eqref{eq:eff_H}, which requires a parameter regime
\beq
\label{pat_regime_3d}
t_{x}, t_{y}, t_{z}, \delta U_{\uparrow\downarrow}=(U_{\uparrow\downarrow}-r_\alpha\Delta\omega_\alpha)\ll U_{\uparrow\downarrow}\approx r_x\Delta\omega_x=r_y\Delta\omega_y=r_z\Delta\omega_z,
\eeq
and yields the following effective Hamiltonian
\beq
\label{eq:H_eff_higher_d}
 H_{\rm eff}=\sum_{{\bf i},\sigma}\epsilon_{{\bf i},\sigma}f_{{\bf i},\sigma}^{\dagger}f_{{\bf i},\sigma}^{\phantom{\dagger}}-\sum_{{\bf i},\alpha}\sum_{\sigma}\left(t_{\alpha}{\rm J}_{r_\alpha\Delta n_{{\bf i+e_\alpha},\overline{\sigma}}}\left(\eta_{\alpha}\right)\ee^{-\ii r_\alpha\varphi_\alpha \Delta n_{{\bf i}+{\bf e}_\alpha,\overline{\sigma}}}f_{{\bf i},\sigma}^{\dagger}f_{{\bf i},+{\bf e}_\alpha,\sigma}^{\phantom{\dagger}}+{\rm H.c.}\right)+\frac{1}{2}\sum_{{\bf i},\sigma}\delta U_{\sigma\overline{\sigma}}f_{{\bf i},\sigma}^{\dagger}f_{{\bf i},\overline{\sigma}}^{\dagger}f_{{\bf i},\overline{\sigma}}^{\phantom{\dagger}}f_{{\bf i},\sigma}^{\phantom{\dagger}},
\eeq
 where $\Delta n_{{\bf i+e_\alpha},\sigma}=n_{{\bf i+e_\alpha},\sigma}-n_{{\bf i},\sigma}$,  the  bare tunnelings are approximated by Eq.~\eqref{eq:gaussian},  and the dressed tunnelings depend on  
\beq
\begin{split}
{\rm J}_{r_\alpha\Delta n_{{\bf i+e_\alpha},\overline{\sigma}}}\left(\eta_\alpha\right)&={\rm J}_0\left(\eta_{\alpha}\right)h_{{\bf i},\overline{\sigma}}h_{{\bf i}+{\bf e}_\alpha,\overline{\sigma}}+{\rm J}_0\left(\eta_{\alpha}\right)n_{{\bf i},\overline{\sigma}}n_{{\bf i}+{\bf e}_\alpha,\overline{\sigma}}+{\rm J}_{r_\alpha}\left(\eta_{\alpha}\right)n_{{\bf i},\overline{\sigma}}h_{{\bf i}+{\bf e}_\alpha,\overline{\sigma}}+{\rm J}_{r_\alpha}\left(\eta_{\alpha}\right)h_{{\bf i},\overline{\sigma}}n_{{\bf i}+{\bf e}_\alpha,\overline{\sigma}},
\end{split}
\eeq
where we have introduced $\eta_{\alpha}=\tilde{V}_{0,\alpha}/\Delta\omega_\alpha$. It is interesting to note that controlling  the intensity difference of each moving optical lattice, we can tune  the spatial anisotropy of the dressed tunnellings.  The possibility of generalizing to 2D is especially interesting in the context of the $t$-$J$ model, and its connection to high-$T_{\rm c}$ cuprate superconductors, as outlined  in Sec.~\ref{tJ_model}.

We have thus seen that the interaction-dependent PAT with a moving optical lattice leads to effective Hubbard models of any dimensionality with dressed tunnelings that are density dependent. In the following section, we will show that by considering a state-dependent  moving optical lattice, the PAT scheme becomes more flexible, which will allow us to target other quantum many-body models, in particular dynamical Gauge fields.

\subsubsection{Two-component fermions in spin-dependent moving optical lattices}
\label{spin_dependent_fermions}

\begin{figure*}
\centering
\includegraphics[width=1\columnwidth]{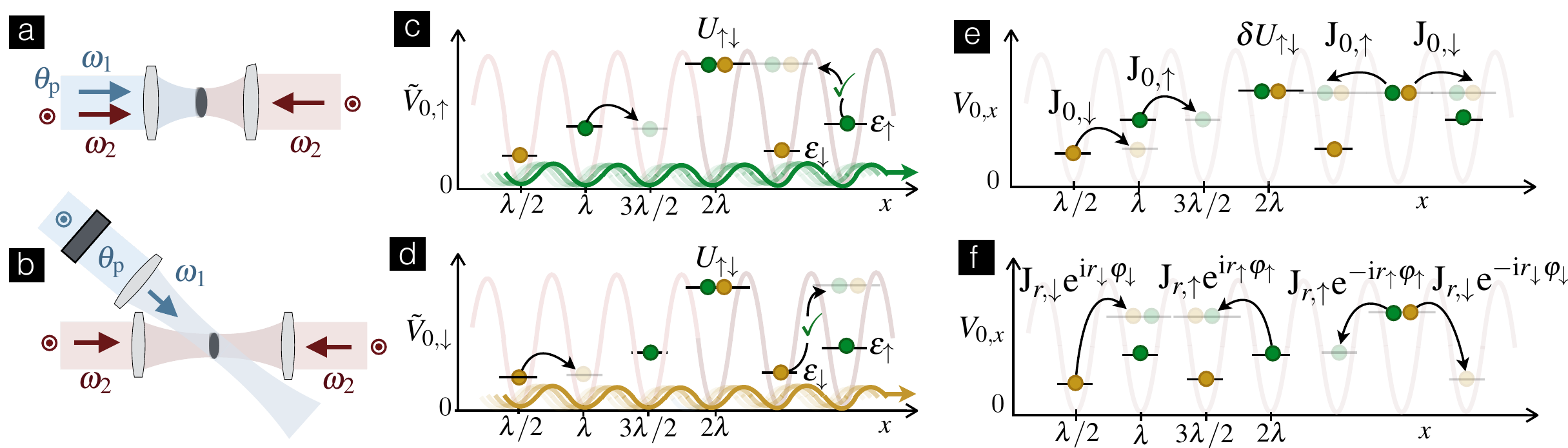}
\caption{ {\bf Scheme of the spin-dependent PAT for fermions: }{\bf (a)} Laser scheme similar to Fig.~\ref{fig_pat_scheme}, but with the linear polarization of the detuned Raman beam (blue arrow) rotated by an angle $\theta_{\rm p}$ with respect to the static lattice laser beams. This gives rise to a spin-dependent moving optical lattice. {\bf (b)} Same as before, but considering that the detuned Raman beam (blue arrow) forms an angle with respect to the static optical lattice beams (red arrows). {\bf (c, d)} The  moving optical-lattice potential  for each pseudospin (green  lines for $\ket{{\uparrow}}$, and orange lines for $\ket{{\downarrow}}$) reactivates the tunneling  involving doubly-occupied sites. The PAT contains diagonal terms {\bf (e)}   corresponding to a spin-dependent dressed tunneling within the subspaces of single- or doubly-occupied sites controlled by the Bessel function ${\rm J}_{0,\sigma}:={\rm J}_0(\eta_{\sigma})$.  The off-diagonal  terms {\bf (f)} contain a spin-dependent tunneling connecting the subspaces of single- and doubly-occupied sites, which are controlled by the Bessel function ${\rm J}_{r,\sigma}:={\rm J}_r(\eta_{\sigma})$ and dressed by the spin-dependent tunneling phase $\ee^{\pm \ii r_\sigma\varphi_{\sigma}}$. }
\label{fig_pat_scheme_spin_dep}
\end{figure*}

{\it (i) One-dimensional scheme:} Let us once again start with the  less demanding   case of 1D.  We note that the  far-detuned moving optical lattice can become state-dependent if the laser-beam  polarizations are not collinear (see Fig.~\ref{fig_pat_scheme_spin_dep} {\bf (a,b)}). This occurs even for detunings that are larger than the Zeeman and hyperfine splittings, as far as they do not exceed the fine-structure splitting~\cite{ol_review}. For fermionic alkali atoms, this could turn to be incompatible with reaching ultracold temperatures,  as the fine-structure splitting is rather small, and  the residual photon scattering may become appreciable~\cite{gd_gauge_fields}. However, we stress that the moving lattice is by construction much weaker than the static spin-independent  one. In fact, we can reduce the residual photon scattering by orders of magnitude by  lowering the intensity of the moving-lattice laser beams, as far as their detuning is simultaneously lowered, such that the ratio $\eta$ controlling the PAT~\eqref{eq:tunn_op} remains constant. We will thus assume that a conservative state-dependent moving optical lattice can be realized without increasing the photon scattering and heating the ultracold atomic gas.

In this case,  we can generalize the driving~\eqref{eq:modulation} by including a state-dependent periodic modulation of the on-site energies 
\beq
\label{eq:modulation_spin_dep}
H_{\rm mod}(t)=\sum_{i,\sigma}\frac{\tilde{V}_{0,\sigma}}{2}\cos\big(\Delta k X_i-\Delta\omega t+\varphi_\sigma\big)n_{i,\sigma}^{\phantom{\dagger}},
\eeq
where the spin-dependent amplitude must again fulfill $\tilde{V}_{0,\sigma}\ll V_{0,x}$, and $\varphi_\sigma$ stands for a phase difference with respect to the static lattice that is generally state dependent~\cite{state_dep_ol_exp,spin_dependent_ol} (see Figs.~\ref{fig_pat_scheme_spin_dep} {\bf (c,d)}). Going back to 
the spin-independent scheme~\eqref{eq:modulation}, the new modulation~\eqref{eq:modulation_spin_dep}  can be achieved by rotating the polarization of the  laser beam that is slightly detuned with respect to the static-lattice lasers~\cite{state_dep_ol_exp}. In this case, the spin-dependent driving amplitudes $\tilde{V}_{0,\sigma}$ can be tuned by controlling such an angle, or instead the direction of propagation of the laser beam with respect to the quantization axis~\cite{spin_dependent_ol}. Another possibility would be  to resolve the hyperfine structure, such that one  can exploit selection rules in the ac-Stark shifts. In fact,  for pseudospins corresponding to the maximally-polarized Zeeman sublevels, it is possible to obtain optical lattices that selectively address a single pseudospin (i.e. $\tilde{V}_{0,\uparrow}=0$,  $\tilde{V}_{0,\downarrow}\neq0$), or vice versa, as realized in ion-trap experiments~\cite{monroe_cats}. This leads to a spin-dependent  driving where the   wavevector $\Delta k_\sigma$, detuning $\Delta\omega_{\sigma}$,   intensity $\tilde{V}_{0,\sigma}$, and relative phase $\varphi_\sigma$ can all be controlled independently for each pseudospin
\beq
\label{eq:modulation_spin_dependent}
H_{\rm mod}(t)=\sum_{i,\sigma}\frac{\tilde{V}_{0,\sigma}}{2}\cos\big(\Delta k_\sigma X_i-\Delta\omega_{\sigma} t+\varphi_\sigma\big)n_{i,\sigma}^{\phantom{\dagger}}.
\eeq

\begin{figure*}
\centering
\includegraphics[width=.8\columnwidth]{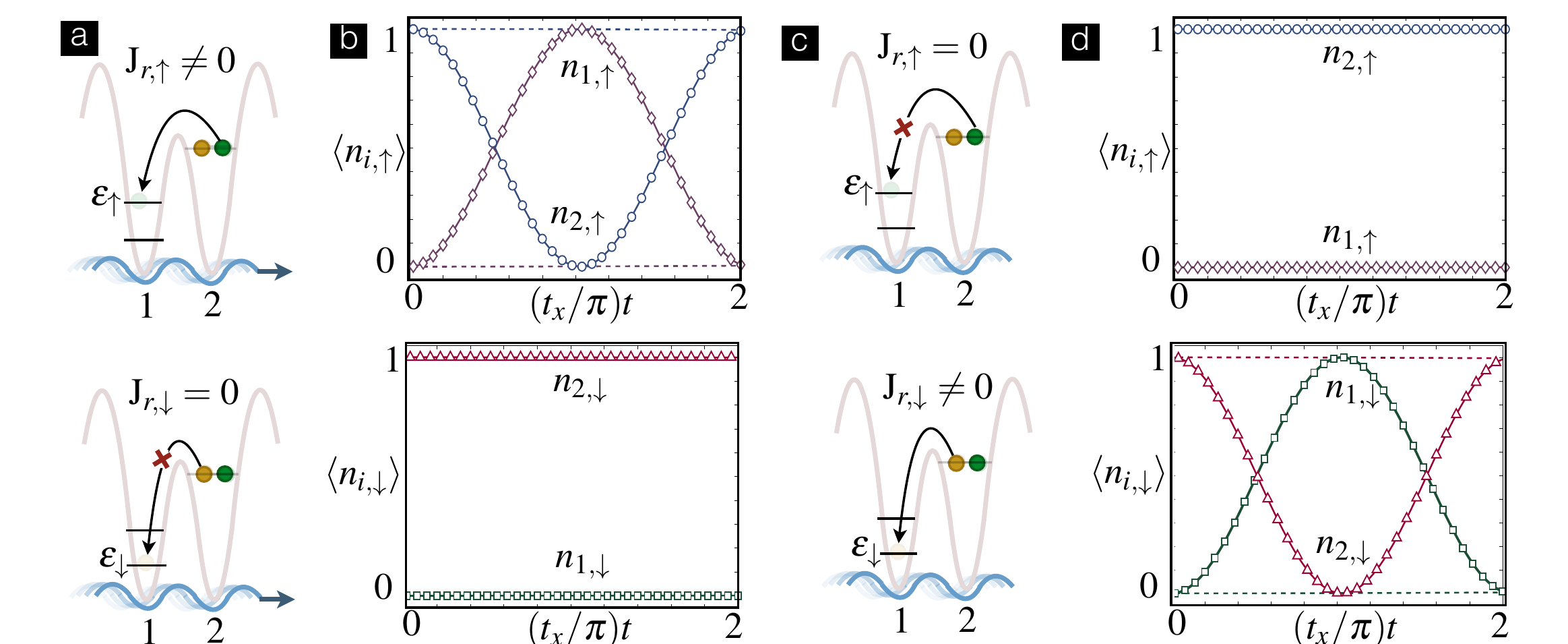}
\caption{ {\bf  Spin-dependent coherent destruction of tunneling  for two-component fermions:} Population dynamics of a periodically-driven Fermi-Hubbard dimer with parameters $t_x=0.1$, $U_{\uparrow\downarrow}=20$,  and subjected to a spin-dependent moving lattice with $\varphi_\uparrow=\varphi_\downarrow=0$, $\Delta\omega_\uparrow=\Delta\omega_\downarrow=U_{\uparrow\downarrow}/2$.   {\bf (a)} Scheme for the spin-dependent PAT for the state $\ket{0_1,\uparrow\downarrow_2}$, showing that the spin-down atoms can be coherently frozen, as shown in {\bf (b)} for $V_{0\uparrow}=3\Delta\omega$, and $V_{0\downarrow}=5.13\Delta\omega$ such that ${\rm J}_2(5.13)=0$. As usual, the solid lines correspond to  the numerical solution of the exact Hamiltonian~\eqref{eq:hubbard1d} with the spin-dependent periodic modulation~\eqref{eq:modulation_spin_dependent}, while the symbols stand for the numerical solution of the effective Hamiltonian~\eqref{eq:state_depn_Heff} with the modified tunnelings~\eqref{eq:tunn_op_state_dep}. {\bf (c)} Scheme for the spin-dependent PAT for the state $\ket{0_1,\uparrow\downarrow_2}$, showing that the spin-down atoms can be coherently frozen, as shown in {\bf (d)} for $V_{0\uparrow}=5.13\Delta\omega$ such that ${\rm J}_2(5.13)=0$, and $V_{0\downarrow}=3\Delta\omega$.  }
\label{fig_pat_fermions_spin_dep}
\end{figure*}

Paralleling the previous section, we will consider   equal wavelengths of the static and moving optical lattices, such that $\Delta k_{\sigma} X_i=(\half \Delta k_\sigma \lambda)i=\pi  i$, although we remark again that the scheme also works for other propagation angles.  Once the new periodic drivings~\eqref{eq:modulation_spin_dep}-\eqref{eq:modulation_spin_dependent} have been discussed, we can address the interaction-dependent PAT they give rise to. We shall use Eq.~\eqref{eq:modulation_spin_dependent}, as the  results also encompass those related to the driving~\eqref{eq:modulation_spin_dep}.
By reproducing the steps that lead  to the effective Hamiltonian~\eqref{eq:eff_H} for the spin-independent driving, we find  a parameter regime analogous to Eq.~\eqref{pat_regime}, namely
\beq
\label{pat_regime_spin_dependent}
t_{x}, \delta U_{\uparrow\downarrow}=(U_{\uparrow\downarrow}-r_\sigma\Delta\omega_\sigma)\ll U_{\uparrow\downarrow}\approx r_\uparrow\Delta\omega_\uparrow=r_\downarrow\Delta\omega_\downarrow,
\eeq
and  a new dressing function of the tunneling that becomes spin-dependent, namely
\beq
\mathrm{f}_{\sigma}(t)=\sum_{m}{\rm J}_m\left(\frac{\eta_\sigma}{2}\right){\rm J}_{m+r_{\sigma}\Delta n_{i+1,\overline{\sigma}}}\left(\frac{\eta_\sigma}{2}\right)\ee^{\ii m\pi}\ee^{-\ii \pi  r_\sigma\Delta n_{i+1,\overline{\sigma}}i}\ee^{-\ii r_\sigma\varphi_\sigma \Delta n_{i+1,\overline{\sigma}}}.
\eeq
where  $\eta_\sigma=\tilde{V}_{0}^{\sigma}/\Delta\omega_\sigma$. In this case, the detunings are chosen such that $r_\sigma$ is  an even integer for both pseudospins, such that we can thus set  $\ee^{-\ii \pi r_\sigma\Delta n_{i+1,\overline{\sigma}}i}=1$. Using the Neuman-Graff addition formula once again, we  find that 
\beq
\label{eq:state_depn_Heff}
 H_{\rm eff}=\sum_{i,\sigma}\epsilon_{i,\sigma}f_{i,\sigma}^{\dagger}f_{i,\sigma}^{\phantom{\dagger}}-\sum_{i,\sigma}\left(t_{x}{\rm J}_{r_\sigma\Delta n_{i+1,\overline{\sigma}}}\left(\eta_{\sigma}\right)\ee^{-\ii r_\sigma\varphi_\sigma\Delta n_{i+1,\overline{\sigma}}}f_{i,\sigma}^{\dagger}f_{i+1,\sigma}^{\phantom{\dagger}}+{\rm H.c.}\right)+\frac{1}{2}\sum_{i,\sigma}\delta U_{\sigma\overline{\sigma}}f_{i,\sigma}^{\dagger}f_{i,\overline{\sigma}}^{\dagger}f_{i,\overline{\sigma}}^{\phantom{\dagger}}f_{i,\sigma}^{\phantom{\dagger}},
\eeq

Remarkably, we find that the amplitude of the density-dependent tunneling can be controlled independently for each pseudospin
\beq
\label{eq:tunn_op_state_dep}
\begin{split}
{\rm J}_{r_\sigma\Delta n_{i+1,\overline{\sigma}}}\left(\eta_{\sigma}\right)&={\rm J}_0\left(\eta_{\sigma}\right)h_{i,\overline{\sigma}}h_{i+1,\overline{\sigma}}+{\rm J}_0\left(\eta_{\sigma}\right)n_{i,\overline{\sigma}}n_{i+1,\overline{\sigma}}+{\rm J}_{r_\sigma}\left(\eta_{\sigma}\right)n_{i,\overline{\sigma}}h_{i+1,\overline{\sigma}}+{\rm J}_{r_\sigma}\left(\eta_{\sigma}\right)h_{i,\overline{\sigma}}n_{i+1,\overline{\sigma}},
\end{split}
\eeq
and  that the tunneling phase of one pseudospin depends on the density of the other pseudospin, which will be crucial for the quantum simulation of dynamical Gauge fields in Sec.~\ref{gauge_fields}. In order to benchmark these predictions, we study numerically a spin-dependent coherent destruction of tunneling in a Fermi-Hubbard dimer subjected to the spin-dependent moving optical lattice. According to Eq.~\eqref{eq:tunn_op_state_dep}, the dressed tunneling of a doubly-occupied half-filled dimer (see Fig.~\ref{fig_pat_fermions_spin_dep}{\bf (a)}) depends on the spin of the atom, such that the spin-up atoms tunneling is controlled by the Bessel function ${\rm J}_{r_\uparrow}(\eta_\uparrow)$, whereas the spin-down atoms  tunneling depends on   the Bessel function ${\rm J}_{r_\downarrow}(\eta_\downarrow)$. Therefore, by controlling the intensities of the the spin-dependent  moving lattice, the dressed tunneling of the spin-down atoms can be coherently destructed ${\rm J}_{r_\downarrow}(\eta_\downarrow^\star)=0$, while the spin-up atoms hop freely in the lattice ${\rm J}_{r_\uparrow}(\eta_\uparrow^\star)\neq0$ (see Figs.~\ref{fig_pat_fermions_spin_dep}{\bf (a)-(b)}). Conversely, we can coherently freeze the spin-up atoms  ${\rm J}_{r_\uparrow}(\eta_\uparrow^\star)=0$, while the spin-down atoms hop freely in the lattice ${\rm J}_{r_\downarrow}(\eta_\downarrow^\star)\neq0$ (see Figs.~\ref{fig_pat_fermions_spin_dep}{\bf (c)-(d)}). Let us emphasize the excellent agreement between our analytical description (symbols), and the exact dynamics of the periodic Hamiltonian (solid lines).

\begin{figure*}
\centering
\includegraphics[width=.9\columnwidth]{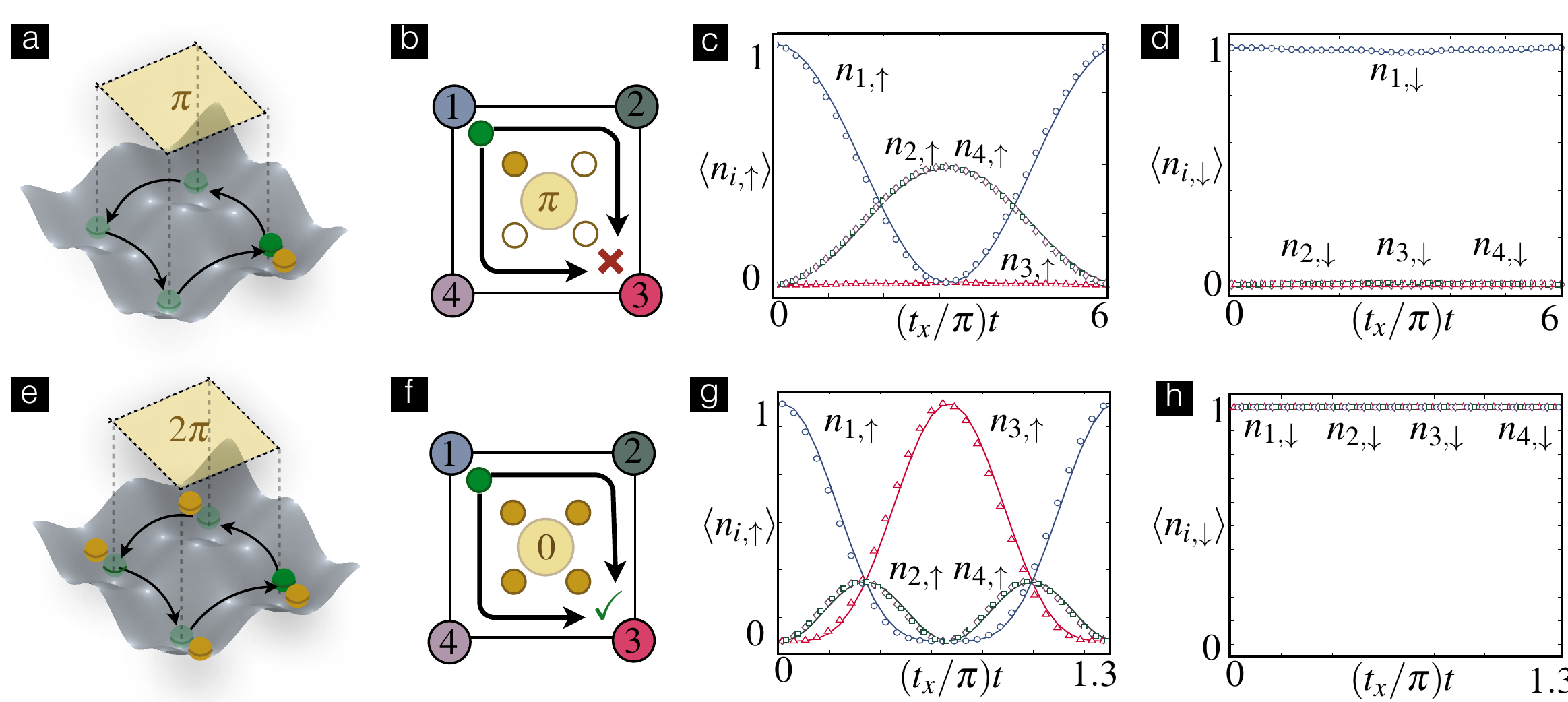}
\caption{ {\bf  Density-dependent Aharonov-Bohm interference  for two-component fermions:} Population dynamics of a periodically-driven Fermi-Hubbard tetramer with parameters $t_x=t_y=0.1$, $U_{\uparrow\downarrow}=20$,  and subjected to a spin-dependent moving lattice with $\varphi_{y,\sigma}=0$, and $\varphi_{x,\sigma}=\pi/2$, and  $\Delta\omega_{\alpha,\sigma}=U_{\uparrow\downarrow}/2$, $\forall\alpha,\sigma$.   {\bf (a)} Scheme for the  PAT for the state $\ket{\uparrow\downarrow_1,0_2,0_3,0_4}$, where the frozen spin-down atom induces a $\pi$-flux in the tunneling of the spin-up atom, leading  to the Aharonov-Bohm destructive interference in {\bf (b)}. {\bf (c,d)} Populations for $V_{0\uparrow}=\Delta\omega$, and $V_{0\downarrow}=5.13\Delta\omega$ such that ${\rm J}_2(5.13)=0$, and the spin-down atom is frozen. As usual, the solid lines correspond to the numerical solution of the exact Hamiltonian~\eqref{eq:hubbard1d} with the spin-dependent periodic modulation~\eqref{eq:modulation_spin_dependent}, while the symbols stand for the numerical solution of the effective Hamiltonian~\eqref{eq:2d_eff_H}. Due to the interference, the spin-up atom cannot reach the opposite corner $\langle n_{3,\uparrow}(t)\rangle=0$. {\bf (e-h)} Same as above, but for the state $\ket{\uparrow\downarrow_1, \downarrow_2,\downarrow_3,\downarrow_4}$, where the spin-down atoms cannot tunnel due to the Pauli exclusion principle. For this density background, the tunneling phase of the spin-up atoms vanishes since there is no spin-down density gradient. Hence, this initial state  does not lead to Aharonov-Bohm destructive interference, and the spin-up atom does indeed reach the opposite corner.    }
\label{fig_pat_square_fermions_spin_dep}
\end{figure*}

\vspace{1ex}
{\it (ii) Two-dimensional scheme:} The scheme presented above can be generalized  beyond 1D. For the sake of concreteness, and for its particular interest in connection to  the $t$-$J$ model in Sec.~\ref{tJ_model}, and the dynamical Gauge fields in Sec.~\ref{gauge_fields}, we will restrict to 2D ($V_{0,z}\gg \{V_{0,x},V_{0,y}\}\gg E_{\rm R}$). The idea is to consider  spin-dependent moving optical lattices along the $x$ and $y$ axes 
\beq
\label{eq:modulation_2d}
H_{\rm mod}(t)=\sum_{{\bf i},\sigma}\sum_{\alpha}\frac{\tilde{V}^{\sigma}_{0,\alpha}}{2}\cos\left(\Delta k_{\sigma, \alpha} r_{{\bf i},\alpha}^0-\Delta\omega_{\sigma,\alpha} t+\varphi_{\sigma,\alpha}\right)n^{\phantom{\dagger}}_{{\bf i},\sigma}, 
\eeq
such that the relative phases of the moving lattices fulfill   $\varphi_{\sigma, x}\geq0$, but $\varphi_{\sigma,y}=0$. In this case, and after following the same steps as above in an analogous parameter regime
\beq
\label{pat_regime_spin_dependent_2d}
t_{x},t_{y}, \delta U_{\uparrow\downarrow}=(U_{\uparrow\downarrow}-r_{\sigma,\alpha}\Delta\omega_{\sigma,\alpha})\ll U_{\uparrow\downarrow}\approx r_{\sigma,\alpha}\Delta\omega_{\sigma,\alpha}=r_{\sigma',\alpha'}\Delta\omega_{\sigma',\alpha'}, \hspace{1ex}\forall \alpha,\alpha',\sigma,\sigma'
\eeq
 one can derive the following effective Hamiltonian   
\beq
\label{eq:2d_eff_H}
 H_{\rm eff}=-\sum_{{\bf i},\sigma}\left(t_{x}{\rm J}_{r_{\sigma,x}\Delta n_{{\bf i}+{\bf e}_x,\overline{\sigma}}}\left(\eta_{\sigma,x}\right)\ee^{-\ii r_{\sigma,x}\varphi_{\sigma,x}\Delta n_{{\bf i}+{\bf e}_x,\overline{\sigma}}}f_{{\bf i},\sigma}^{\dagger}f_{{\bf i}+{\bf e}_x,\sigma}^{\phantom{\dagger}}+t_{y}{\rm J}_{r_{\sigma,y}\Delta n_{{\bf i}+{\bf e}_y,\overline{\sigma}}}\left(\eta_{\sigma,y}\right)f_{{\bf i},\sigma}^{\dagger}f_{{\bf i}+{\bf e}_y,\sigma}^{\phantom{\dagger}}+{\rm H.c.}\right)+\sum_{{\bf i}}\delta U_{\uparrow\downarrow} n_{{\bf i},\uparrow}n_{{\bf i},\downarrow},
\eeq
where $\eta_{\sigma,\alpha}=\tilde{V}^{\sigma}_{0,\alpha}/\Delta\omega_{\sigma,\alpha}$. We thus see that when atoms tunnel along the $x$-axis, they acquire a dynamical  phase that depends on the density of the other pseudospin, whereas they experience a vanishing  phase when  tunneling along the $y$-axis. This will be equivalent to the so-called Landau Gauge  in Sec.~\ref{gauge_fields}, which is accompanied by a  non-vanishing dynamical Wilson loop.

So far, all of our numerical tests have been independent of the phase of the moving optical lattices, as we have only addressed a Fermi-Hubbard dimer. By considering the simplest 2D case, a Fermi-Hubbard tetramer forming a square plaquette, we can already test numerically the predicted effect of the  moving  lattice phase, which according to Eq.~\eqref{eq:2d_eff_H}, induces a density-dependent Peierls phase in the tunneling. To observe the effects of such a Peierls phase, we shall first exploit the above spin-dependent destruction of tunneling, or the Pauli exclusion principle, to freeze the dynamics of the spin-down atoms (see Figs.~\ref{fig_pat_fermions_spin_dep}{\bf (a-b)}). Then, the immobile spin-down atoms yield a density background that modifies the tunneling phase of the spin-up atoms. We explore this possibility in Fig.~\ref{fig_pat_square_fermions_spin_dep} for two different density distributions of the spin-down atoms, which lead to the presence/absence of an Aharonov-Bohm destructive interference in the tunneling dynamics of the spin-up atom. Besides confirming the validity of our analytical description~\eqref{eq:2d_eff_H} in a transparent scenario, let us note that by lifting the coherent destruction of tunneling, and allowing the spin-down atoms to tunnel, the Peierls phase will acquire its own dynamics, which will be crucial for the quantum simulation of dynamical Gauge fields in Sec.~\ref{gauge_fields}. At this point, the reader may skip the following sections, and move directly to the quantum simulator applications dealing with fermionic models in Sec.~\ref{qs_applications}.


\subsection{Scheme for a periodically-modulated  ultracold Bose gas}

Let us now turn our attention to a single-species gas of  bosonic atoms, and consider again two hyperfine states $\ket{{\uparrow}}=\ket{F,M}, \ket{{\downarrow}}=\ket{F',M'}$, such that we have a unique mass $m_\uparrow=m_\downarrow=:m$ and recoil energy $E_{\rm R,\uparrow}=E_{\rm R,\downarrow}=:E_{\rm R}$. In the Wannier basis, $\Psi_\sigma({\bf r})=\sum_{{\bf i}}w({\bf r}-{\bf r}^0_{\bf i})b_{{\bf i},\sigma}$, the general Hamiltonian~\eqref{eq:originalH} can be expressed in terms of  the  bosonic operators $b_{{\bf i},\sigma}$ as a two-component Bose-Hubbard model~\cite{bose_hubbard_ol}, namely
\beq
\label{eq:hubbard_bose}
H_{\rm BH}=H_{\rm loc}+H_{\rm kin}+V_{\rm int}=\sum_{{\bf i},\sigma}\epsilon_{{\bf i},\sigma}b_{\bf i,\sigma}^{\dagger}b_{{\bf i},\sigma}^{\phantom{\dagger}}-\sum_{{\bf i},\sigma}\sum_\alpha\left(t_{\alpha}b_{{\bf i},\sigma}^{\dagger}b_{{\bf i+e_\alpha},\sigma}^{\phantom{\dagger}}+{\rm H.c.}\right)+\frac{1}{2}\sum_{{\bf i}}\sum_{\sigma,\sigma'}U_{\sigma{\sigma'}}b_{{\bf i},\sigma}^{\dagger}b_{{\bf i},{\sigma'}}^{\dagger}b_{{\bf i},{\sigma'}}^{\phantom{\dagger}}b_{{\bf i},\sigma}^{\phantom{\dagger}},
\eeq
where the Hamiltonian parameters coincide with the fermionic ones~\eqref{eq:gaussian} under similar approximations. However, due to the different statistics, s-wave scattering among pairs of atoms with the same electronic state are now allowed (i.e. $U_{\uparrow\uparrow},U_{\downarrow\downarrow}\neq0$), which gives more freedom for the interaction-dependent PAT. We shall discuss different regimes of interest for the quantum simulation of bosonic quantum many-body models, which can be achieved by tuning the Feshbach resonances.

\subsubsection{Two-component hardcore bosons in spin-independent moving optical lattices} Let us start from the 1D Bose-Hubbard model obtained from Eq.~\eqref{eq:hubbard_bose} for $\{V_{0,y},V_{0,z}\}\gg V_{0,x}$, such that only tunneling along the $x$-axis is relevant. In the hard-core limit,  double occupancies of bosons with the same pseudospin are energetically forbidden (i.e. $U_{\uparrow\uparrow},U_{\downarrow\downarrow}\gg U_{\uparrow\downarrow}\gg t_{x}$). In this limit, we can project out all states with sites occupied by more than one boson of the same pseudospin provided that the filling is $\langle n_{i,\sigma}\rangle\leq 1$.  By using the corresponding projector $\mathcal{P}_{\rm s}$, we can map the bosonic creation-annihilation operators onto an $\mathfrak{su}(2)$ spin algebra
\beq
\mathcal{P}_{\rm s}b_{i,\sigma}\mathcal{P}_{\rm s}=\ket{0_{i,\sigma}}\bra{1_{i,\sigma}}=:\tilde{b}^{\phantom{\dagger}}_{i,\sigma}, \hspace{2ex} \mathcal{P}_{\rm s}b^\dagger_{i,\sigma}\mathcal{P}_{\rm s}=\ket{1_{i,\sigma}}\bra{0_{i,\sigma}}=:\tilde{b}^\dagger_{i,\sigma}, \hspace{2ex} \mathcal{P}_{\rm s}{b}^\dagger_{i,\sigma}{b}^{\phantom{\dagger}}_{i,\sigma}\mathcal{P}_{\rm s}=\ket{1_{i,\sigma}}\bra{1_{i,\sigma}}=:\tilde{n}_{i\sigma}.
\eeq
In such a hardcore limit, the Bose-Hubbard model~\eqref{eq:hubbard_bose} only contains the  on-site Hubbard interactions for two bosons of opposite pseudospin, namely
\beq
\label{eq:hubbard_bose_hc}
\tilde{H}_{\rm hBH}=\tilde{H}_{\rm loc}+\tilde{H}_{\rm kin}+\tilde{V}_{\rm int}:=\mathcal{P}_{\rm s}{H}_{\rm HB}\mathcal{P}_{\rm s}=\sum_{i,\sigma}\epsilon_{i,\sigma}\tilde{b}_{i,\sigma}^{\dagger}\tilde{b}_{i,\sigma}^{\phantom{\dagger}}-\sum_{i,\sigma}\left(t_{x}\tilde{b}_{i,\sigma}^{\dagger}\tilde{b}_{i+1,\sigma}^{\phantom{\dagger}}+{\rm H.c.}\right)+\frac{1}{2}\sum_{i,\sigma}U_{\sigma{\overline{\sigma}}}\tilde{b}_{i,\sigma}^{\dagger}\tilde{b}_{i,\overline{\sigma}}^{\dagger}\tilde{b}_{i,\overline{\sigma}}^{\phantom{\dagger}}\tilde{b}_{i,\sigma}^{\phantom{\dagger}}.
\eeq
Analogously, we must also project the spin-independent periodic modulation due to the moving optical lattice which, in the same regime as discussed for fermions~\eqref{eq:modulation}, yields
\beq
\tilde{H}_{\rm mod}(t):=\mathcal{P}_{\rm s}{H}_{\rm mod}(t)\mathcal{P}_{\rm s}=\sum_{i,\sigma}\frac{\tilde{V}_{0}}{2}\cos\big(\Delta k X_i-\Delta\omega t+\varphi\big) \tilde{n}_{i,\sigma}^{\phantom{\dagger}}
\eeq
We can now proceed in analogy to  the fermionic gas, as the difference between the fermionic operators and the hardcore-boson ones does not change any of the steps of the derivation.
We move to the interaction picture with respect to the projected driving and Hubbard interactions $U_0(t)=\mathcal{T}\left(\exp\{\ii \int_0^t{\rm d}\tau(\tilde{V}_{\rm int}+\tilde{H}_{\rm mod}(\tau))\}\right)$, and follow the same steps as in the fermionic case to express the time-evolution operator $U(t)=U_0^{\dagger}(t)\ee^{-\ii t\sum_{i}\delta U_{\uparrow\downarrow}\tilde{n}_{i\uparrow}\tilde{n}_{i\downarrow}}\ee^{-\ii \tilde{H}_{\rm eff}t},$ in terms of 
\beq
\label{eq:eff_H_hc}
 \tilde{H}_{\rm eff}=\sum_{i,\sigma}\epsilon_{i,\sigma}\tilde{b}_{i,\sigma}^{\dagger}\tilde{b}_{i,\sigma}^{\phantom{\dagger}}-\sum_{i,\sigma}\left(t_{x}{\rm J}_{r\Delta \tilde{n}_{i+1,\overline{\sigma}}}\left(\eta\right)\ee^{-\ii r\varphi \Delta \tilde{n}_{i+1,\overline{\sigma}}}\tilde{b}_{i,\sigma}^{\dagger}\tilde{b}_{i+1,\sigma}^{\phantom{\dagger}}+{\rm H.c.}\right)+\frac{1}{2}\sum_{i,\sigma}\delta U_{\sigma\overline{\sigma}}\tilde{b}_{i,\sigma}^{\dagger}\tilde{b}_{i,\overline{\sigma}}^{\dagger}\tilde{b}_{i,\overline{\sigma}}^{\phantom{\dagger}}\tilde{b}_{i,\sigma}^{\phantom{\dagger}},
\eeq
where the dressed tunnelling strengths and phases are again density dependent. Given the one-to-one correspondence with the fermionic scheme, the numerical results to test  the validity of our derivation by comparing the full and effective dynamics would not add anything different from Fig.~\ref{fig_pat_fermions_spin_indep}, and we shall not include them here. 

Since the hardcore-boson and the fermionic number operators have the same algebraic properties, we can rewrite this density-dependent tunnelling strength in complete analogy to the fermionic case
\beq
\label{eq:tunn_op_hc}
\begin{split}
{\rm J}_{r\Delta \tilde{n}_{i+1,\overline{\sigma}}}\left(\eta\right)&={\rm J}_0\left(\eta\right)\tilde{h}_{i,\overline{\sigma}}\tilde{h}_{i+1,\overline{\sigma}}+{\rm J}_0\left(\eta\right)\tilde{n}_{i,\overline{\sigma}}\tilde{n}_{i+1,\overline{\sigma}}+{\rm J}_{r}\left(\eta\right)\tilde{n}_{i,\overline{\sigma}}\tilde{h}_{i+1,\overline{\sigma}}+{\rm J}_{r}\left(\eta\right)\tilde{h}_{i,\overline{\sigma}}\tilde{n}_{i+1,\overline{\sigma}},
\end{split}
\eeq
where the hardcore hole operator is $\tilde{h}_{i,\overline{\sigma}}=1-\tilde{n}_{i,\overline{\sigma}}$. Therefore, the dressed tunnelings for hardcore bosons can be described pictorially  by Figs.~\ref{fig_pat_scheme}{\bf (e,f)}, which distinguish  events that preserve/modify the double occupancy of the tunneling sites.

Let us also note that the fermionic schemes for spin-dependent drivings in Sec.~\ref{sec:fermions}, and the generalization to higher dimensions,  equally applies to the bosonic gas in this hardcore limit. Although we shall focus on the fermionic applications of the quantum simulator in Sec.~\ref{qs_applications}, we emphasize that all the quantum many-body models discussed there have a hardcore-boson counterpart, including the dynamical Gauge fields.

\begin{figure*}
\centering
\includegraphics[width=0.75\columnwidth]{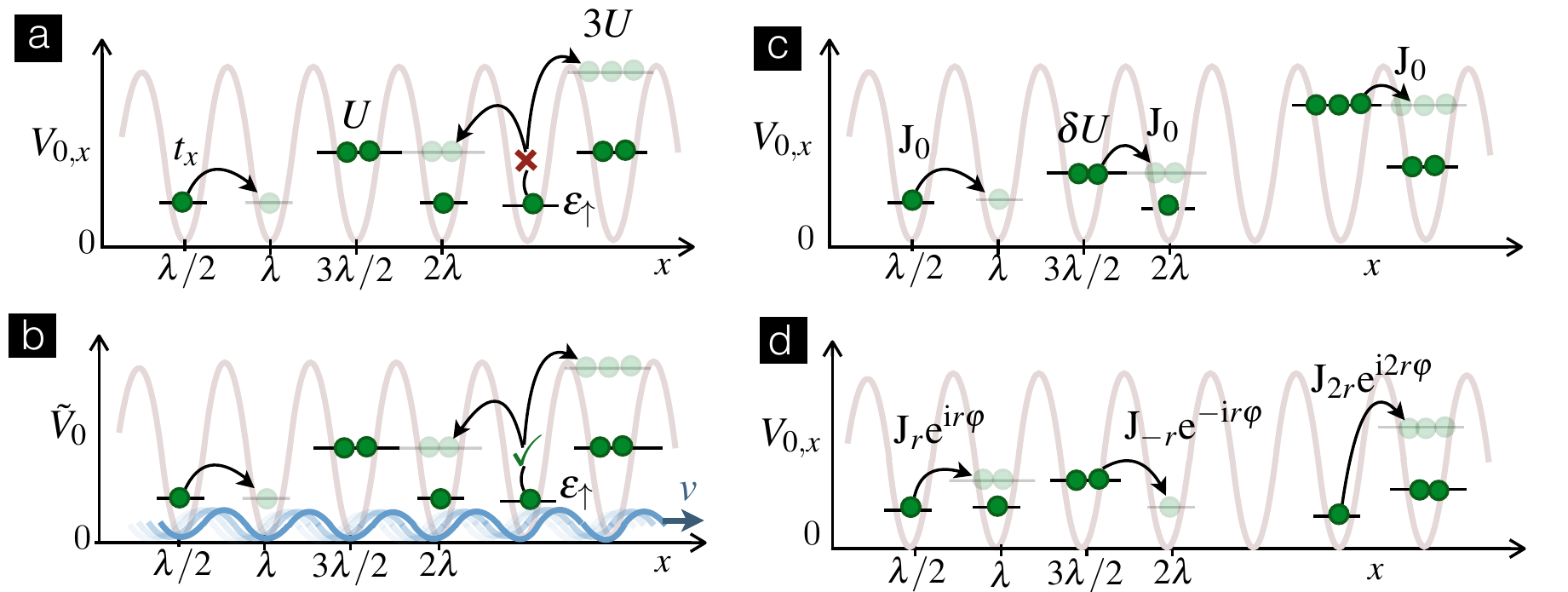}
\caption{ {\bf Scheme of the   spin-independent PAT for bosons: }{\bf (a)} Bosonic atoms in a single hyperfine state $\ket{{\uparrow}}$ (green circles) are trapped at the nodes of a static optical lattice potential (red lines). Tunneling of one atom to an already-occupied site is inhibited by the Hubbard blockade $t_{x}\ll U$. {\bf (b)} Moving optical lattice potential  (blue lines) reactivates the tunneling by  providing the required energy for the atoms to overcome the interaction penalty. {\bf (c,d)} The PAT  can be divided into events that conserve the Hubbard interaction energy  {\bf (c)} controlled by the Bessel function ${\rm J}_0:={\rm J}_0(\eta)$, and those that change it {\bf (d)}, and are controlled by various Bessel functions ${\rm J}_\ell:={\rm J}_\ell(\eta)$ and the corresponding tunneling phases $\ee^{\pm \ii \ell\varphi}$. Also note that the residual dressed  interaction $\delta U$ is changed with respect to the bare original one $U$.}
\label{fig_pat_scheme_spin_indep_boson}
\end{figure*}

\vspace{1ex}
\subsubsection{Two-component softcore bosons in spin-independent moving optical lattices}  The objective of this section is to relax the hardcore constraint  $U_{\uparrow\uparrow},U_{\downarrow\downarrow}\gg U_{\uparrow\downarrow}\gg t_{x}$, which forbids   double occupancies of bosons with the same pseudospin. Let us, however,  start by  understanding the PAT  of a single-component bosonic gas described by Eq.~\eqref{eq:hubbard_bose}, but restricted  to a single pseudospin (e.g. $\sigma=\uparrow$). For notational convenience, we drop the pseudospin index, such that the Hamiltonian corresponds to the standard Bose-Hubbard model
\beq
\label{eq:hubbard_bose_single}
{H}_{\rm BH}=H_{\rm loc}+H_{\rm kin}+V_{\rm int}=\sum_{i}\epsilon_{i}{b}_{i}^{\dagger}{b}_{i}^{\phantom{\dagger}}-\sum_{i}\left(t_{x}{b}_{i}^{\dagger}{b}_{i+1}^{\phantom{\dagger}}+{\rm H.c.}\right)+\sum_{i}\frac{U}{2}n_{i}(n_{i}^{\phantom{\dagger}}-1).
\eeq
According to our  discussion of the Hubbard blockade $t_x\ll U$, the tunnelling of bosons that changes the parity of the  occupation number is energetically forbidden (see Fig.~\ref{fig_pat_scheme_spin_indep_boson}{\bf (a)}). As customary, we activate this tunnelling by means of the periodic modulation
\beq
\label{eq:single_boson_driving}
H_{\rm mod}(t)=\sum_{i}\frac{\tilde{V}_{0}}{2}\cos\big(\Delta k X_i-\Delta\omega t+\varphi\big)n^{\phantom{\dagger}}_{i},
\eeq
given by a moving optical lattice acting on the bosonic atoms (see Fig.~\ref{fig_pat_scheme_spin_indep_boson}{\bf (b)}).
Due to the different Hubbard interaction, which now involves a single pseudospin, one cannot use  the expression in Eq.~\eqref{eq:int_picture}. Instead, we consider  the interaction picture of a bond operator $B_{i,i+1}=b^{{\dagger}}_{i}b^{\phantom{\dagger}}_{i+1}+b^{{\dagger}}_{i+1}b^{\phantom{\dagger}}_{i}$ which, again up to an irrelevant Gauge transformation, can be shown to be 
\beq
U_0(t)B_{i,i+1}U^\dagger_0(t)=\ee^{-\ii t U(n_{i+1}-n_i+1)}\ee^{-\ii\frac{\tilde{V}_{0}}{2\Delta\omega}\sin(\Delta k X_i-\Delta\omega t+\varphi)}\ee^{\ii\frac{\tilde{V}_{0}}{2\Delta\omega}\sin(\Delta k X_{i+1}-\Delta\omega t+\varphi)}b^{{\dagger}}_{i}b^{\phantom{\dagger}}_{i+1}+{\rm H.c.},
\eeq
where $U_0(t)=\mathcal{T}\left(\exp\{\ii \int_0^t{\rm d}\tau({V}_{\rm int}+{H}_{\rm mod}(\tau))\}\right)$ is the interaction-picture unitary.
After defining the bosonic population difference operator $\Delta n_{i+1}=n_{i+1}-n_i$, and using the Jacobi-Auger expansion for each of the time-dependent exponentials, the expression of the kinetic energy, 
$H_{\rm kin}(t)=U_0(t)H_{\rm kin}U^{\dagger}_0(t)$, becomes
 \beq
 H_{\rm kin}(t)=-\sum_{i,\sigma}\left(t_{x}(t)b_{i}^{\dagger}b_{i+1}^{\phantom{\dagger}}+{\rm H.c.}\right),\hspace{2ex}t_{x}(t)=t_{x}\ee^{-\ii t U(\Delta n_{i+1}+1)}\mathrm{f}(t),
\eeq
with exactly the same modulation function as in Eq.~\eqref{eq:mod_function}. 

\begin{figure*}
\centering
\includegraphics[width=.9\columnwidth]{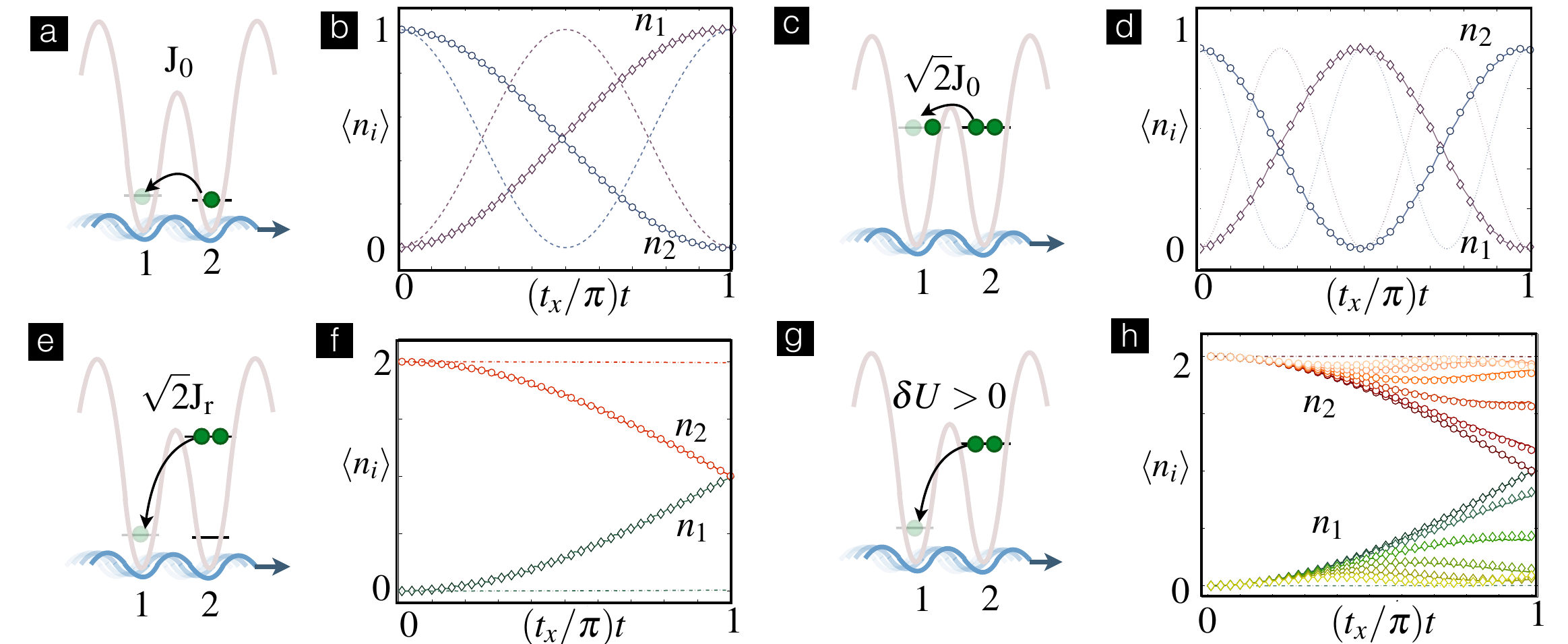}
\caption{ {\bf Interaction-dependent  PAT for single-component softcore bosons:} Population dynamics of a periodically-driven Bose-Hubbard dimer with  $t_x=0.1$, $U=20$, and $\varphi=0$ for different density distributions.  {\bf (a,b)} Initial state with one atom on the right well $\ket{0,1}$. The dashed lines correspond to the tunneling for the undriven $V_0=0$ dimer, while the resonant driving  $\Delta\omega=U/2$ (i.e. two-photon assisted tunneling $r=2$) with $V_0=1.5\Delta\omega$ corresponds to the numerical solution of the exact Hamiltonian~\eqref{eq:hubbard_bose_single}-\eqref{eq:single_boson_driving} (solid lines), and the effective one~\eqref{eq:eff_H_singel_species}-\eqref{dressed_tunn_scbosons} (symbols). We use the same criterion dashed-line/solid-line/symbols in all the figures. {\bf (c,d)} Same as above for an initial state with two atoms on the right well  and one atom on the left well $\ket{1,2}$. {\bf (e,f)} Same as above for an initial state $\ket{0,2}$.  {\bf (g,h)} Off-resonant effects on the PAT for the initial state $\ket{0,2}$ by modifying the detuning $\delta U/t_x\in\{0, 0.5, 1, 1.5, 2, 2.5\}$ (dark-to-bright coloring implies increasing the detuning $\delta U$).  }
\label{fig_pat_bososn_1_spin}
\end{figure*}
We can then proceed by following the same steps as for the fermionic case, to find a parameter regime
\beq
\label{pat_regime_bosons}
t_{x}, \delta U=(U-r\Delta\omega)\ll U\approx r\Delta\omega,
\eeq
and finally obtain the effective interaction-dependent PAT Hamiltonian
\beq
\label{eq:eff_H_singel_species}
 H_{\rm eff}=\sum_{i}\epsilon_{i}b_{i}^{\dagger}b_{i}^{\phantom{\dagger}}-\sum_{i}\left(t_{x}{\rm J}_{r(\Delta n_{i+1}+1)}\left(\eta\right)\ee^{-\ii r\varphi(\Delta n_{i+1}+1)}b_{i}^{\dagger}b_{i+1}^{\phantom{\dagger}}+{\rm H.c.}\right)+\sum_{i}\frac{\delta U}{2}n_{i}(n_i-1).
\eeq
Here, we observe that the dressed tunnelling depends on the density of bosons of the same pseudospin at the sites involved in the tunnelling event (see Figs.~\ref{fig_pat_scheme_spin_indep_boson}{\bf (c,d)}). Since the boson number per lattice site  is not restricted anymore by the hardcore constraint, we cannot express it as the simple  polynomial~\eqref{eq:tunn_op_hc} quadratic in the density operators, but rather as a highly non-linear term.  Using the orthogonal projectors onto subspaces with a well-defined difference number of bosons $\mathcal{P}_{\Delta n_{i+1}=\pm \ell}$, this non-linearity becomes apparent
\beq
\label{dressed_tunn_scbosons}
{\rm J}_{r(\Delta n_{i+1}+1)}\left(\eta\right)=\sum_{\ell'=0}^{N_{\rm b}+1}{\rm J}_{r\ell'}(\eta)\mathcal{P}_{\Delta n_{i+1}+1=\pm \ell'}=\sum_{\ell'=0}^{N_{\rm b}+1}{\rm J}_{r\ell'}(\eta)\prod_{\ell\neq\ell'}\frac{\ell^2-(\Delta n_{i+1}+1)^2}{\ell^2-\ell'^2},
\eeq
where $N_{\rm b}$ is the total number of bosons loaded in the optical lattice, and we have again assumed that $r$ is an even integer. In analogy to the studies for the phase-modulation driving~\cite{interaction_pat_theory,interaction_resonances_cdt_theory}, we observe that there will be interaction-shifted resonances that correspond to the zeros of the  Bessel functions for different density backgrounds. In fact, using our formalism, one could derive similar analytic expressions for a phase-modulation driving~\cite{interaction_pat_theory,interaction_resonances_cdt_theory}. For the intensity-modulated lattices of the recent experiments~\cite{pat_bhm, superexchange_pat}, the situation is  simpler  as there can only be one particular occupation that is resonant with the periodic modulation of the tunneling matrix element, and no Bessel functions arise. The possibility of engineering Bose-Hubbard models with a density-dependent tunneling strength and phase by our PAT scheme can be considered as an alternative to the proposals based on periodic modulations of the $s$-wave scattering length~\cite{interaction_modulations, peierls_cold_atoms_int_modulation}. Let us finally note that  it is possible to generalize this scheme to higher dimensions, paralleling  the fermionic case Sec.~\ref{sec:fermions}.

In order to test the validity of our derivations, we have numerically compared the time-evolution predicted by either the effective Hamiltonian~\eqref{eq:eff_H_singel_species}-\eqref{dressed_tunn_scbosons}, or the periodically driven one~\eqref{eq:hubbard_bose_single}-\eqref{eq:single_boson_driving} for a Bose-Hubbard dimer in the regime of interaction-dependent PAT. In Fig.~\ref{fig_pat_bososn_1_spin}, we explore the real-time dynamics for different configurations of atoms in the initial state: {\it (i) Non-blockaded regime}: Figs.~\ref{fig_pat_bososn_1_spin}{\bf (a-b)} represent the dynamics of the initial atomic configurations $\ket{0,1}=b_2^\dagger\ket{\rm vac}$, which does not display Hubbard blockade as it consists of a single atom. Yet, the bare tunneling (see dashed lines of Fig.~\ref{fig_pat_bososn_1_spin}{\bf (b)}) is renormalized due to the periodic driving, as shown by the different population dynamics displayed by the solid lines (exact) and the symbols (effective). The same renormalization occurs for the configuration $\ket{1,2}$ in Figs.~\ref{fig_pat_bososn_1_spin}{\bf (c-d)}, which is not blockaded as the overall occupation parity is conserved in the tunneling process.  On top of the dressing of the tunneling, we observe a bosonic enhancement which leads to a doubled tunneling rate with respect to Fig.~\ref{fig_pat_bososn_1_spin}{\bf (b)}. {\it (ii) Hubbard-blockaded regime}: Figs.~\ref{fig_pat_bososn_1_spin}{\bf (e-h)} represent the dynamics of the initial atomic configuration $\ket{0,2}$, which suffers a Hubbard blockade as the tunneling must change the overall occupation parity. Hence, in the absence of the driving, the atomic tunneling is totally forbidden (see dashed lines of Fig.~\ref{fig_pat_bososn_1_spin}{\bf (f)}). By switching on the driving, we observe that the tunneling is reactivated, as shown  by the solid lines (exact) and the symbols (effective). So far, all these simulations correspond to the resonant PAT, where the parameter regime~\eqref{pat_regime_bosons} is achieved for $\delta U=0$. In Figs.~\ref{fig_pat_bososn_1_spin}{\bf (g-h)}, we explore the off-resonant case $\delta U>0$, and the possibility of describing the driving detuning as a residual Hubbard interaction. The agreement between the solid lines (exact) and the symbols (effective) in Fig.~\ref{fig_pat_bososn_1_spin}{\bf (h)}, shows that this is indeed the case. We observe that, as  $\delta U$ is increased, the periodic exchange of particles is gradually inhibited, as one would expect since the single- and doubly-occupied Hubbard bands become more and more separated in energy.
 
\begin{figure*}
\centering
\includegraphics[width=.95\columnwidth]{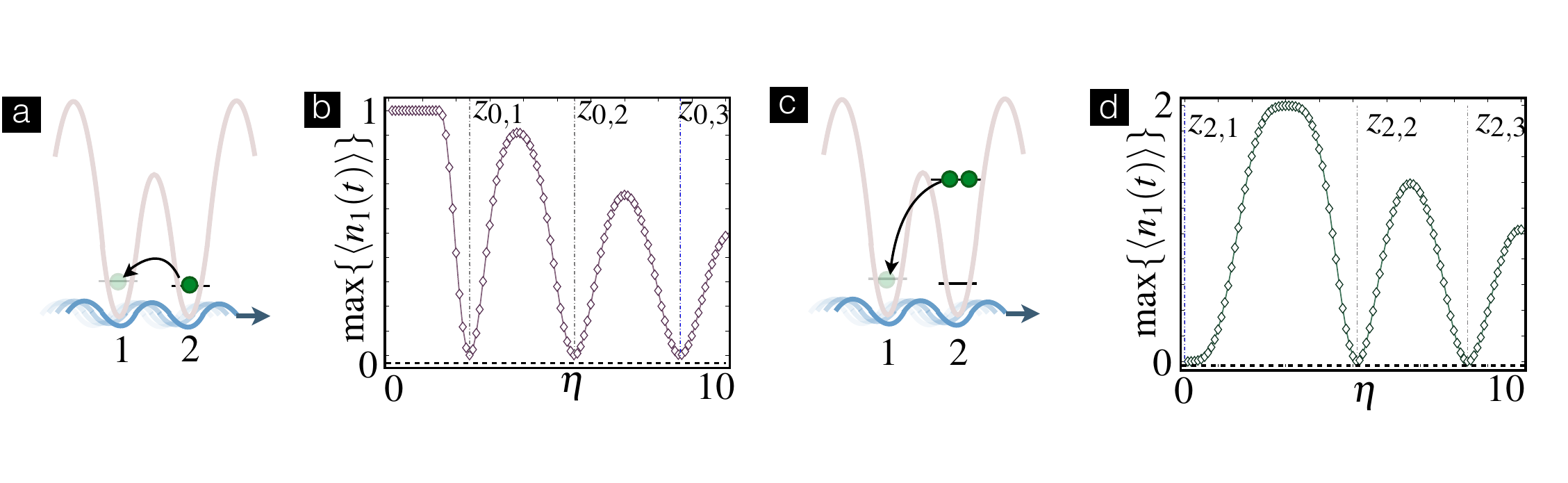}
\caption{ {\bf Density-dependent coherent destruction of tunneling for single-component softcore bosons:} Maximum population reaching the left well ${\rm max}\{\langle n_1(t):0<t<\pi t_x\}$ for a periodically-driven Bose-Hubbard dimer with parameters $t_x=0.1$, $U=20$, for a resonant drying $\Delta\omega=U/2$ (i.e. two-photon assisted tunneling $r=2$) with $\varphi=0$, as a function of the driving ratio $\eta$ for different density distributions. {\bf (a-b)} Initial state $\ket{0,1}$, which displays minima in ${\rm max}\{\langle n_1(t):0<t<\pi /t_x\}$ exactly at  the zeros of the Bessel function ${\rm J}_0(z_{0,n})=0$. {\bf (a-b)} Initial state $\ket{0,2}$, which displays minima in ${\rm max}\{\langle n_1(t):0<t<\pi /t_x\}$ exactly at  the zeros of the Bessel function ${\rm J_2}(z_{2,n})=0$.   }
\label{fig_pat_scheme_bosons_cdt}
\end{figure*}

As a further numerical proof of the consistency of  our effective description, let us explore the phenomenon of coherent destruction of tunneling. According to Eq.~\eqref{dressed_tunn_scbosons}, the effective tunneling is dressed by a different Bessel function depending on the densities of the bosonic atoms. For instance, the PAT tunneling of Fig.~\ref{fig_pat_scheme_bosons_cdt}{\bf (a)} is controlled by ${\rm J}_0(\eta)$, whereas the tunneling of Fig.~\ref{fig_pat_scheme_bosons_cdt}{\bf (c)} is controlled by ${\rm J}_r(\eta)$. Therefore, whenever the driving parameter $\eta$ coincides with a zero of the corresponding Bessel function, the tunneling should get coherently suppressed. In Fig.\ref{fig_pat_scheme_bosons_cdt}{\bf (b)}, we observe this effect at $\eta=z_{0,n}$ for $n=1,2,3$ zeros of the Bessel function ${\rm J}_0(z_{0,n})=0$, which are displayed by the  dashed dotted lines. We see how the maximal average population that reaches the left site of the Hubbard dimer vanishes when the driving ratio coincides with any of the zeros.  In Fig.\ref{fig_pat_scheme_bosons_cdt}{\bf (d)}, we see that for a different atomic density distribution, the coherent destruction of tunneling takes place at the zeros of a different Bessel function, namely  $\eta=z_{r,n}$ for $n=1,2,3$ fulfilling  ${\rm J}_r(z_{r,n})=0$ for the chosen $r=2$.

\begin{figure*}
\centering
\includegraphics[width=.95\columnwidth]{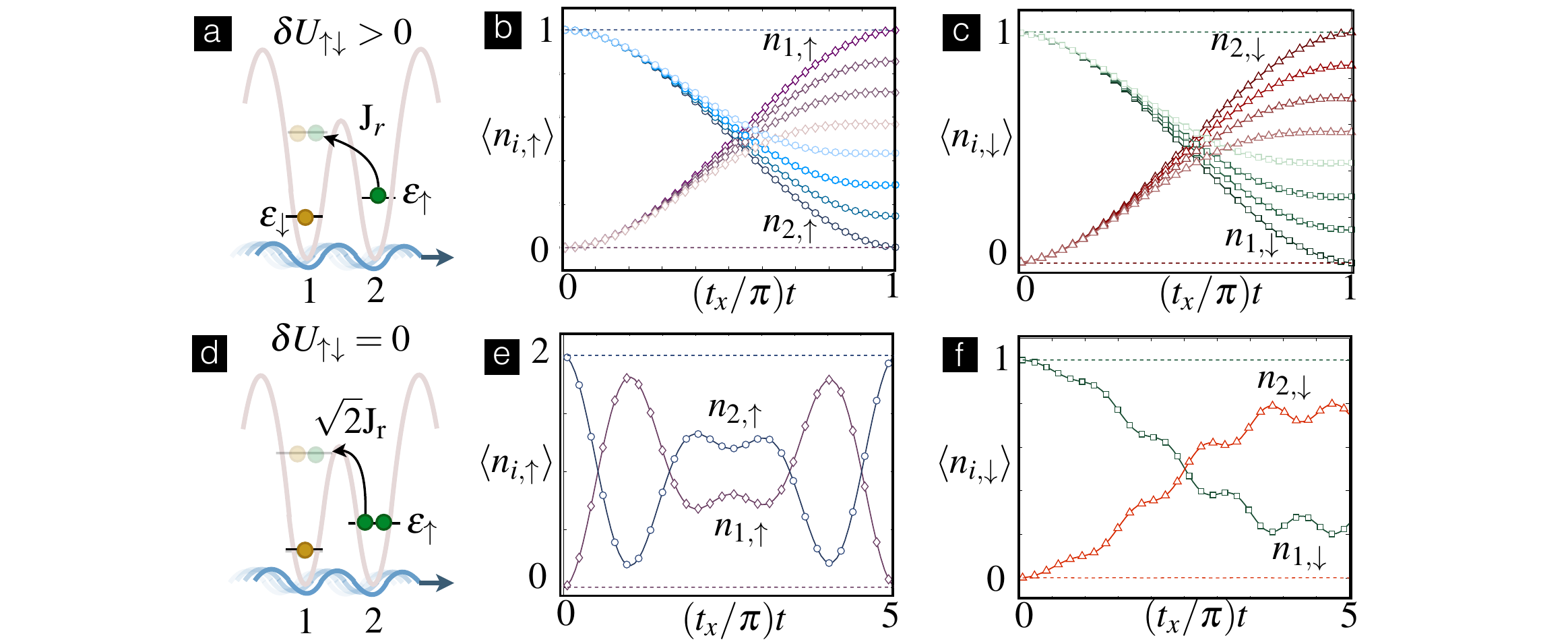}
\caption{ {\bf Interaction-dependent PAT for two-component softcore bosons:} Population dynamics of a periodically-driven two-component Bose-Hubbard dimer with parameters $t_x=0.1$, $U_{\uparrow\downarrow}=20$,  $U_{\uparrow\uparrow}=U_{\downarrow\downarrow}=0$, and $\varphi=0$,  for different density distributions.  {\bf (a,b,c)} Initial state with one spin-up atom on the right well, and one spin-down atom on the left well $\ket{1_\downarrow,1_\uparrow}$. The dashed lines correspond to the tunneling for the undriven $V_0=0$ dimer, which shows the Hubbard blockade as a consequence of $t_x\ll U_{\uparrow\downarrow}$. Switching on the  off-resonant  driving  $\Delta\omega\approx U_{\uparrow\downarrow}/2$  (i.e. two-photon assisted tunneling $r=2$) with detunings $ \delta U_{\uparrow\downarrow}/t_x\in\{0, 0.5, 0.75, 1\}$ (dark-to-bright transition implies increasing the detuning $\delta U_{\uparrow\downarrow}$), we observe that the tunneling for the spin-up atoms {\bf (b)} (and spin-down atoms {\bf (c)}) is reactivated, and depends on the residual Hubbard repulsion $ \delta U_{\uparrow\downarrow}$. Here, the solid lines correspond to the numerical solution of the exact dynamics~\eqref{eq:2_bh}-\eqref{eq:two_spin_boson_driving}, while the symbols represent  the effective dynamics~\eqref{eq:eff_H_two_species}-\eqref{dressed_tunn_scbosons_2_spins}, for a periodic driving of intensity $V_0=3\Delta\omega$. {\bf (d,e,f)} Same as above, but for  an initial state $\ket{1_\downarrow,2_\uparrow}$ and resonant driving $\delta U_{\uparrow\downarrow}=0$. Due to the bosonic enhancement of the dressed tunneling of the spin-up atoms {\bf (d)}, the tunneling of each spin-species takes place at a different rate. As a consequence, the dynamical Hubbard blockade leads to a more complex tunneling dynamics. }
\label{fig_pat_bososn_2_spin}
\end{figure*}

Once  the interaction-dependent  PAT for the single-pseudospin bosons has been understood, and its validity has been checked numerically,  we can  turn our attention to the situation of two-pseudospin bosons without the hardcore constraint
\beq
\label{eq:2_bh}
H_{\rm BH}=H_{\rm loc}+H_{\rm kin}+V_{\rm int}=\sum_{ i,\sigma}\epsilon_{{ i},\sigma}b_{i,\sigma}^{\dagger}b_{{ i},\sigma}^{\phantom{\dagger}}-\sum_{{ i},\sigma}\left(t_{x}b_{{ i},\sigma}^{\dagger}b_{{ i+1},\sigma}^{\phantom{\dagger}}+{\rm H.c.}\right)+\frac{1}{2}\sum_{{ i}}\sum_{\sigma,\sigma'}U_{\sigma{\sigma'}}b_{{ i},\sigma}^{\dagger}b_{{ i},{\sigma'}}^{\dagger}b_{{ i},{\sigma'}}^{\phantom{\dagger}}b_{{ i},\sigma}^{\phantom{\dagger}}.
\eeq
We shall be interested in the regime of $U_{\uparrow\downarrow}\gg U_{\uparrow\uparrow}, U_{\downarrow\downarrow},t_x$, where 
double occupancy of bosons of the same (different) pseudospin is allowed (forbidden).  In this regime, as the intra-spin interactions do not blockade the tunneling, we only include the inter-spin interactions in  the interaction picture  $U_0(t)=\mathcal{T}\left(\exp\{\ii \int_0^t{\rm d}\tau({V}_{\rm int,\uparrow\downarrow}+{H}_{\rm mod}(\tau))\}\right)$, where a spin-independent  moving optical lattice is applied to both pseudospins
\beq
\label{eq:two_spin_boson_driving}
H_{\rm mod}(t)=\sum_{i,\sigma}\frac{\tilde{V}_{0}}{2}\cos\big(\Delta k X_i-\Delta\omega t+\varphi\big)n^{\phantom{\dagger}}_{i,\sigma},
\eeq
 One can then see that, in the following parameter regime
\beq
\label{pat_regime_bosons_soft_core}
t_{x}, U_{\uparrow\uparrow}, U_{\downarrow\downarrow}, \delta U_{\uparrow\downarrow}=(U_{\uparrow\downarrow}-r\Delta\omega)\ll U_{\uparrow\downarrow}\approx r\Delta\omega,
\eeq
the dressed tunnelings will only depend on the density of atoms of the opposite pseudospin, as occurs for the hardcore bosons or the fermions. Therefore, the effective Hamiltonian becomes
\beq
\label{eq:eff_H_two_species}
 H_{\rm eff}=\sum_{i,\sigma}\epsilon_{i,\sigma}b_{i,\sigma}^{\dagger}b_{i,\sigma}^{\phantom{\dagger}}-\sum_{i,\sigma}\left(t_{x}{\rm J}_{r\Delta n_{i+1,\overline{\sigma}}}\left(\eta\right)\ee^{-\ii r\varphi \Delta n_{i+1,\overline{\sigma}}}b_{i,\sigma}^{\dagger}b_{i+1,\sigma}^{\phantom{\dagger}}+{\rm H.c.}\right)+\frac{1}{2}\sum_{i}\sum_{\sigma,\sigma'}\tilde{U}_{\sigma\sigma'}{b}_{i,\sigma}^{\dagger}{b}_{i,\sigma'}^{\dagger}{b}_{i,{\sigma'}}^{\phantom{\dagger}}{b}_{i,\sigma}^{\phantom{\dagger}},
\eeq
where we have introduced the residual interactions $\tilde{U}_{\uparrow\uparrow}={U}_{\uparrow\uparrow}$, $\tilde{U}_{\downarrow\downarrow}={U}_{\downarrow\downarrow}$, and $\tilde{U}_{\uparrow\downarrow}=\tilde{U}_{\downarrow\uparrow}=\delta{U}_{\uparrow\downarrow}$. For $\varphi=0$, we get an exotic Bose-Hubbard model with the analogue of the fermionic bond-charge interactions, whereby the dressed tunneling of one pseudospin depends on all the possible density backgrounds of the other pseudospin through the corresponding Bessel functions. The difference with respect to the fermionic bond-charge interactions~\eqref{eq:tunn_op} is  the  highly non-linear function of the bosonic densities, namely
\beq
\label{dressed_tunn_scbosons_2_spins}
{\rm J}_{r\Delta n_{i+1,\overline{\sigma}}}\left(\eta\right)=\sum_{\ell'=0}^{N_{\rm b}}{\rm J}_{r\ell'}(\eta)\mathcal{P}_{\Delta n_{i+1,\overline{\sigma}}=\pm \ell'}=\sum_{\ell'=0}^{N_{\rm b}}{\rm J}_{r\ell'}(\eta)\prod_{\ell\neq\ell'}\frac{\ell^2-(\Delta n_{i+1,\overline{\sigma}})^2}{\ell^2-\ell'^2}.
\eeq
 It is also worth commenting that, had we set $U_{\uparrow\downarrow}=U_{\uparrow\uparrow}=U_{\downarrow\downarrow}\gg t_x$ in the blockade conditions, the dressed tunneling would depend on the density of both pseudospins ${\rm J}_{r\Delta n_{i+1,\overline{\sigma}}}\left(\eta\right)\ee^{-\ii r\varphi \Delta n_{i+1,\overline{\sigma}}}\to{ \rm J}_{r(\Delta n_{i+1,{\sigma}}+\Delta n_{i+1,\overline{\sigma}}+1)}\left(\eta\right)\ee^{-\ii r\varphi( \Delta n_{i+1,{\sigma}}+\Delta n_{i+1,\overline{\sigma}}+1)}$, which might also be interesting regarding exotic Bose-Hubbard models. Let us note that, once again, this effective description matches perfectly the dynamics of the driven two-component  Bose-Hubbard dimer (see Fig.~\ref{fig_pat_bososn_2_spin}).

Once again, we could generalize  to higher dimensions, or to spin-dependent drivings, and  study  the strongly-correlated  models that arise. However, the  properties particular for the restricted number of particles of fermions and hardcore bosons, which allow for instance to build a  quantum simulator of dynamical Gauge fields (see Sec.~\ref{qs_applications}), cannot be generalized to the soft-core boson case. Let us finally note that, by considering the hardcore-boson limit on Eq.~\eqref{eq:eff_H_two_species}, one recovers the previous result~\eqref{eq:eff_H_hc}. 

\subsection{Scheme for a periodically-modulated  ultracold Fermi-Bose mixture}

Let us now turn our attention to a mixture of  bosonic and fermionic atoms, and consider that the two hyperfine states of the original formulation~\eqref{eq:originalH}   correspond to   the fermionic and bosonic atoms $\ket{{\uparrow}}=\ket{F,M}=:\ket{\rm f}, \ket{{\downarrow}}=\ket{F',M'}=:\ket{\rm b}$, respectively. In the Wannier basis, $\Psi_\uparrow({\bf r})=\sum_{{\bf i}}w({\bf r}-{\bf r}^0_{\bf i})f_{{\bf i}}$ for the fermions, and $\Psi_\downarrow({\bf r})=\sum_{{\bf i}}w({\bf r}-{\bf r}^0_{\bf i})b_{{\bf i}}$ for the bosons, the general Hamiltonian~\eqref{eq:originalH} can be expressed in terms of  the fermionic $f_{{\bf i}}$ and  bosonic  $b_{{\bf i}}$  operators~\cite{bose_fermi_hubbard_ol} as a  Bose-Fermi Hubbard model
\beq
\label{eq:hubbard_bose}
\begin{split}
H_{\rm BFH}&= \sum_{{\bf i}}\epsilon_{{\bf i},\rm b}b_{\bf i}^{\dagger}b_{{\bf i}}^{\phantom{\dagger}}-\sum_{{\bf i},\alpha}\left(t_{\alpha,{\rm b}}b_{{\bf i}}^{\dagger}b_{{\bf i+e_\alpha}}^{\phantom{\dagger}}+{\rm H.c.}\right)+\sum_{{\bf i}}U_{\rm bb}b_{{\bf i}}^{\dagger}b_{{\bf i}}^{\dagger}b_{{\bf i}}^{\phantom{\dagger}}b_{{\bf i}}^{\phantom{\dagger}}\\
&+\sum_{{\bf i}}\epsilon_{{\bf i},\rm f}f_{\bf i}^{\dagger}f_{{\bf i}}^{\phantom{\dagger}}-\sum_{{\bf i},\alpha}\left(t_{\alpha,{\rm f}}f_{{\bf i}}^{\dagger}f_{{\bf i+e_\alpha}}^{\phantom{\dagger}}+{\rm H.c.}\right)+\sum_{{\bf i}}U_{\rm bf}b_{{\bf i}}^{\dagger}f_{{\bf i}}^{\dagger}f_{{\bf i}}^{\phantom{\dagger}}b_{{\bf i}}^{\phantom{\dagger}}.
\end{split}
\eeq
Here, the Hamiltonian parameters now depend on the fermionic/bosonic nature of the atoms, such as the tunnelings
\beq
\label{eq:parameters_fb}
 t_{\alpha,\rm b}=\frac{4}{\sqrt{\pi}}E_{\rm R,\rm b}\left(\frac{V_{0,\alpha}}{E_{\rm R, b}}\right)^{3/4}\ee^{-2\sqrt{\frac{V_{0,\alpha}}{E_{\rm R,b}}}},\hspace{2ex}t_{\alpha,\rm f}=\frac{4}{\sqrt{\pi}}E_{\rm R,\rm f}\left(\frac{V_{0,\alpha}}{E_{\rm R, f}}\right)^{3/4}\ee^{-2\sqrt{\frac{V_{0,\alpha}}{E_{\rm R,f}}}},
\eeq
which differ due to the different masses $m_{\rm f}, m_{\rm b}$ through the corresponding recoil energies $E_{\rm R,f},E_{\rm R,b}$. The Hubbard interactions  
\beq
U_{\rm bb}=\sqrt{\frac{8}{\pi}}ka_{\rm bb}E_{\rm R,b}\left(\frac{V_{0,x}V_{0,y}V_{0,z}}{E_{\rm R,b}^3}\right)^{1/4},\hspace{2ex} U_{\rm bf}=U_{\rm bb}\frac{a_{\rm bf}}{a_{\rm bb}}\sqrt{2}\left(\frac{m_{\rm f}}{m_{\rm b}}\right)^{3/4}\left(1+\frac{m_{\rm f}}{m_{\rm b}}\right)\left(1+\sqrt{\frac{m_{\rm f}}{m_{\rm b}}}\right)^{-3/2},
\eeq
not only differ by the mass ratio, but also by the different scattering lengths for a boson-fermion  $a_{\rm bf}$, and a boson-boson  $a_{\rm bb}$ collision.
Once again, the fermion-fermion collisions are forbidden by the Pauli exclusion principle. We shall discuss different regimes of interest for the quantum simulation of Bose-Fermi quantum many-body models, which can be achieved by tuning the Feshbach resonances, and controlling these scattering lengths appropriately.

Let us focus on the 1D case. The moving optical lattices can be  made dependent on the bosonic/fermionic atomic species, as the corresponding atoms usually  have a very different atomic level structure. Therefore, in analogy to Eq.~\eqref{eq:modulation_spin_dependent}, we consider the following periodic driving
\beq
H_{\rm mod}(t)=\sum_{i}\frac{\tilde{V}_{0,\rm f}}{2}\cos\big(\Delta k_{\rm f} X_i-\Delta\omega_{\rm f} t+\varphi_{\rm f}\big)n_{i,\rm f}^{\phantom{\dagger}}+\sum_{i}\frac{\tilde{V}_{0,\rm b}}{2}\cos\big(\Delta k_{\rm b} X_i-\Delta\omega_{\rm b} t+\varphi_{\rm b}\big)n_{i,\rm b}^{\phantom{\dagger}},
\eeq
where the amplitudes $\tilde{V}_{0,\rm f/b}$ wave vectors $\Delta k_{\rm f/b}$, frequencies $\Delta\omega_{\rm f/b}$, and relative phases $\varphi_{\rm f/b}$, depend on the particular atomic species, and can be controlled experimentally. These periodic drivings will assist the tunneling against an energy penalty given by  the Bose-Fermi interaction, which is very large 
\beq
\label{pat_regime_bosons_soft_core}
t_{x}, U_{\rm bb} , \delta U_{\rm bf}=(U_{\rm bf}-r_{\rm b}\Delta\omega_{\rm b})\ll U_{\rm bf}\approx r_{\rm b}\Delta\omega_{\rm b}=r_{\rm f}\Delta\omega_{\rm f},
\eeq
where we have introduced the residual Bose-Fermi interactions $\delta U_{\rm bf}$, and two integers $r_{\rm b},r_{\rm f}\in\mathbb{Z}$ that determine how many photons are absorbed from the periodic driving to overcome the energy penalty, and assist the tunneling. 

In analogy with the two-component softcore bosons, only the interspecies Hubbard interactions inhibits the tunneling, such that   the interaction picture  must be $U_0(t)=\mathcal{T}\left(\exp\{\ii \int_0^t{\rm d}\tau({V}_{\rm int,bf}+{H}_{\rm mod}(\tau))\}\right)$, leading to the time-evolution operator $U(t)=U_0^{\dagger}(t)\ee^{-\ii t\sum_{i}\delta U_{\rm bf}{n}_{i,{\rm b}}{n}_{i,{\rm f}}}\ee^{-\ii {H}_{\rm eff}t},$ with the following effective Bose-Fermi Hubbard Hamiltonian
\beq
\label{eq:state_depn_Heff_bf}
\begin{split}
 H_{\rm eff}&=\sum_{i}\epsilon_{i,\rm b}b_{i}^{\dagger}b_{i}^{\phantom{\dagger}}-\sum_{i}\left(t_{x}{\rm J}_{r_{\rm b}\Delta n_{i+1,{\rm f}}}\left({\eta}_{\rm b}\right)\ee^{-\ii r_{\rm b}\varphi_{\rm b}\Delta n_{i+1,{\rm f}}}b_{i}^{\dagger}b_{i+1}^{\phantom{\dagger}}+{\rm H.c.}\right)+\frac{1}{2}\sum_{i}U_{\rm bb}b_{i}^{\dagger}b_{i}^{\dagger}b_{i}^{\phantom{\dagger}}b_{i}^{\phantom{\dagger}}\\
 &+\sum_{i}\epsilon_{i,\rm f}f_{i}^{\dagger}f_{i}^{\phantom{\dagger}}-\sum_{i}\left(t_{x}{\rm J}_{r_{\rm f}\Delta n_{i+1,{\rm b}}}\left({\eta}_{\rm f}\right)\ee^{-\ii r_{\rm f}\varphi_{\rm f}\Delta n_{i+1,{\rm b}}}f_{i}^{\dagger}f_{i+1}^{\phantom{\dagger}}+{\rm H.c.}\right)+\frac{1}{2}\sum_{i}\delta U_{\rm bf}b_{i}^{\dagger}f_{i}^{\dagger}f_{i}^{\phantom{\dagger}}b_{i}^{\phantom{\dagger}}.
 \end{split}
\eeq
We thus observe that the bosonic/fermionic dressed tunnelings, both the amplitude and phase, depend on the fermionic/bosonic densities. Moreover, they can be independently tuned by controlling the parameters of the bosonic/fermionic moving lattice, which will be interesting for the quantum simulator of dynamical Gauge fields (see Sec.~\ref{qs_applications}). We also note that the generalization to 2D yields a Bose-Fermi analogue of the effective Hamiltonian~\eqref{eq:2d_eff_H}.

\section{Applications of the  photon-assisted tunneling quantum simulator}
\label{qs_applications}

We have introduced a scheme to implement an interaction-dependent PAT with trapped ultracold atoms. In this way, we have obtained a toolbox with different effective Hamiltonians that depend on  the quantum statistics, dimensionality, and spin-dependent/spin-independent nature of the periodic driving used to assist the tunnelling. In this section, we will discuss how such a toolbox can be exploited for the quantum simulation of interesting problems in condensed matter and high-energy physics.  Instead of focusing on a particular application, we have decided to stress the wide scope of the proposed quantum simulator by describing a variety of interesting quantum many-body models. All these problems share a common feature, they are described by complex quantum many-body models, either on the lattice or in the continuum, which still present open questions that defy the capabilities of existing numerical methods.
We will describe the context of the particular models that can be targeted with the quantum simulator, and try to discuss the essence of the phenomena that they try to capture. Moreover, we will  highlight what we believe are  open questions of the models that have been  studied in detail over the years, and point to future work required to understand the models that have not been explored in such  a detail.

\subsection{Bond-charge interactions and correlated electron-hole tunnelings}
\label{bond_charge_model}
In the original derivation of the Hubbard model for electrons in transition metals~\cite{hubbard},  it was shown that a variety of longer-range terms also arise in the Hamiltonian. In addition to  the  tunnelling and the on-site density-density coupling,  Coulomb interactions also yield  nearest-neighbor terms that can be described as the repulsive interaction between: {\it (i)} charges localized at neighboring ions $V,$ {\it (ii)} charges localized at an ion and a neighboring bond/link $X,$ and {\it (iii)} charges localized at two neighboring bonds/links $W.$ These bond-charge interactions  can be responsible for a host of interesting effects in the context of charge density waves~\cite{charge_density_waves}, ferromagnetism in itinerant electron systems~\cite{ferromagnetism}, or alternative mechanisms of superconductivity based on electron  holes~\cite{hole_supercond}, and other interesting effects~\cite{bc_phases}. However, it has been argued that the required values of $X, W$ with respect to $U,V$ to observe such effects are not likely to be realized in standard solid-state materials~\cite{bond_charge_real_materials}. On the other hand, the possibility of controlling these terms in the synthetic solids offered by   ultracold atoms in optical lattices has recently raised some  interest in the community~\cite{non_standard_review}.

From the perspective of ultracold trapped atoms, one can evaluate these bond-charge terms for the contact interaction in Eq.~\eqref{eq:originalH}.  In the Wannier basis, one can introduce the  bond operators, $B^{\sigma}_{{\bf i, j}}=f_{{\bf i},\sigma}^\dagger f_{{\bf j},\sigma}^{\phantom{\dagger}}+f_{{\bf j},\sigma}^\dagger f_{{\bf i},\sigma}^{\phantom{\dagger}},$  to account for the  state-dependent density located at the bond $({\bf i,j})$.  Then, the bond-charge terms modify the standard Hubbard Hamiltonian~\eqref{eq:hubbard} by introducing 
\beq
\label{bond_charge}
\Delta H^{(1)}_{\rm FH}=\sum_{\bf i,\alpha}\sum_{\sigma}\bigg(\frac{V_\alpha}{2} n_{{\bf i},\sigma}n_{{\bf i+e_\alpha},\overline{\sigma}}+X_\alpha B^{\sigma}_{{\bf i, i+e_\alpha}}\left(n_{\bf i,\overline{\sigma}}+n_{\bf i+e_\alpha,\overline{\sigma}}\right)+W_\alpha B^{\sigma}_{{\bf i, i+e_\alpha}}B^{\overline{\sigma}}_{{\bf i, i+e_\alpha}}\bigg),
\eeq
where we can express the different interaction strengths in the Gaussian approximation as follows
\beq
\label{extended_parameters}
V_\alpha=\sqrt{\frac{8}{\pi}}ka_{\uparrow\downarrow}E_{\rm R}\left(\frac{V_{0,x}V_{0,y}V_{0,z}}{E_{\rm R}^3}\right)^{1/4}\ee^{-\frac{\pi^2}{4}\sqrt{\frac{V_{0,\alpha}}{E_{\rm R}}}},\hspace{3ex}
X_\alpha=\sqrt{\frac{8}{\pi}}ka_{\uparrow\downarrow}E_{\rm R}\left(\frac{V_{0,x}V_{0,y}V_{0,z}}{E_{\rm R}^3}\right)^{1/4}\ee^{-\frac{3\pi^2}{16}\sqrt{\frac{V_{0,\alpha}}{E_{\rm R}}}}, \hspace{3ex} W_{\alpha}=V_\alpha.
\eeq
From these expressions, we observe that the ratios $V_\alpha/U,W_\alpha/U$, and $X_{\alpha}/U$ are vanishingly small for deep optical lattices~\cite{comment_bond_charge_bosons}. 
An interesting possibility to reach regimes $\{V_\alpha,W_\alpha,X_{\alpha}\}\sim U$, where the bond-charge interactions can lead to new phases of matter, is to consider ultracold dipolar  gases~\cite{non_standard_review} or, as we show in this section, to exploit the  PAT toolbox.

At this point, we introduce an additional modification of the standard Hubbard model~\eqref{eq:hubbard}, whereby the tunnelling is not only modified by the density at separate sites (i.e. bond-charge interaction $X_\alpha$ in Eq.~\eqref{bond_charge}), but also by the density-density correlations
\beq
\label{bond_density_density}
\Delta H^{(2)}_{\rm FH}=\sum_{\bf i,\alpha}\sum_{\sigma}\tilde{X}_\alpha B^{\sigma}_{{\bf i, i+e_\alpha}}\left(n_{\bf i,\overline{\sigma}}\hspace{0.25ex}n_{\bf i+e_\alpha,\overline{\sigma}}\right).
\eeq
Here, $\tilde{X}_\alpha$ is the strength of this tunnelling, which cannot be obtained from any two-body interaction in the Wannier basis. In fact this term would require  rather exotic three-body interactions, which cannot be realized even with ultracold dipolar  gases~\cite{non_standard_review}. These terms are  interesting in a condensed-matter context, where they appear after reducing models with hybridized   bands to single-band Hubbard models,  as occurs in intermediate-valence solids~\cite{intermediate_valence}, and the high-$T_{\rm c}$ cuprates~\cite{cuprates}. As shown below, the PAT toolbox can  control all these terms in the effective cold-atom Hamiltonian.

\subsubsection{Correlated electron-hole tunnelings: bond-ordered waves and triplet pairing}
 In this section, we  focus on a quantum simulator of the Fermi-Hubbard model with  tunable ratios of $t_{\alpha}/U, X_{\alpha}/U,  \tilde{X}_{\alpha}/U$, which  may lead to very interesting many-body effects.  Such a Fermi-Hubbard model $H_{\rm FH}=H_{\rm loc}+H^{\rm corr}_{\rm kin}+V_{\rm int}$ can be rewritten in terms of asymmetric tunnelings that are correlated to the electron/hole occupation 
\beq
\label{eq:as_tunn}
H^{\rm corr}_{\rm kin}=\sum_{{\bf i,\alpha}}\sum_\sigma \left(t_{\rm hh}^{\alpha}h_{{\bf i},\overline{\sigma}}h_{{\bf i}+{\bf e}_\alpha,\overline{\sigma}}+t_{\rm eh}^{\alpha}(n_{{\bf i},\overline{\sigma}}h_{{\bf i}+{\bf e}_\alpha,\overline{\sigma}}+h_{{\bf i},\overline{\sigma}}n_{{\bf i}+{\bf e}_\alpha,\overline{\sigma}})+t_{\rm ee}^{\alpha}n_{{\bf i},\overline{\sigma}}n_{{\bf i}{\bf e}_\alpha,\overline{\sigma}}\right)f_{{\bf i},\sigma}^{\dagger}f_{{\bf i},+{\bf e}_\alpha,\sigma}^{\phantom{\dagger}}+{\rm H.c.},
\eeq
where we have introduced tunnelings in a hole-hole background $t_{\rm hh}^{\alpha}=-t_{\alpha}$,  in an electron-hole background $t_{\rm eh}^{\alpha}=-t_{\alpha}+X_{\alpha}$, and in an electron-electron background $t_{\rm ee}^{\alpha}=-t_{\alpha}+\tilde{X}_\alpha+2X_\alpha$. 

Let us focus for simplicity on the 1D limit of Eq.~\eqref{eq:as_tunn}.  The effective cold fermion Hamiltonian~\eqref{eq:eff_H}-\eqref{eq:tunn_op}, which is obtained through the   PAT by a spin-independent moving lattice, already contains these correlated particle-hole tunnelings. For the parameter regime fulfilling~\eqref{pat_regime} with $r=2$, and setting the moving lattice relative phase to $\varphi=0$, one finds
\beq
\label{eq:tunn_rates}
t^x_{\rm hh}=-t_x{\rm J}_0(\eta),\hspace{2ex} t^x_{\rm eh}=-t_x{\rm J}_2(\eta),\hspace{2ex}  t_{\rm ee}^x=-t_x{\rm J}_0(\eta).
\eeq
Accordingly, the correlated tunneling asymmetry can be tuned by  modifying the intensity of the moving optical lattice $\tilde{V}_0$, or its frequency $\Delta\omega$, such that $\eta=\tilde{V}_0/\Delta\omega$ is varied. This effect could also be achieved through a periodic modulation of the $s$-wave scattering length as considered in other recent  schemes~\cite{cold_atoms_int_modulation_fermions,cold_atoms_double_int_modulation_fermions}. 

The 1D Fermi-Hubbard model with the asymmetric correlated tunnelling strengths~\eqref{eq:tunn_rates}    hosts a variety of  quantum many-body phases,  contrasting the situation of the standard 1D Fermi-Hubbard model, where only insulating and Luttinger-liquid phases occur. We now comment on the phases 
 that could be explored with cold atoms given the constraints imposed by the specific tunnelling rates~\eqref{eq:tunn_rates}. This model was initially studied for $t^x_{\rm ee}=t^x_{\rm hh}<t_{\rm eh}^x$, where a groundstate with spin-density-wave  order can be found for sufficiently strong repulsion at half-filling~\cite{sdw_phase}. Later on, it was realized that for $t^x_{\rm ee}=t^x_{\rm hh}>t_{\rm eh}^x$, a new density wave where  charge alternates on the bonds (i.e. {\it bond-ordered wave}) can be stabilized for not too large interactions~\cite{bow_phase}. Interestingly,  it has been recently  shown that, by modifying the filling factor,  the phase diagram of the model is much richer even for vanishing Hubbard interactions~\cite{cold_atoms_double_int_modulation_fermions}. For instance, a superconducting phase with unconventional {\it triplet pairing}  was identified. In addition to the interest of exploring these phases with  cold atoms using the above  PAT scheme~\eqref{eq:eff_H}-\eqref{eq:tunn_op},  this could serve to benchmark the proposed quantum simulator, since these results rest on very accurate and efficient analytical and numerical methods that exist in 1D. Moreover, the quantum simulator could be used to study the fate of the predicted phases for different chemical potentials~\cite{cold_atoms_double_int_modulation_fermions}, and the appearance of new ones,  as the   Hubbard repulsion is  switched on.
 
  Once the quantum simulator has been verified,  it would be very interesting to consider  the 2D Fermi-Hubbard model with  correlated tunnelings. Although the phase diagram is  to the best of our knowledge mostly unknown, the results on  the 1D model suggests that it can host  a variety of new phases with respect to the standard  2D Fermi-Hubbard model, the understanding of which  defies  analytical and numerical methods. We believe that the possibility of finding interesting phases of matter, even above the stringent temperatures to observe magnetic ordering in the 2D Fermi-Hubbard model, is certainly worth exploring.

To introduce the topic of the following subsection,  we emphasize that it is not possible to set $\tilde{X}_\alpha=0$  without making  ${X}_\alpha=0$ simultaneously, as would be required to study solely the effects of  bond-charge interactions~\eqref{bond_charge}. This also occurs for the schemes based on a  periodic modulation of the $s$-wave scattering length~\cite{cold_atoms_int_modulation_fermions,cold_atoms_double_int_modulation_fermions}. In the following subsection, we shall show that this becomes possible by introducing an additional linear  gradient in our scheme.

\subsubsection{Bond-charge interactions: hole superconductivity and $\eta$-pairing}
 In this section, we focus on a quantum simulator of the Fermi-Hubbard model with bond-charge interactions, whose importance can be controlled by tuning   the ratios of $t_{\alpha}/U, X_{\alpha}/U$. For simplicity, we focus on the 1D case, and consider the PAT by a spin-independent moving lattice~\eqref{eq:eff_H} in the presence of an additional linear gradient, which  leads to the dressed tunneling in Eq.~\eqref{eq:tunn_op_grad}. For the sake of concreteness, we consider a parameter regime fulfilling Eq.~\eqref{pat_regime} for $r=2$, and set  $\varphi=0$. By adjusting the intensity of the moving lattice to $\tilde{V}_{0}=1.56\Delta\omega$ (i.e. $\eta_\star=1.56$), such that terms like~\eqref{bond_density_density} in the effective Hamiltonian~\eqref{eq:eff_H} vanish $\tilde{X}=0$,  we obtain\beq
\label{eq:eff_H_bond_charge}
 H_{\rm eff}=\sum_{i,\sigma}(\epsilon_{i,\sigma}+\delta U_{\sigma\bar{\sigma}}i)f_{i,\sigma}^{\dagger}f_{i,\sigma}^{\phantom{\dagger}}-\sum_{i,\sigma}\left({t}f_{i,\sigma}^{\dagger}f_{i+1,\sigma}^{\phantom{\dagger}}+{\rm H.c.}\right)+\sum_{i,\sigma} X B^{\sigma}_{{ i, i+1}}\left(n_{i+1,\overline{\sigma}}-n_{i,\overline{\sigma}}\right) +\frac{1}{2}\sum_{i,\sigma}\delta U_{\sigma\overline{\sigma}}f_{i,\sigma}^{\dagger}f_{i,\overline{\sigma}}^{\dagger}f_{i,\overline{\sigma}}^{\phantom{\dagger}}f_{i,\sigma}^{\phantom{\dagger}}.
\eeq
where ${t}=t_{x}{\rm J}_{2}\left(\eta_\star\right)$, and the bond-charge interaction $X=t_{x}({\rm J}_{0}\left(\eta_\star\right)-{\rm J}_{2}\left(\eta_\star\right))$ alternates between neighboring sites. Other possible values of the moving-lattice intensity fulfilling $\tilde{X}_\alpha=0$ are $\eta_\star=\{4.89,8.29,11.53,\ldots\}$, and correspond to solutions of the equation ${\rm J}_{0}\left(\eta_\star\right)+{\rm J}_{4}\left(\eta_\star\right)-2{\rm J}_{2}\left(\eta_\star\right)=0$. Moreover, one can also change the moving-lattice detuning such that $r=\{4,6,8,\ldots\}$, and look for the solutions of ${\rm J}_{0}\left(\eta_\star\right)+{\rm J}_{2r}\left(\eta_\star\right)-2{\rm J}_{r}\left(\eta_\star\right)=0$. This will yield different values of $\eta_\star$, and allow for 
the tunability of the ratio of $X/t$, and the signs of $X, t$. This quantum simulator can explore the phenomenon of hole superconductivity~\cite{hole_supercond}, where the bond-charge interaction $0<X\ll t$ can be responsible for a superconducting phase even in the presence of a repulsive interaction $\delta U_{\uparrow\downarrow}>0$, as has been predicted using a mean-field approximation for  2D~\cite{bcs_hole_superconductivity}.   Since our effective Hamiltonian~\eqref{eq:eff_H_bond_charge} can be generalized to higher dimensions following our results in Sec.~\ref{scheme}, the quantum simulator could test the correctness of such mean-field predictions~\cite{hole_supercond}. From a broader perspective, the quantum simulator can explore  the phase diagram of the model in different regimes, such as $X>t$, and study the effects of the bond-charge alternation in Eq.~\eqref{eq:eff_H_bond_charge}

Since the residual Hubbard interactions in~\eqref{eq:eff_H_bond_charge} depend on the detuning of the photon-assisted scheme, one can also study the attractive case $\delta U_{\uparrow\downarrow}<0$. It has been  shown~\cite{bond_charge_eta} that a  bond-charge interaction $X=t$  stabilizes an $\eta$-pairing groundstate~\cite{yang} for finite attraction, which for $X=0$ only occurs for  strictly infinite interactions  $\delta U_{\uparrow\downarrow}\to\infty$~\cite{eta_hubbard_infinite}. Although our quantum simulator cannot reach the exact condition of  $X=t$ (e.g. for $\eta_\star=1.56$, $X/t\approx0.94$), it has been argued that relaxing these conditions may still host the $\eta$-pairing groundstate, even if the exact methods of~\cite{bond_charge_eta}  cannot be applied any longer. It would be very interesting to explore this possibility with the quantum simulator,   addressing the role of the finite residual gradient $\delta U_{\sigma\bar{\sigma}}$ in Eq.~\eqref{eq:eff_H_bond_charge},  the bond-charge alternation, and  the possibility of achieving a regime $X>t$  that is not feasible for the standard Hubbard model.

Before closing this section, we note  that this can also be addressed  with   hardcore bosons,  following our results in Sec.~\ref{scheme}.

\subsection{High-T$_{\rm c}$ superconductivity and itinerant ferromagnetism}
\label{tJ_model}

To introduce the topic of this section, let us consider again the Fermi-Hubbard model~\eqref{eq:hubbard} with  additional bond-charge terms~\eqref{bond_charge}. The interaction between charges localized in neighboring  bonds $W_\alpha$ leads to a couple of terms: {\it (i)} a  pair tunnelling  for fermions of opposite pseudospin $\Delta H_{\rm pt}=\sum_{\bf i, \alpha}2W_\alpha f_{{\bf i},\uparrow}^{\dagger}f_{{\bf i},\downarrow}^{\dagger}f_{{\bf i+e_\alpha},\uparrow}f_{{\bf i+e_\alpha},\downarrow}+{\rm H.c.}$, and {\it (ii)} a  direct exchange interaction $\Delta H_{\rm de}=-\sum_{\bf i, \alpha}J^{\rm de}_\alpha S_{{\bf i}}^{+}S_{{\bf i+e_\alpha}}^{-}+{\rm H.c.}$  between neighboring pseudospin excitations, where we have introduced the spin-$1/2$ operators 
\beq
S_{{\bf i}}^{+}=\frac{1}{2}f_{{\bf i},\uparrow}^{\dagger}f_{{\bf i},\downarrow}^{\phantom{\dagger}},\hspace{2ex}S_{{\bf i}}^{-}=\frac{1}{2}f_{{\bf i},\downarrow}^{\dagger}f_{{\bf i},\uparrow}^{\phantom{\dagger}},\hspace{2ex}S_{{\bf i}}^{z}=\frac{1}{2}\left(f_{{\bf i},\uparrow}^{\dagger}f_{{\bf i},\uparrow}^{\phantom{\dagger}}-f_{{\bf i},\downarrow}^{\dagger}f_{{\bf i},\downarrow}^{\phantom{\dagger}}\right).
\eeq
Regarding the direct exchange,  the possibility of observing magnetic ordering of localized pseudospins in the cold-atom scenario  would require a direct exchange  $J^{\rm de}_\alpha=4W_\alpha$ that is much larger than the attainable values~\eqref{extended_parameters}. The same occurs for transition metals, where  the direct exchange cannot explain the appearance of magnetic ordering. As first pointed out by P. W. Anderson~\cite{superexchange_anderson}, magnetic ordering can also arise as a consequence of the strong Hubbard interactions that prevent conduction. Moreover, in the presence of doping, the interplay of this tendency towards  magnetic  ordering with the correlated dynamics of holes is one of the proposed mechanisms that could explain high-$T_{\rm c}$ superconductivity~\cite{rvb_cuprates}.

In this section, we show that the effective Hamiltonian~\eqref{eq:H_eff_higher_d} for the PAT scheme with $\varphi=0$ may open a new route to study models of itinerant quantum magnetism and high-$T_{\rm c}$  superconductivity, provided that  the limit of large repulsive interactions $\delta U_{\uparrow\downarrow}$ is considered. In this regime,  the subspaces of single-occupied $\mathcal{H}_{\rm s}$ and doubly-occupied $\mathcal{H}_{\rm d}$ lattice sites become well-separated Hubbard sub-bands. The kinetic energy in Eq.~\eqref{eq:H_eff_higher_d} contains terms that act within each of these  sub-bands
\beq
\label{eq:hopping_ss_dd}
K_0=\sum_{\bf i,\alpha}\sum_{\sigma}h_{{\bf i},\overline{\sigma}}\left(-t_{\alpha}{\rm J}_{0}\left(\eta_\alpha\right)f_{{\bf i},\sigma}^{\dagger}f_{{\bf i+e_\alpha},\sigma}\right)h_{{\bf i+e_\alpha},\overline{\sigma}}^{\phantom{\dagger}}+\sum_{\bf i,\alpha}\sum_{\sigma}n_{{\bf i},\overline{\sigma}}\left(-t_{\alpha}{\rm J}_{0}\left(\eta_\alpha\right)f_{{\bf i},\sigma}^{\dagger}f_{{\bf i+e_\alpha},\sigma}\right)n_{{\bf i+e_\alpha},\overline{\sigma}}^{\phantom{\dagger}}+{\rm H.c.},
\eeq
and terms that connect the two sub-bands, such as the operator $K_{\rm s\to d}:\mathcal{H}_{\rm s}\to\mathcal{H}_{\rm d}$ expressed as
\beq
\label{eq:hopping_sd}
K_{\rm s\to d}=\sum_{\bf i,\alpha}\sum_{\sigma}n_{{\bf i},\overline{\sigma}}\left(-t_{\alpha}{\rm J}_{r_\alpha}\left(\eta_\alpha\right)f_{{\bf i},\sigma}^{\dagger}f_{{\bf i+e_\alpha},\sigma}\right)h_{{\bf i+e_\alpha},\overline{\sigma}}^{\phantom{\dagger}}+\sum_{\bf i,\alpha}\sum_{\sigma}n_{{\bf i+e_\alpha},\overline{\sigma}}\left(-t_{\alpha}{\rm J}_{r_\alpha}\left(\eta_\alpha\right)f_{{\bf i+e_\alpha},\sigma}^{\dagger}f_{{\bf i},\sigma}\right)h_{{\bf i},\overline{\sigma}}^{\phantom{\dagger}},
\eeq
or the operator $K_{\rm d\to s}:\mathcal{H}_{\rm d}\to\mathcal{H}_{\rm s}$, which can be expressed as 
\beq
\label{eq:hopping_ds}
K_{\rm d\to s}=\sum_{\bf i,\alpha}\sum_{\sigma}h_{{\bf i},\overline{\sigma}}\left(-t_{\alpha}{\rm J}_{r_\alpha}\left(\eta_\alpha\right)f_{{\bf i},\sigma}^{\dagger}f_{{\bf i+e_\alpha},\sigma}\right)n_{{\bf i+e_\alpha},\overline{\sigma}}^{\phantom{\dagger}}+\sum_{\bf i,\alpha}\sum_{\sigma}h_{{\bf i+e_\alpha},\overline{\sigma}}\left(-t_{\alpha}{\rm J}_{r_\alpha}\left(\eta_\alpha\right)f_{{\bf i+e_\alpha},\sigma}^{\dagger}f_{{\bf i},\sigma}\right)n_{{\bf i},\overline{\sigma}}^{\phantom{\dagger}}.
\eeq
We shall use this formulation to propose a quantum simulator that explores {\it Nagaoka ferromagnetism} even in the limit of finite repulsive interactions, and the full phase diagram of the $t$-$J$ {\it model} in 1D and, more interestingly, in 2D.

\subsubsection{Itinerant ferromagnetism: correlated  destruction of tunnelling and Nagaoka ferromagnetism}

For a strong Coulomb repulsion, electrons in undoped transition metals  lower their energy by displaying antiferromagnetic ordering. In the standard Fermi-Hubbard model, the origin of this effect can be traced back to the so-called super-exchange interaction, whereby an antiferromagnetic spin pattern  allows for virtual electron tunnelling between neighboring sites that lowers the kinetic energy~\cite{superexchange_anderson}. 
The situation is utterly different for ferromagnetism, where the reliability of initial mean-field predictions of large ferromagnetic regions in the phase diagram  is highly questionable~\cite{fazekas_book}. One of the few rigorous results on the existence of ferromagnetism in the Fermi-Hubbard model is due to Y. Nagaoka~\cite{nagaoka_ferro}, who showed that a single hole in a large class of  half-filled Hubbard models can lead to a fully-polarized ferromagnetic groundstate  for infinite repulsion.

The stability of this {\it Nagaoka ferromagnet} for different regimes has been a topic of  recurrent interest in the literature. For two holes~\cite{nagaoka_two_holes}, the  ferromagnet is no longer the groundstate. Nonetheless, for finite hole densities, the fully-polarized  Nagaoka ferromagnet can be stable up to a critical hole doping~\cite{nagaoka_finite_doping}. Despite  some initial  discrepancy regarding its full polarization~\cite{nagaoka_finite_doping_finite_interaction}, more recent numerical results based on quantum Monte Carlo~\cite{nagaoka_qmc} and density-matrix renormalization group~\cite{nagaoka_dmrg} agree with the above scenario~\cite{nagaoka_finite_doping}. Since the task of doping a transition metal with exactly one hole seems quite daunting, these results are crucial for an experimental realization of the Nagaoka ferromagnet. Another obstacle for the realization of a Nagaoka ferromagnet is the  requirement of infinite repulsion. In a cold-atom context, alternatives exploiting a long-range double-exchange interaction in a two-band Hubbard model~\cite{nagaoka_optical_ladder_superlattice}, or a large spin-imbalance in an optical lattice with a ladder structure~\cite{nagaoka_optical_ladder_spin_imbalanced} have been considered. 
We show below that our PAT scheme allows to access the physics of the infinite-repulsion Hubbard model, and thus Nagaoka ferromagnetism, even for finite Hubbard interactions.

Let us consider the effective kinetic energy  obtained for the 2D or 3D scheme~\eqref{eq:H_eff_higher_d}. By looking at the tunnelings in Eqs.~\eqref{eq:hopping_ss_dd}-\eqref{eq:hopping_ds}, one notices that by modifying the intensity of the moving optical lattice such that ${\rm J}_r(\eta_{\star})=0$, while ${\rm J}_0(\eta_{\star})\neq0$, we can exactly cancel the terms that do not preserve the parity in the occupation number, namely Eqs.~\eqref{eq:hopping_sd}-\eqref{eq:hopping_ds}. For concreteness, we consider  a parameter regime fulfilling~\eqref{pat_regime} for $r=2$, such that the moving lattices have the same intensity and detuning, and  $\eta_\star=\tilde{V}_{0,\alpha}/\Delta\omega_\alpha=5.135$. Accordingly, only the parity-conserving tunnelling~\eqref{eq:hopping_ss_dd} is preserved. This effect is similar to the so-called coherent destruction of tunnelling~\cite{coherent_destruction_tunneling}, where the tunnelling of electrons subjected to a periodically-modulated  force is totally suppressed for certain parameters of the force. In our case, the suppressed tunnelling is fully correlated to  a particular electron-hole background (see Fig.~\ref{fig_pat_fermions_spin_indep}), such that the effect can be understood as a {\it correlated  destruction of tunnelling}.

For large but finite Hubbard repulsion $\delta U_{\uparrow\downarrow}$, the Hubbard sub-bands are separated in energies and no term in the effective Hubbard Hamiltonian can connect them. By controlling the atomic filling factor such that $\langle n_{{\bf i},\uparrow}+n_{{\bf i},\downarrow}\rangle < 1$, all the dynamics takes place within the single-occupation subspace, and is controlled by 
\beq
\label{eq:Nagaoka}
H_{\rm eff}=\mathcal{P}_{\rm s}\sum_{{\bf i},\alpha}\sum_{\sigma}\left(-\tilde{t}_\alpha f_{{\bf i},\sigma}^{\dagger}f_{{\bf i+e_\alpha},\sigma}^{\phantom{\dagger}}+{\rm H.c.}\right)\mathcal{P}_{\rm s},
\eeq
where $\tilde{t}_\alpha=t_\alpha {\rm J}_0(\eta_\star)$, and we have introduced the Gutzwiller projector onto the single-occupied sub-band $\mathcal{P}_{\rm s}=\prod_{\bf i}(1-n_{{\bf i},\uparrow}n_{{\bf i},\downarrow})$.   Interestingly enough, the Hamiltonian~\eqref{eq:Nagaoka} corresponds to the infinitely-repulsive Hubbard model supporting Nagaoka ferromagnetism. However, in our scheme, only a finite repulsion is required to allow for the adiabatic loading of the lower Hubbard sub-band. By varying the filling factor, the stability of the ferromagnetic phase and the full phase diagram can be explored experimentally. In particular, it can serve as a benchmark of the phase diagram presented~\cite{nagaoka_dmrg}, which is based on extrapolating numerical results for ladders with increasing number of legs, and has predicted an intermediate phase-separation region between the fully-polarized Nagaoka ferromagnet and the paramagnetic phase. 
 
\subsubsection{High-T$_{\rm c}$ superconductivity: tunable $t$-$J$ and $t$-$XXZ$ models}

In the previous section, we have considered an alternative route to access the physics of the limit of infinite repulsion in the hole-doped  Fermi-Hubbard model (i.e. $\langle n_{{\bf i},\uparrow}+n_{{\bf i},\downarrow}\rangle < 1$). However,  a large but finite repulsion can also lead to very interesting physics. In this regime,  the competition  of the projected kinetic energy~\eqref{eq:Nagaoka} with the antiferromagnetic super-exchange~\cite{superexchange_anderson} leads to  the so-called $t$-$J$ {\it model}
\beq
\label{eq:tJ}
H_{tJ}=\mathcal{P}_{\rm s}\tilde{H}_{tJ}\mathcal{P}_{\rm s},\hspace{2ex}\tilde{H}_{tJ}= \sum_{{\bf i},\alpha}\sum_{\sigma}\left(-\tilde{t} f_{{\bf i},\sigma}^{\dagger}f_{{\bf i+e_\alpha},\sigma}^{\phantom{\dagger}}+{\rm H.c.}\right)+\sum_{{\bf i},\alpha}\tilde{J}\left(\boldsymbol{S}_{\bf i}\cdot \boldsymbol{S}_{\bf i+e_\alpha}-\fourth n_{\bf i}n_{\bf i+e_\alpha}\right),
\eeq
where $\tilde{t}$ is the tunnelling within the single-occupied subspace, and $\tilde{J}>0$ is the strength of the antiferromagnetic super-exchange interaction. In the case of  square lattices, this Hamiltonian~\eqref{eq:tJ} has been  considered as  the canonical effective model in the theory of the high-$T_{\rm c}$ cuprates by part of the scientific community~\cite{rvb_cuprates,review_tj,review_tj_2}. In this context,  the $t$-$J$ model arises   after 
mapping a more microscopic three-band Hubbard model of the cuprates~\cite{3_band_model_cuprates} onto a single-band one \cite{tj_cuprates,cuprates}. By using the microscopic parameters of the three-band  model~\cite{three_band_parameters}, one finds that $\tilde{J}/\tilde{t}\sim 0.3$ is the typical regime that can be realized in these materials. For the Fermi-Hubbard model  with ultracold atoms~\eqref{eq:hubbard}, one can obtain an effective $t$-$J$ model in the limit of very strong repulsion $\tilde{t}\ll U_{\uparrow\downarrow}$. Then, one finds  $\tilde{J}=4\tilde{t}^2/U_{\uparrow\downarrow}$, which cannot attain values larger than $\tilde{J}<0.4 \tilde{t}$ since the tunnelling must be at least  $\tilde{t}<0.1U_{\uparrow\downarrow}$ to allow for the perturbative process underlying the super-exchange. Therefore, the standard Fermi-Hubbard model is almost at the verge of the regime of importance for the hole-doped  cuprates $\tilde{J}/\tilde{t}\sim 0.3$~\cite{comment_hubbard_electron_doping}. At this point, it  should  be mentioned that reaching the low temperatures required to observe the effect of the antiferromagnetic super exchange at equilibrium is a great challenge that has only been achieved recently~\cite{af_exp}.  However, from a broader perspective,  the rest of the  rich phase diagram of the $t$-$J$ model cannot be explored with these experiments. A possibility to attain tunability over these parameters, while at the same time controlling and homogeneous atomic doping from half-filling, would be to consider  composite-fermion quasiparticles whose tunneling corresponds to a correlated  tunneling of a boson-fermion pair in a Bose-Fermi mixtures~\cite{tJ_eckardt}.  We show below that our PAT scheme leads to a standard fermionic $t$-$J$ model where the ratio $\tilde{J}/\tilde{t}$ can attain any desired value by controlling the intensity of the moving optical lattice. In this way, the full phase diagram of the $t$-$J$ model may become accessible to cold-atom quantum simulators. 

In our opinion, the realization

Let us consider the effective Hamiltonian obtained by the PAT scheme~\eqref{eq:H_eff_higher_d} for any dimensionality. We  assume  equal tunnelings $t_\alpha=:{t}$,  driving parameters $r_\alpha=:{r}$, and driving ratios $\eta_\alpha=:{\eta}$, in all directions $\forall\alpha$, and set $\varphi_\alpha=0$. The 
limit of strong Hubbard repulsion in this case corresponds to ${t}{\rm J}_{r}\left({\eta}\right)\ll\delta U_{\uparrow\downarrow}$, where the parity-violating tunnelings described by the terms $K_{\rm s\to d}$ and $K_{\rm d\to s}$ in Eqs.~\eqref{eq:hopping_sd}-\eqref{eq:hopping_ds} can only take place virtually.  As in the standard Hubbard model~\cite{tj_hubbard,tj_hubbard_unitary}, such virtual tunnelings can be calculated by a Schrieffer-Wolff-type unitary transformation $\tilde{H}_{\rm eff}=\ee^{\ii S}H_{\rm eff}\ee^{-\ii S}$, where $S=-\ii(K_{\rm s\to d}-K_{\rm d\to s})/\delta U_{\uparrow\downarrow}$ is responsible for eliminating the energetically forbidden tunnelings to second order in the small expansion parameter $\xi={t}{\rm J}_{r}\left({\eta}\right)/\delta U_{\uparrow\downarrow}$. Considering the commutation properties of the different operators defined so far, one finds that $\tilde{H}_{\rm eff}=H_{\rm loc}+K_0+V_{\rm int}+([K_{\rm s\to d},K_{\rm d\to s}]+[K_{\rm s\to d},K_0]+[K_0,K_{\rm d\to s}])/\delta U_{\uparrow\downarrow}+\mathcal{O}(\xi^2)$. For hole-doping about half-filling, one can project onto the single-occupancy sub-band, which yields the aforementioned  $t$-$J$ model~\eqref{eq:tJ} with an additional density-dependent next-nearest-neighbor tunnelling
 \beq
 \label{eq:eff_tj}
 \tilde{H}_{\rm eff}=H_{tJ}+\mathcal{P}_{\rm s}\Delta H\mathcal{P}_{\rm s},\hspace{2ex} \Delta H=\sum_{{\bf i},\sigma}\sum_{\alpha, n}\frac{J}{4}\left(-f_{{\bf i},\sigma}^{\dagger}n_{{\bf i+e_\alpha},\overline{\sigma}}^{\phantom{\dagger}}f_{{\bf i+u^{\alpha}_n},\sigma}^{\phantom{\dagger}}+f_{{\bf i},\overline{\sigma}}^{\dagger}f_{{\bf i+e_\alpha},{\sigma}}^\dagger f_{{\bf i+e_\alpha},\overline{\sigma}}^{\phantom{\dagger}}f_{{\bf i+u^{\alpha}_n},\sigma}^{\phantom{\dagger}}+{\rm H.c.}\right),
 \eeq
where we have introduced the effective  cold-atom parameters
\beq 
\label{tj_couplings}
 \tilde{t}={t}{\rm J}_0({\eta}), \hspace{2ex}  \tilde{J}=4{t}^2{\rm J}_r({\eta})^2/\delta U_{\uparrow\downarrow},
 \eeq
 and  the next-nearest-neighbor vectors ${\bf u}^{\alpha}_n$ (e.g. for 1D ${\bf u}^x_1=2{\bf e}_x$, for 2D $\{{\bf u}^x_1=2{\bf e}_x, {\bf u}^x_2={\bf e}_x-{\bf e}_y,  {\bf u}^y_1=2{\bf e}_y, {\bf u}^y_2={\bf e}_x+{\bf e}_y\}$). Let us note that this additional tunnelling requires that the target site is populated with a hole, as otherwise $\mathcal{P}_{\rm s}f_{{\bf i},\sigma}^{\dagger}\ket{\Psi}=0$. Hence, close to half-filling $\langle n_i\rangle \approx 1$, this term is reduced by a factor $(1-\langle n_i\rangle)/4$ with respect  to the Heisenberg super-exchange, and is typically neglected in the literature~\cite{review_tj}. 

As announced above, we have obtained an effective $t$-$J$ model where the ratio of the  coupling constants~\eqref{tj_couplings}, namely  $\tilde{J}/\tilde{t}=4\xi {\rm J}_r({\eta})/{\rm J}_0({\eta})$, can be tuned by  modifying the intensity of the moving optical lattice. Even if $\xi\leq 0.1$ in the regime of validity of the $t$-$J$ model, we can make   ${\rm J}_r({\eta})\gg {\rm J}_0({\eta})$, such that exchange coupling is not required to be much smaller than the tunnelling as in the standard Fermi-Hubbard model.  In this way, we can explore the full phase diagram of the $t$-$J$ model. 
  
In the 1D case,  theoretical predictions about the phase diagram  are supported by very accurate numerical methods~\cite{tj_1d_numerics}. Such  results could serve to benchmark the accuracy of the proposed quantum simulator, which can be prepared in a regime corresponding to a metallic phase (i.e. a repulsive  Luttinger liquid~\cite{luttinger_haldane}), a gapless superconductor (i.e. an attractive Luttinger liquid~\cite{luttinger_haldane}), or the so-called spin-gap phase~\cite{luther_emery_tj}, which consists of a quantum fluid of bound singlets with gapless density excitations, a gapped spin sector, and enhanced superconducting correlations (i.e. a Luther-Emery liquid~\cite{luther_emery}). Finally, for sufficiently strong super-exchange interactions,   the antiferromagnetic order expels the doped holes leading to separated hole-rich and hole-poor regions (i.e. phase separation~\cite{phase_separation_tj}). The richness of this phase diagram highlights  the potential of our PAT scheme, and contrasts with the standard 1D Fermi-Hubbard model where  only the repulsive Luttinger liquid can be achieved. For instance, the Luther-Emery liquid, which has eluded experimental confirmation so far, requires $\tilde{J}\approx 2.5 \tilde{t}$ and thus lies out of the range of parameters that can be obtained from the repulsive Hubbard model.
    Moreover, our quantum simulator would allow to test the numerical results~\cite{3_site_terms}  predicting the disappearance  of the phase separation in favor of  enlarged superconducting and spin-gap regions, once the next-nearest-neighbor tunnelling terms in Eq.~\eqref{eq:eff_tj} are considered.

In the 2D case, a detailed understanding of certain regions of the phase diagram is still an open problem, and the subject of considerable debate. As emphasized in~\cite{kivelson_review}, theoretical predictions are difficult to verify due to {\it (i)} the absence of controlled analytical methods, and {\it (ii)} the limitation of numerical methods to small system sizes where finite-size effects can affect the predicting power. From this perspective, the proposed quantum simulator may eventually address some of the  following open questions regarding the properties of the $t$-$J$ model. Variational methods  based on a resonating-valence-bond trial state~\cite{rvb_cuprates} (i.e. a linear superposition of all possible configurations of singlet pairs with a weight that depends on the pairing symmetry), have predicted a rich phase diagram~\cite{review_tj} with  regions of {\it (i)} ferromagnetism, {\it (ii)} $s$-wave pairing, {\it (iii)} $d$-wave pairing, {\it (iv)} coexistent antiferromagnetism  and superconductivity, and {\it (v)} phase separation.   However, this  variational approach  introduces a certain bias through the choice of the particular set of ansatzs, and this compromises its reliability leading to considerable controversy in the community~\cite{review_high_Tc}. In particular, there are contradictory predictions for  low dopings and not too large ratios of $J/t$, which turns out to be the regime of interest for the high-$T_{\rm c}$ cuprates. For instance, the results of~\cite{review_tj,tj_high_T_expansion} contradict those of~\cite{phase_separation_tj,tj_monte_carlo}, regarding the onset of  the phase separation in this low-doping region. There has also been some disagreement regarding the regions of  superconductivity predicted by  the variational approach~\cite{review_tj}, exact diagonalization~\cite{tj_d_wave_lanczos}, and quantum Monte Carlo~\cite{tj_d_wave_qmc}.
From the perspective of the  high-$T_{\rm c}$ cuprates, addressing the conflicting predictions in~\cite{stripes_tj_mf,stripes_tj_dmrg} and~\cite{review_tj,no_stripes_tj} about the existence of stripe phases (i.e. inhomogeneous charge and spin distributions) in the $t$-$J$ model is even more compelling, as these have been measured experimentally in the cuprates. If the $t$-$J$ model is to function as a canonical model of the cuprates, as advocated in~\cite{rvb_cuprates,review_tj,review_tj_2}, it is important to settle this dispute and determine if it admits stripe phases. We believe that the proposed cold-atom experiment could be helpful in this respect.

Before closing this section, let us comment on two additional possibilities for the $t$-$J$ model quantum simulator. The first, and most obvious one, is the possibility of controlling the spatial anisotropy of the parameters of the effective Hamiltonian~\eqref{eq:tJ} by simply exploiting  the dependence of the dressed tunnelings  on the lattice axes. Accordingly, the effective $t$-$J$ model becomes anisotropic, such that the anisotropy of the tunnelings  $\tilde{t}_\alpha={t}_\alpha{\rm J}_0({\eta}_\alpha)$, and the super-exchange couplings $\tilde{J}_\alpha=4{t}_\alpha^2{\rm J}_r({\eta}_\alpha)^2/\delta U_{\uparrow\downarrow}$, can be controlled through the intensities of the static and moving optical lattices along the different axes. Such anisotropy becomes especially interesting in the context of  certain cuprate ladder compounds~\cite{ladders}, which can be modeled by a number  $\ell\in\{1,\dots,n_{\ell}\}$ of one-dimensional $t$-$J$ chains that are coupled to each other by the transverse tunneling $t'$ and super-exchange coupling $J'$. This defines the so-called rungs of the $t$-$J$ {\it ladder} Hamiltonian $H_{tJ}^{\rm ladder}=\mathcal{P}_{\rm s}\tilde{H}_{tJ}^{\rm ladder}\mathcal{P}_{\rm s}$ , where
\beq
 \label{eq:tJ_ladder}
 \begin{split}
\tilde{H}_{tJ}^{\rm ladder}&= \sum_{ i,\sigma}\sum_{\ell=1}^{n_{\ell}}\left(-\tilde{t} f_{{ i},\ell,\sigma}^{\dagger}f_{{ i+1},\ell\sigma}^{\phantom{\dagger}}+{\rm H.c.}\right)+\sum_{{ i}}\sum_\ell \tilde{J}\left(\boldsymbol{S}_{ i,\ell}\cdot \boldsymbol{S}_{ i+1,\ell}-\fourth n_{ i,\ell}n_{ i+1,\ell}\right)\\
&+\sum_{ i,\sigma}\sum_{\ell=1}^{n_{\rm l}}\left(-{t'} f_{{ i},\ell,\sigma}^{\dagger}f_{{ i},\ell+1\sigma}^{\phantom{\dagger}}+{\rm H.c.}\right)+\sum_{{ i}}\sum_\ell {J}'\left(\boldsymbol{S}_{ i,\ell}\cdot \boldsymbol{S}_{ i,\ell+1}-\fourth n_{ i,\ell}n_{ i,\ell+1}\right).
\end{split}
\eeq
Such ladder Hamiltonians can be implemented in our quantum simulator if we supplement  the above scheme~\eqref{eq:eff_tj} with  additional static lattices along the $y$-axis with   commensurate wavelengths with respect to the original lattice. For instance,  combining two lattices with doubled wavelengths  $\tilde{\lambda}_y=2\lambda_y$, leads to an array of decoupled ladders with $n_{\ell}=2$ legs~\cite{bichromatic_lattice}. By adding further harmonics, one may create ladders with other numbers of legs, at least in principle (e.g. the first $n_{\ell}$ harmonics of the Fourier series of the square-wave function  yield an approximation to an array of decoupled $n_\ell$-legged ladders). Hence, the interaction-dependent PAT scheme leads to Eq.~\eqref{eq:tJ_ladder} with  tunable number of legs and Hamiltonian parameters
\beq
\label{tJ_parameters}
\tilde{t}={t}_x{\rm J}_0({\eta}_x),\hspace{2ex} \tilde{J}=\frac{4{t}_x^2{\rm J}_r({\eta}_x)^2}{\delta U_{\uparrow\downarrow}},\hspace{2ex} t'={t}_y{\rm J}_0({\eta}_y),\hspace{2ex} {J'}=\frac{4{t}_y^2{\rm J}_r({\eta}_y)^2}{\delta U_{\uparrow\downarrow}},
\eeq
together with the corresponding density-dependent next-to-nearest-neighbor tunneling~\eqref{eq:eff_tj}, typically neglected for small dopings. These $t$-$J$ ladders provide a very interesting interpolation between the  well-understood 1D $t$-$J$ model, and the more intriguing 2D case full of open questions of relevance to high-$T_{\rm c}$ superconductivity. In fact, already for $n_{\ell}=2$ legs, the doped holes tend to pair~\cite{tj_ladders_hole_pairing} developing a superconducting $d$-$wave$-like order~\cite{tj_ladders_d_wave_mf}. In addition to the phases that also occur for  the 1D $t$-$J$ model~\cite{tj_1d_numerics},  this $d$-wave superconductivity takes place in a wide  region of the phase diagram~\cite{tj_ladder_phase_diagram}, which includes  the  parameters relevant for the cuprates. Also, a very interesting even-odd effect reminiscent of the spin-gap presence/absence in the undoped system has been identified~\cite{tj_lader_even_odd}, whereby the hole $d$-wave  pairing disappears for ladders with an odd numbers of legs. We note that some of these results~\cite{tj_ladders_hole_pairing}  depend on ratios $t'/J'<1$ that cannot be reached from standard ladder Hubbard models where $t'\gg J'$. Likewise, the regime $J'\gg \tilde{t},t',\tilde{J}$, and its connection to short-range resonating valence bond states~\cite{hcb_ladder,dimer_Rvb}, cannot be reached from standard  Hubbard models. It would be very interesting to test these predictions with our quantum simulator, which allows exploring all these parameter regimes.

Let us move to the last possibility of the $t$-$J$ model quantum simulator: inducing a Heisenberg-Ising anisotropy in the super-exchange interactions. This leads to a very interesting $t$-$XXZ$ {\it model} described by the Hamiltonian  $H_{tXXZ}=\mathcal{P}_{\rm s}\tilde{H}_{tXXZ}\mathcal{P}_{\rm s}$
\beq
\label{eq:tXXZ}
\tilde{H}_{tXXZ}= \sum_{{\bf i},\alpha}\sum_{\sigma}\left(-\tilde{t}_\sigma f_{{\bf i},\sigma}^{\dagger}f_{{\bf i+e_\alpha},\sigma}^{\phantom{\dagger}}+{\rm H.c.}\right)+\sum_{{\bf i},\alpha}\left(\tilde{J}_\perp\big({S}^x_{\bf i} {S}^x_{\bf i+e_\alpha}+{S}^y_{\bf i} {S}^y_{\bf i+e_\alpha}\big)+ \tilde{J}_z\big({S}^z_{\bf i} {S}^z_{\bf i+e_\alpha}-\fourth n_{\bf i}n_{\bf i+e_\alpha}\big)\right),
\eeq
where the Ising $\tilde{J}_z$ and flip-flop $\tilde{J}_{\perp}$ interaction strengths are generally   different, such that the $t$-$J$ model is recovered when $\tilde{J}_{\perp}=\tilde{J}_z$. To achieve such an effective model with our quantum simulator, we must employ the PAT by a state-dependent moving optical lattice, which yields the effective Hamiltonian~\eqref{eq:state_depn_Heff} in 1D, and 
~\eqref{eq:2d_eff_H} in 2D. We assume  equal tunnelings $t_\alpha=:{t}$,  spin-dependent driving parameters $r_{\sigma,\alpha}=:r_\sigma$, and ratios $\eta_{\sigma,\alpha}=:{\eta_\sigma}$, in all directions $\forall\alpha$, and set $\varphi_{\sigma,\alpha}=0$. One may observe  in Fig.~\ref{fig_pat_scheme_spin_dep}{\bf (f)} that  second-order super-exchange interactions will  depend on the spin configuration. In fact, we find 
\beq
\tilde{t}_\sigma=t{\rm J}_0(\eta_\sigma),\hspace{2ex} \tilde{J}_{\perp}=\frac{4{t}^2 {\rm J}_{r_\uparrow}(\eta_\uparrow){\rm J}_{r_\downarrow}(\eta_\downarrow)}{\delta U_{\uparrow\downarrow}},\hspace{2ex} \tilde{J}_{z}=\frac{2{t}^2 \big({\rm J}_{r_\uparrow}(\eta_\uparrow)^2+{\rm J}_{r_\downarrow}(\eta_\downarrow)^2\big)}{\delta U_{\uparrow\downarrow}},
\eeq 
such that the Heisenberg-Ising anisotropy $\zeta=\tilde{J}_{\perp}/\tilde{J}_z$ can be tuned all the way from small spin quantum  fluctuations  $\zeta\to 0$ (i.e. $t$-$J_z$ model), to the isotropic regime where spin quantum fluctuations play an important role  $\zeta\to 1$ (i.e.  $t$-$J$ model). These Hamiltonians have been studied in the context of the propagation of a single hole in an antiferromagnetic matrix. For instance, in the $t$-$J_z$ model, the tunneling of the  hole leaves behind a string of flipped Ising spins that costs an energy proportional to the string length, such that the holes are almost~\cite{trugman} localized to the site where they were doped~\cite{single_hole_t_J_z}. The situation is considerably more complex as the spin fluctuations are switched on $\zeta>0$, and the $t$-$J$ model is approached $\zeta\to 1$~\cite{single_hole_fluctuations}, and some controversy regarding the limitations of the different analytical or numerical methods has been identified~\cite{single_hole_qmc}. The possibility of controlling the amount of spin fluctuations in our quantum simulator, together with the possibility of using the high-resolution optics of quantum gas microscopes~\cite{boson_gas_microscope,fermion_gas_microscope}  to create localized holes and watch them propagate in real time, opens a very nice perspective in accessing this quantum many-body effect with ultracold atoms in optical lattices.

\subsection{Synthetic dynamical Gauge fields}
\label{gauge_fields}

Gauge theories play a prominent role in several areas of modern theoretical physics, such as the strong interactions between quarks and gluons in quantum chromodynamics. Although perturbative predictions are reliable at short distances, the low-energy regime eludes a perturbative treatment and leads to numerous unsolved questions such as the phase diagram of quark matter~\cite{qcd_phase_diagram}. Therefore, a quantum simulator for quantum field theories of coupled Gauge and matter fields would indeed be very useful.

From a general perspective, this would require {\it (i)} Gauge (matter) degrees of freedom evolving under the Hamiltonian field theory $H_{\mathcal{G}}$ ($H_{\mathcal{M}}$), and {\it (ii)} a tunable  
interaction $H_{\mathcal{M}\mathcal{G}}$ introduced by the so-called minimal coupling. So far, most of the theoretical and experimental progress has considered  static/background Gauge fields $H_{\mathcal{G}}=0$, where the Gauge is fixed~\cite{goldman_review}. There is however an increasing interest in promoting this situation to a regime of dynamical Gauge fields $H_{\mathcal{G}}\neq0$, as reviewed in~\cite{wiese_review}. To preserve the Gauge symmetry, one parallels the construction of lattice Gauge theories~\cite{wilson_lgt}. In the Hamiltonian formulation~\cite{hamiltonian_lgt},  the fermionic matter field $\psi({\bf r})\to \psi_{\bf i}/ a^{d/2} $ is defined on the  sites of a $d$-dimensional lattice ${\bf r}\to {\bf r}_{\bf i}={\bf i }a$,  where  $\bf i$  is a vector of integers and  $a$ the lattice spacing, whereas the Gauge degrees of freedom are defined in terms of unitary matrices $ U_{{\bf i},{\bf j}}$. Such unitaries can be expressed in terms of a Gauge field $ U_{{\bf i},{\bf j}}=\ee^{\ii a A_{\boldsymbol{\mu}}({\bf r}_{{\bf ij}})}$ defined on the links   ${\bf r}_{{\bf ij}}=({\bf r}_{\bf i}+{\bf r}_{\bf j})/2$ of two neighboring  sites connected by  $\boldsymbol{\mu}=({\bf r}_{\bf j}-{\bf r}_{\bf i})/a$. The complete Hamiltonian for the lattice Gauge theory 
\beq
\label{lgt}
H=H_{\mathcal{G}}-\sum_{\langle {\bf i},\bf {j}\rangle}t_{\bf ij}\left(\psi_{\bf i}^\dagger \ee^{\ii a A_{\boldsymbol{\mu}}({\bf r}_{{\bf ij}})}\psi_{\bf j}^{\phantom{\dagger}}+{\rm H.c.}\right)+\sum_{\bf i}\epsilon_{\bf i}\psi_{\bf i}^{{\dagger}}\psi_{\bf i}^{\phantom{\dagger}},
\eeq
requires a particular Gauge-invariant construction of $H_{\mathcal{G}}$, and a particular choice of tunnelings $t_{\bf ij}$, and on-site energies $\epsilon_{\bf i}$, which lead to the corresponding Gauge and matter field theories in the continuum limit $a\to 0$ (e.g. minimally-coupled Maxwell and Dirac quantum field theories). Moreover, one must enforce Gauss law by considering only the physical states of a  sector of the Hilbert space. Although there are very interesting  proposals to accomplish this goal in the pure Gauge sector~\cite{lgt_pure_gauge}, and the complete lattice Gauge theory~\cite{lgt_fermions}, we will restrict to a  simpler quantum simulator of Eq.~\eqref{lgt} with a Hamiltonian $H_{\mathcal{G}}$ that is not Gauge invariant. Although  departing form the standard formulation of lattice Gauge  theories, the dynamical character of the fields brought by $H_{\mathcal{G}}\neq 0$ can lead to very interesting phenomena, which  might be still associated with a dynamical Gauge field theory within a fixed gauge~\cite{gauge_fixed_qed}. Alternatively, one can use focus on  lattice field theories that lead to interacting relativistic  quantum field theories at low energies, such as the Thirring and Gross-Neveu models~\cite{gross_neveu_cold_atoms}. 

\subsubsection{ Interacting relativistic quantum field theories: Yukawa-type couplings}

Let us focus on the 1D case, and consider the PAT of a Fermi gas by a spin-dependent moving lattice~\eqref{eq:state_depn_Heff} in a parameter regime fulfilling Eq.~\eqref{pat_regime_spin_dependent} for $r_\sigma=2$, and $\delta U_{\uparrow\downarrow}=0$. Let us note that the same can be obtained for a Bose-Fermi mixture~\eqref{eq:state_depn_Heff_bf}, provided that the hardcore constraint is considered. In order to obtain a quantum simulator of quantum matter coupled to dynamical gauge fields~\eqref{lgt},  one needs to find  a particular set of  parameters such that: ${\it (i)}$ the tunnelling amplitude~\eqref{eq:tunn_op_state_dep} becomes a simple $c$-number, and $\it (ii)$ the tunnelling phase in Eq.~\eqref{eq:state_depn_Heff} is non-vanishing  only for one of the pseudospin states. In order  to  fulfill $\it (i)$, the ratio of the moving-lattice intensities  with respect to the detuning must fulfill ${\rm J}_{0}\left({\eta}^\sigma_\star\right)={\rm J}_{2}\left({\eta}^\sigma_\star\right)$,  which can be achieved with any of the following values ${\eta}^\sigma_\star\in\{4.89,8.29,11.53,\ldots\}$. By direct substitution in Eq.~\eqref{eq:tunn_op_state_dep}, one finds that  the dressed tunnelling  strength becomes the desired  $c$-number $t_x{\rm J}_{r_\sigma\Delta n_{i+1,\overline{\sigma}}}\left({\eta}_\star^{\sigma}\right)=t_x{\rm J}_0\left(\eta_{\sigma,\star}\right)=:t_{\rm eff,}^{\sigma}$. In order to fulfill $\it (ii)$, it suffices to set the relative phase of the moving optical lattice  $\varphi_\downarrow=0$ for the pseudospins that will play the role of the Gauge field. Hence, the tunnelling phases in Eq.~\eqref{eq:state_depn_Heff} fulfill $\varphi_\downarrow=0$, and $2\varphi_\uparrow=:-\theta\neq0$, such that  the atoms with  pseudospin  $\ket{{\downarrow}}$ play the role of the dynamical Gauge field for the atoms with pseudospin  $\ket{{\uparrow}}$ (see Fig.~\ref{fig_pat_square_fermions_spin_dep}), namely
\beq
\label{H_eff_gauge}
H_{\rm eff}=\sum_{ i}\epsilon_{i,\downarrow}f^\dagger_{{ i},\uparrow}f^{\phantom{\dagger}}_{{ i},\downarrow}-\sum_it_{\rm eff}^{\downarrow}\left(f^\dagger_{{ i},\downarrow}f^{\phantom{\dagger}}_{{ i+1},\downarrow}+{\rm H.c.}\right)-\sum_i t_{\rm eff}^{\uparrow}\left(f^\dagger_{{ i},\uparrow}\ee^{\ii\theta\left(n^{\phantom{\dagger}}_{{ i+1},\downarrow}-n^{\phantom{\dagger}}_{{ i},\downarrow}\right)}f^{\phantom{\dagger}}_{{ i+1},\uparrow}+{\rm H.c.}\right)+\sum_{ i}\epsilon_{i,\uparrow}f^\dagger_{{ i},\uparrow}f^{\phantom{\dagger}}_{{ i},\uparrow}.
\eeq
This is the general result of this section. The PAT scheme has allowed us to build an effective Hamiltonian where the atoms with one of the pseudospins hop freely in the lattice and play the role of a dynamical 'Gauge' field for the atoms with the remaining pseudospin. This must be contrasted with other interesting proposals~\cite{density_dependent_gauge_fields, peierls_cold_atoms_int_modulation}, where the tunnelling Peierls phase for atoms  depends on their own density, such that the roles of matter and Gauge fields cannot be distinguished.  We believe that these type of Hamiltonians~\eqref{H_eff_gauge}, and their straightforward generalization to higher dimensions, to other lattices, to hardcore bosons,  or to slightly modified PAT schemes~\cite{comment_staggered}, will lead to several interesting many-body phenomena that deserve to be studied in further detail. To illustrate this richness, we  describe a  particular example that leads to an interesting relativistic quantum field theory in the continuum.

The effective Hamiltonian~\eqref{H_eff_gauge} corresponds exactly to the structure of the dynamical gauge field theory in Eq.~\eqref{lgt} if we make the following identifications: {\it (i)} The fermionic quantum matter is represented by the atoms with pseudospin  $\ket{{\uparrow}}$, namely $\psi_{ i}:=f_{{ i},\uparrow}$. {\it (ii)} the Gauge degrees of freedom will be some collective low-energy excitations of the atoms with pseudospin  $\ket{{\downarrow}}$. In the  half-filled 1D case, it is well-known that the particle-hole excitations of the free tight-binding Hamiltonian can be mapped onto a  pair of bosonic branches~\cite{luttinger_haldane, bosonization_delft}, which will play the role of the gauge degrees of freedom
\beq
K_{\rm eff,\downarrow}=\sum_{ i}\epsilon_{i,\downarrow}f^\dagger_{{ i},\downarrow}f^{\phantom{\dagger}}_{{ i},\downarrow}-t_{\rm eff}^{\downarrow}\sum_i\left(f^\dagger_{{ i},\downarrow}f^{\phantom{\dagger}}_{{ i+1},\downarrow}+{\rm H.c.}\right)\to H_{\mathcal{G}}=\sum_q(\epsilon_\downarrow+c_\downarrow q)\left(b_{q,{\rm R}}^\dagger b_{q,{\rm R}}^{\phantom{\dagger}}+b_{q,{\rm L}}^\dagger b_{q,{\rm L}}^{\phantom{\dagger}}\right),
\eeq
where we have introduce  the quasi momentum $q=2\pi n/L$ for $n\in\mathbb{Z}^+$,  the effective speed of light $c_\downarrow=2t_{\rm eff}^{\downarrow}a=t_{\rm eff}^{\downarrow}\lambda$, and the bosonic operators for the particle-hole excitations $b_{q,{\rm R}}$ ($b_{q,{\rm L}}$) around the right (left) Fermi point.
Moreover, we have assumed that the modifications with respect to half-filling coming from the  weak parabolic trapping, encoded in $\epsilon_{i,\downarrow}=\epsilon_\downarrow+\half m \omega_{{\rm t},x}^2X_{i}^2$, can be accounted for using a local chemical potential $\delta \mu_i$. Note that, in the continuum limit, this bosonic Hamiltonian becomes a scalar field theory with the energy zero set at $\epsilon_\downarrow$, namely
\beq
\label{massless_scalar}
H_{\mathcal{G}}=\int\frac{{\rm d}x}{2\pi}c_\downarrow\bigg(\frac{1}{2}\big(\pi(x)\big)^2+\frac{1}{2}\big(\partial_x\phi(x)\big)^2\bigg),
\eeq
where the scalar field $\phi(x)=\phi_{\rm R}(x)-\phi_{\rm L}(x)$ is expressed in terms of the inverse Fourier transform of the bosonic operators $b_{q,{\rm R}}$, $b_{q,{\rm L}}$ and their Hermitian adjoints, and $\pi(x)$ is its canonically-conjugate momentum. Although there is no Gauge invariance in the massless scalar field theory~\eqref{massless_scalar}, the dynamical bosonic field will interact with the fermionic matter, and can still lead to interesting phenomena. The particular form of the interaction brings us to the final identification {\it (iii)} the gauge field that dresses the tunnelling of the quantum matter in Eq.~\eqref{lgt} becomes $a A_{x}({ x}_{i,i+1})=\theta(n_{i+1,\downarrow}-n_{i,\downarrow})$, which amounts to the density difference of the 'Gauge' species. In order to express the gauge unitaries in terms of the bosonic  particle-hole excitations, we use
\beq
 a A_{x}({ x}_{i,i+1})=a\theta \frac{ n_{i+1,\downarrow}- n_{i,\downarrow}}{a}=a^2\theta \partial_x\left(\psi^\dagger(x)\psi(x)\right)\to A_{x}({ x})=\theta a\partial_x^2\phi(x),\hspace{1ex}U_{i,i+1}=1+\ii aA_x(x)+\mathcal{O}(a^2A_x^2),
\eeq
where we have again applied the continuum limit. To be consistent with such a limit, we note that the bare tunnelling of the fermionic quantum matter corresponds to the 1D version~\cite{ks_fermions} of the so-called Kogut-Susskind fermions~\cite{hamiltonian_lgt}. At half-filling,  one obtains the massless Dirac field theory  in the continuum limit, which minimally couples to a derivative of the scalar field
\beq
\label{eq:eff_rqft}
H_{\rm RQFT}=H_{\mathcal{G}}+\int\frac{{\rm d}x}{2\pi}\bigg(\tilde{\Psi}_{\rm L}^\dagger(x)(\delta+c_\uparrow(-\ii\partial_x-A_x(x)))\tilde{\Psi}_{\rm L}^{\phantom{\dagger}}(x)+\tilde{\Psi}_{\rm R}^\dagger(x)(\delta+c_\uparrow(+\ii\partial_x+A_x(x)))\tilde{\Psi}_{\rm R}^{\phantom{\dagger}}(x)\bigg),
\eeq
where we have introduced the fermionic field operators $\tilde{\Psi}_{\rm R}(x)$ ($\tilde{\Psi}_{\rm L}(x)$) for the right (left) moving fermions with pseudospin $\sigma=\uparrow$,  the energy difference due to the different Zeeman shifts $\delta=\epsilon_\uparrow-\epsilon_\downarrow$, and the effective speed of light $c_\uparrow=t_{\rm eff}^{\uparrow}\lambda$. We have obtained a peculiar relativistic theory of interacting quantum fields. Instead of the standard Yukawa coupling between scalar and fermionic fields $\overline{\psi}\phi{\psi}$, we get a minimal coupling with the derivative of the scalar field  $\overline{\psi}\gamma^1\partial_x^2\phi{\psi}$. Moreover, the effective speeds of light of the scalar $c_\downarrow$ and fermionic $c_\uparrow$ particles can be tuned independently. Anyhow,  scattering between the fermions will occur due to the exchange of  scalar particles, such that the cold-atom experiment could be exploited to calculate scattering amplitudes in the spirit of~\cite{scatteing_qft}. However, it is not clear how one would  create the initial incoming particles and measure the outgoing scattering probabilities in our scheme. A  simpler goal would be to study collective properties of the model~\eqref{eq:eff_rqft}. For instance, for very large $\delta$, fermion-fermion interactions will be mediated by the virtual exchange of scalar particles, such that the fermionic properties of the groundstate will be modified (e.g. correlation functions).  Increasing the  flux $\theta$ could lead to  new phases departing from the  Luttinger-liquid phase of the free Kogut-Susskind fermions. All these  questions could be studied with the proposed quantum simulator. 

\subsubsection{Correlated topological insulators: Hofstadter-type models with dynamical Gauge fields}

Topological insulators represent a family of holographic phases of matter that are insulating in the bulk and conducting at the boundaries~\cite{top_ins_review}. These states are topologically different from trivial band insulators, as the bulk is characterized by a finite topological invariant that cannot be changed unless the bulk energy gap is closed (i.e. phase transition). Another difference occurs at the boundaries, where gapless edge states are responsible for  the conductance. In the absence of a boundary energy gap, it is the chirality, or some additional symmetry of the problem, which underlies the robustness of the conductance.  This is clearly exemplified by the so-called Hofstadter model~\cite{hofstadter}, which describes fermions in a square lattice subjected to a perpendicular magnetic field, and displays the aforementioned bulk~\cite{tkkn} and edge~\cite{edge_hofstadter} properties. Similar phenomenology also arises in the Haldane model~\cite{haldane_model}, which is a topological insulator in the same symmetry class (i.e. time-reversal symmetry breaking of the integer quantum Hall effect). Remarkably, other instances of topological insulators belonging to different symmetry classes have also been found, such as the Kane-Mele model~\cite{kane_mele} built from two time-reversed copies of the Haldane model, or the time-reversal Hofstadter model~\cite{tr_hofstadter} built from two time-reversed copies of the Hofstadter model.

Paralleling the effect of interactions in the quantum Hall effect~\cite{fractional_qhe}, one expects that even more exotic phases of matter will appear when considering the effect of correlations in the above models~\cite{interacting_ti_review}. So far, the typical route to introduce such correlation effects has been to include the effect of on-site and nearest-neighbor Hubbard interactions in the Haldane~\cite{haldane_hubbard}, Kane-Mele~\cite{kane_mele_hubbard}, or time-reversal Hofstadter~\cite{tr_hofstadter_hubbard} models. These studies show that the topological features are robust to interactions, but no other exotic phases such as topological Mott insulators~\cite{top_mott_honeycomb_hubbard} (i.e. interaction-induced bulk gap and protected edge states) or topological fractional insulators~\cite{fti} (i.e. fractional excitations and protected edge states)  were found. In this section, we consider explicitly the 2D PAT by a spin-dependent moving lattice~\eqref{eq:modulation_2d},  and  discuss how this scheme may be used to explore a new type of correlation effects introduced by substituting the fixed background gauge field by a dynamical one in  the standard~\cite{hofstadter}, and  time-reversal~\cite{tr_hofstadter}, Hofstadter models. Given the  recent  realizations of both models with non-interacting cold atoms~\cite{gauge_ol}, we believe that future experiments will be able to explore the full phase diagram, and the possibility of finding more exotic phases brought  by the interactions with the dynamical Gauge field.

\vspace{1ex}
{\it (i) Hofstadter model in a dynamical Gauge field.--} Let us consider the 2D scheme~\eqref{eq:modulation_2d} leading to the effective Hamiltonian~\eqref{eq:2d_eff_H}. We generalize  our prescription for the 1D case, and set for $r^\alpha_\sigma=2$, and $\delta U_{\uparrow\downarrow}=0$. In order to make the dressed  tunnelling amplitude a $c$-number, we tune again the   ratio of the moving-lattice intensities  with respect to the detunings to ${\eta}^\sigma_x={\eta}^\sigma_y={\eta}^\sigma_\star\in\{4.89,8.29,11.53,\ldots\}$, such that ${\rm J}_{0}\left({\eta}^\sigma_\star\right)={\rm J}_{2}\left({\eta}^\sigma_\star\right)$. We can thus define the effective tunnelling amplitudes along the $x$- and $y$-axes as $t_{{\rm eff},x}^\sigma:=t_x{\rm J}_0\left(\eta_{\sigma,\star}\right)$, and $t_{{\rm eff},y}^\sigma:=t_y{\rm J}_0\left(\eta_{\sigma,\star}\right)$. Additionally, we need to control the relative phases of the moving lattices, such that only one of the pseudospins develops a non-vanishing Peierls phase  $\varphi_{\downarrow, x}=0$, but $2\varphi_{\uparrow, x}:-\theta\neq0$. Accordingly, the effective Hamiltonian~\eqref{eq:2d_eff_H} becomes a Hofstadter model in the Landau gauge for the atoms with pseudospin $\ket{{\uparrow}}$
\beq
\label{eq:dyn_hofstadter}
H_{\rm eff}=H_{\mathcal{G}}-\sum_{{\bf i}}\left(t_{{\rm eff},x}^\uparrow\ee^{\ii\theta\left(n_{{\bf i}+{\bf e}_x,\downarrow}-n_{{\bf i},\downarrow}\right)}f_{{\bf i},\uparrow}^{\dagger}f_{{\bf i}+{\bf e}_x,\uparrow}^{\phantom{\dagger}}+t_{{\rm eff},y}^\uparrow f_{{\bf i},\uparrow}^{\dagger}f_{{\bf i}+{\bf e}_y,\uparrow}^{\phantom{\dagger}}+{\rm H.c.}\right).
\eeq
Here, the synthetic Gauge field according to Eq.~\eqref{lgt} corresponds to $a{\bf A}(x_{{\bf i},{\bf i}+{\bf e}_x})=\theta(n_{{\bf i}+{\bf e}_x,\downarrow}-n_{{\bf i},\downarrow}){\bf e}_x$, and thus depends on the atoms with the remaining pseudospin $\ket{{\downarrow}}$, which evolve under  the free tight-binding Hamiltonian 
\beq
H_{\mathcal{G}}=-\sum_{{\bf i}}\left(t_{{\rm eff},x}^\downarrow f_{{\bf i},\downarrow}^{\dagger}f_{{\bf i}+{\bf e}_x,\downarrow}^{\phantom{\dagger}}+t_{{\rm eff},y}^\downarrow f_{{\bf i},\downarrow}^{\dagger}f_{{\bf i}+{\bf e}_y,\downarrow}^{\phantom{\dagger}}+{\rm H.c.}\right).
\eeq
The same can be obtained for a Bose-Fermi mixture~\eqref{eq:state_depn_Heff_bf}, provided that the hardcore constraint is considered, such that the hardcore bosons play the role of the Gauge field. 
Let us note that the dynamical Peierls phase cannot be gauged away due to its inhomogeneity, and thus corresponds to a non-trivial 'Gauge' field. Alternatively, one can compute the Wilson loop operator around a square plaquette,
\beq
W_{\circlearrowleft}=U_{{\bf i}+{\bf e}_y,{\bf i}}U_{{\bf i}+{\bf e}_x+{\bf e}_y,{\bf i}+{\bf e}_y}U_{{\bf i}+{\bf e}_x,{\bf i}+{\bf e}_x+{\bf e}_y}U_{{\bf i},{\bf i}+{\bf e}_x}=\ee^{\ii\theta(n_{{\bf i},\downarrow}-n_{{\bf i}+{\bf e}_x,\downarrow}-n_{{\bf i}+{\bf e}_y,\downarrow}+n_{{\bf i}+{\bf e}_x+{\bf e}_y,\downarrow})}=\ee^{\ii\oint_{\circlearrowleft} {\bf A}(x)\cdot{\rm d}{\bf l}}
\eeq
and check that it is a non-trivial operator for $\theta\in(0,2\pi)$. As the 'Gauge' fields commute at any position, one can regard the Hamiltonian~\eqref{eq:dyn_hofstadter} as a dynamical Abelian Hofstadter model. The quantum simulator will be able to explore this interesting model, and the fate of the  Hofstadter quantum Hall phase, in the light of the aforementioned  correlated topological insulators.

\vspace{1ex}
{\it (ii) Time-reversal Hofstadter model in a dynamical gauge field.--} We can now generalize the above construction to a time-reversal invariant situation, such as having two copies of the Hofstadter model subjected to anti-parallel magnetic fields. In our case, this is straightforward if we tune the relative phases as follows $2\varphi_{\downarrow, x}=-2\varphi_{\uparrow, x}=2\theta\neq0$.  Moreover, we also  set $t_{{\rm eff},\alpha}^\uparrow=t_{{\rm eff},\alpha}^\downarrow=:t_{{\rm eff},\alpha}$ by controlling the moving optical lattices appropriately. Accordingly, the effective Hamiltonian~\eqref{eq:2d_eff_H} becomes a doubled Hofstadter model in the Landau gauge 
\beq
\begin{split}
H_{\rm eff}=&-\sum_{{\bf i}}\left(t_{{\rm eff},x}\ee^{+\ii\theta\left(n_{{\bf i}+{\bf e}_x,\downarrow}-n_{{\bf i},\downarrow}\right)}f_{{\bf i},\uparrow}^{\dagger}f_{{\bf i}+{\bf e}_x,\uparrow}^{\phantom{\dagger}}+t_{{\rm eff},y} f_{{\bf i},\uparrow}^{\dagger}f_{{\bf i}+{\bf e}_y,\uparrow}^{\phantom{\dagger}}+{\rm H.c.}\right)\\
&-\sum_{{\bf i}}\left(t_{{\rm eff},x}\ee^{-\ii\theta\left(n_{{\bf i}+{\bf e}_x,\uparrow}-n_{{\bf i},\uparrow}\right)}f_{{\bf i},\downarrow}^{\dagger}f_{{\bf i}+{\bf e}_x,\downarrow}^{\phantom{\dagger}}+t_{{\rm eff},y} f_{{\bf i},\downarrow}^{\dagger}f_{{\bf i}+{\bf e}_y,\downarrow}^{\phantom{\dagger}}+{\rm H.c.}\right),
\end{split}
\eeq
where  atoms of any pseudospin play the role of the dynamical 'Gauge' field for the atoms with the remaining  pseudospin, and the magnetic fluxes are opposite for each pseudospin. Accordingly, the time-reversal symmetry should invert the flux $\theta\to-\theta$, and flip the pseudospin $f_{{\bf i},\uparrow}\to f_{{\bf i},\downarrow}$. Hence, the quantum simulator can explore the fate of the  fate of the  time-reversal Hofstadter quantum spin-Hall phase due to the presence of strong correlations, and dynamical effects of the Gauge field.

\section{Conclusions and Outlook}
\label{conclusions}

We have introduced an interaction-dependent photon-assisted tunneling by combining the strong Hubbard interactions of cold atoms in optical lattices with a periodic driving stemming from a moving optical lattice. This effect leads to  exotic Bose-, Fermi-, and Bose-Fermi Hubbard models with generalized  tunnelings whose strength depends on the atomic density through a Bessel function that is controlled by the intensity of the moving lattice. Additionally, the effective Peierls phase of the tunnelling also  depends on the atomic density, but  through the phase of the moving lattice. We have argued that this effect can be exploited as a flexible tool to implement a variety of quantum simulations of quantum many-body models in the context of strongly-correlated electrons and high-energy physics. In particular, our scheme may allow to explore paradigmatic models, such as the $t$-$J$ model, in regimes that were previously inaccessible  to cold-atom experiments. Such an experiment would be very relevant to test the accuracy of current approximate methods that study the phase diagram of the model, and the possibility of displaying d-wave superconductivity mediated by strong correlations. Moreover, this proposal introduces a new perspective in the realization of dynamical Gauge fields departing from the lattice Gauge theory approach, which can also lead to interesting quantum many-body models. We once again remark that such the dynamics of such synthetic Gauge fields is not itself Gauge invariant, and thus such models  cannot be related to a  lattice Gauge theory.

From an experimental point of view, the quantum simulation of the models that does not require super-exchange (e.g. phases related to the bond-charge interactions, and Nagaoka ferromagnetism), will require less stringent time-scales as the dressed tunneling can be made of the same order of magnitude as the bare one  by choosing the right parameters. Conversely, the quantum simulation of the $t$-$J$ model will require much slower timescales of the adiabatic protocol, and also stringent cooling conditions to guarantee that the adiabatic protocol is closer to idealized one. However, the dressed super-exchange~\eqref{tJ_parameters} can be made also on the same order of magnitude of the bare super-exchange if the right parameters are chosen.  For the dynamical Gauge fields, although the timescales would be favorable as one is interested in the dressed tunneling and not in second-order processes, the scheme gets complicated by the requirements of the photon-assisted-tunneling scheme (e.g. state-dependent moving lattices).

Although we have focused on cold atoms, the scheme can be applied to other setups, provided that the relevant dynamics can be described by a lattice model, and one can control the periodic driving and the strong interactions. This might be the case of trapped-ion crystals, where the local vibrations and electronic states lead to a lattice model with bosonic and pseudospin degrees of freedom, and the interactions and periodic drivings are provided by their interaction with laser beams.A similar situation arises for superconducting circuits by considering photons in arrays of microwave cavities and superconducting qubits leading to the aforementioned lattice models.

{\it Acknowledgements.-} A.B. acknowledges support from the Spanish MINECO Project FIS2012-33022, and CAM regional research consortium QUITEMAD S2009-ESP-1594. D.P. is supported by the EU Marie Curie Career Integration Grant 630955 NewFQS.



\begin{references}
\bibitem{comment_gauge}
The fermion operators are obtained after  a trivial $U(1)$ gauge transformation $f_{i,\sigma}\to{\rm exp}\big\{\ii\frac{\tilde{V}_{0}}{2\Delta\omega}\sin(\Delta k x_i^0+\varphi)\big\}f_{i,\sigma}$.
\bibitem{comment_bond_charge_bosons} 
For very deep optical lattices $U\gg\{t_{\alpha}, X_{\alpha}\}$, $X_{\alpha}$ can approach $X_{\alpha}\sim t_\alpha/10$. Taking into account the additional bosonic enhancement due to the density, the effects of the bond-charge interaction can have observable consequences on the  tunnelling dynamics of bosons~\cite{non_standard_review,bond_charge_bosons}.
\bibitem{comment_hubbard_electron_doping}
For electron doping, the additional electrons only populate the copper orbitals, and a single-band Fermi-Hubbard model may be directly an adequate description of the high-$T_{\rm c}$ cuprates.  Nonetheless, some of the most exciting effects that defy the standard Fermi liquid picture in the hole-doped cuprates (e.g. pseudo-gap phase), seem to be absent in the electron-dopped case.
\bibitem{comment_staggered}
For instance, by considering that the 'Gauge' atoms occupy the sites of a static optical lattice with a doubled wavelength, one could realise a staggered Peierls phase  that only depends on the local density ${\rm exp}\{\pm \ii \theta n_{i,\downarrow}\}$, rather than on the density difference~\eqref{H_eff_gauge}.
\end{references}
\end{document}